\documentclass[a4paper,12pt]{article}
\pdfoutput=1

\usepackage{jheppub}
\usepackage[utf8]{inputenc}
\usepackage{amsmath,amssymb,amsthm}
\usepackage{physics,mathtools}
\usepackage{tikz,float}

\makeatletter
\def\@fpheader{\vspace{0.1mm}}
\makeatother
\newcommand\tvev[1]{{\ensuremath{\left\langle{#1}\right\rangle}}}

\newcommand{\Zb}{\overline{Z}}
\newcommand{\bw}{\begin{widetext}}
\newcommand{\ew}{\end{widetext}}
\newcommand{\bea}{\begin{eqnarray}}
\newcommand{\eea}{\end{eqnarray}}
\newcommand{\be}{\begin{equation}}
\newcommand{\ee}{\end{equation}}
\newcommand{\bca}{\begin{cases}}
\newcommand{\eca}{\end{cases}}

\newcommand{\p}{\partial}
\def\et{{\widetilde E}}
\def\LL{{\cal L}}

\def\b{{\beta}}

\newcommand\lam{\lambda}
\newcommand\Lam{\Lambda}
\newcommand\om{\omega}

\newcommand\de{{\ensuremath{{\delta}}}}
\newcommand\De{{\ensuremath{{\Delta}}}}

\def\th{{\theta}}

\newcommand\ov{\over}

\newcommand\apr{{\ensuremath{{\alpha'}}}}
\def\le{\left}
\def\ri{\right}
\newcommand{\ben}{\begin{enumerate}}
\newcommand{\een}{\end{enumerate}}
\newcommand{\coo}[1]{{\mathcal{O}\left(#1\right)}}

\def\cL{{\cal L}}

\def\wlo{\widetilde{\Lambda}^{(1)}_n}
\usetikzlibrary{patterns}
\makeatletter
\let\over\@@over
\makeatother

\title{{{Black Hole Singularity from OPE}}}
\author{Nejc \v{C}eplak,$^a$}
\emailAdd{ceplakn@tcd.ie}
\author{Hong Liu,$^b$}
\emailAdd{hong\_liu@mit.edu}
\author{Andrei Parnachev,$^a$}
\emailAdd{parnachev@maths.tcd.ie}
\author{Samuel Valach$^a$}
\emailAdd{valachs@tcd.ie}

\affiliation{
$^a$School of Mathematics and Hamilton Mathematics Institute,
Trinity College, Dublin 2, Ireland}
\affiliation{$^b$Center for Theoretical Physics, Massachusetts Institute of Technology, Cambridge, MA 02139}

\abstract{
Eternal asymptotically AdS black holes are dual to thermofield double states  in the boundary CFT. It has long been known that  black hole singularities have certain signatures in boundary thermal two-point functions related to null geodesics bouncing
off the singularities (bouncing geodesics).
In this paper we shed light on the manifestations of black hole
singularities in the dual CFT.
We decompose the boundary CFT correlator of scalar operators using the Operator Product Expansion (OPE) and focus on the contributions from the identity, the stress tensor, and its products. We show that this part of the correlator develops singularities precisely at the points that are connected by bulk bouncing geodesics. 
Black hole singularities are thus encoded in the analytic behavior of the boundary correlators determined by multiple stress tensor exchanges. Furthermore, we show that in the limit where the conformal dimension of the operators is large, the sum of multi-stress-tensor contributions develops a branch point singularity as predicted by the geodesic analysis. We also argue that the appearance of complexified geodesics, which play an important role in computing the full correlator, is related to the contributions of the double-trace operators in the boundary CFT.}

\begin{document}

\maketitle


\section{Introduction and summary}
\label{sec.introduction}
 
{The AdS/CFT duality}~\cite{Maldacena:1997re, Gubser:1998bc, Witten:1998qj} provides a powerful laboratory for  understanding quantum gravity in the bulk and conformal field theory (CFT) on the boundary.
{In particular,} it relates  black holes in AdS$_{d+1}$ to the boundary CFT$_d$ at finite 
temperature~\cite{Witten:1998zw,Mal01}.  {CFT observables} can hence be used to probe the
interior of black holes and possibly the black hole singularity, see e.g.~\cite{Louko:2000tp,Kraus:2002iv,Fidkowski:2003nf,Festuccia:2005pi,Festuccia:2006sa,Hartman:2013qma,LiuSuh13a,LiuSuh13b,Grinberg:2020fdj,Rodriguez-Gomez:2021pfh,Leutheusser:2021frk,deBoer:2022zps,Horowitz:2023ury,Parisini:2023nbd,Dodelson:2023nnr}. 

In~\cite{Fidkowski:2003nf,Festuccia:2005pi}, for $d \geq 3$, signatures of black hole singularities were identified in the boundary thermal two-point functions\footnote{For simplicity, the spatial separation of the operators has been set to zero.}
\be \label{Tcor}
G (t) = \tvev{\phi(t,\vec 0)\phi(0,\vec 0)}_\beta
\ee
where  $\phi$ is a scalar CFT operator 
of conformal dimension $\Delta$ and
$\b$ is the inverse temperature.
In the limit of large $\Delta$,  $G(t)$ can be computed using 
bulk geodesics connecting the two boundary points where the operators are inserted.
In~\cite{Fidkowski:2003nf} it was found that a specific analytically continued $G(t)$, which we denote as $\hat G(t)$,  exhibits singularities of the form 
\be \label{Tcor1}
\hat G (t) \propto {1 \over (t-t_c^\pm)^{2 \De}}, \quad t \to t_c^\pm = \pm  {\beta e^{\mp {i \pi\over d}} \over 2 \sin {\pi\over d}}
 \equiv \pm {\tilde \b \over 2}- i {\beta \over 2}  , \quad \De \to \infty  \ .
\ee
On the gravity side, considering complex time of the form $t=t_L-i{\b}/{2}$, with $t_L \in \mathbb{R}$, corresponds to analysing geodesics in the two-sided eternal AdS-Schwarzschild black hole. The behavior~\eqref{Tcor1} arises because the spacelike geodesics which connect the two asymptotic regions of the black hole,  approach a null geodesic bouncing off the future (past) black hole singularity
as $t \to t_c^+$ ($t \to t_c^-$), see right of Figure~\ref{fig:Penrose}.
We will refer to the singular behavior~\eqref{Tcor1} as {the \textit{bouncing singularities}.}
 
The correlator  $G(t)$  can only have singularities  at $t =0$ and $t=- i \b$. 
To obtain $\hat G (t)$,
which exhibits bouncing singularities,
one observes~\cite{Fidkowski:2003nf} that%
\footnote{To be more precise, $\cL(t)$ is obtained by an analytic continuation from the analogous expression using the Euclidean correlator $G(\tau)$, with $t = i\tau$. 
This analytic continuation is subtle and as a consequence the limit in \eqref{eq:cllim} is not always well defined.}
\be
\label{eq:cllim}
\LL (t) =  -\lim_{\Delta \to \infty} \frac{1}{\Delta} \log G(t), 
\ee
develops a branch point singularity at $t = - i {\b \ov 2}$.
Analytically continuing $\LL(t)$ through a branch cut emanating from this branch point to the second sheet gives $\hat \LL (t)$.
The latter is given by the proper length of the bouncing geodesic and can be used to define
\be\label{ghas}
\hat G(t) \equiv e^{-\De \hat \LL (t)}\,,
\ee
which exhibits~\eqref{Tcor1}. 
In contrast,  for $t=t_L-i\b/2$, 
$G(t)$ is given by a sum of two complex geodesics which are are regular for all $t_L$.

In~\cite{Festuccia:2005pi} it was found that the same bouncing geodesics have direct signatures in the Fourier transform $G (\om)$ of~\eqref{Tcor},\footnote{For notational simplicity, we use the same notation $G(\om)$ for the Fourier transform of $G(t)$, distinguishing them only by the arguments. Similarly, below we will use the same notation $G (\tau)$ for the Euclidean analytic continuation of $G(t)$.}  albeit for large imaginary frequencies, 
\be \label{Tcor2}
G (\om)  \propto \om^{2 \De} e^{i t_c^\pm \om} , \quad \om \to \pm i \infty  , \quad \De \to \infty \ .
\ee

The results~\eqref{Tcor1} and~\eqref{Tcor2} raised a number of questions whose boundary understanding has been elusive: 

\ben 
\item What is the CFT origin of the singular behavior in~\eqref{Tcor1}? 
\item
What is CFT origin of the branch point singularity of $\LL (t)$
at $t = -{i \b} /{2}$ in the large $\De$ limit?
\item
What is the boundary interpretation of the pair of complex geodesics
which dominate the correlator at  $t =t_L- i{\b}/{ 2} $
in the large $\De$ limit?
\item What is the CFT origin of~\eqref{Tcor2}?
\item  The behavior~\eqref{Tcor1} and~\eqref{Tcor2} applies only to $d \geq 3$,
as for $d=2$ (the BTZ black hole) the black hole singularities are  orbifold singularities rather than curvature singularities.
What is the boundary origin of this difference? 
\een

In this paper we  address these questions by investigating the behavior of thermal correlators of scalar operators by performing the Operator Product Expansion (OPE) in
the boundary theory.  We will restrict our discussion to the  boundary theory on flat space.

The OPE of two  operators $\phi$ separated in Euclidean time $\tau = i t$ 
can be written schematically as 
\be \label{eyh}
\phi (\tau,0) \phi (0,0) = \sum_n C_n \tau^{\De_n - 2 \De} O_n (0), 
\ee
where $n$ collectively labels all operators $O_n$ with $\De_n$ being their conformal dimension.
The Euclidean analytic continuation of $G(t)$ can be written in terms of OPE as 
\be \label{eun}
G (\tau)  \equiv G (t=-i \tau)  = {1 \ov \tau^{2 \De}} \sum_n C_n v_n \le({\tau \ov \b} \ri)^{\De_n} , \quad \tvev{O_n}_\b = v_n \b^{-\De_n}  \ .
\ee
Note that the sums in (\ref{eyh}), (\ref{eun}) may have operators of the same conformal dimension which only differ by their spin.
In this case, each such operator contributes a separate term in (\ref{eyh}) and (\ref{eun}).

In holographic theories that are dual to Einstein gravity in the bulk,%
\footnote{For example, in the case of $\mathcal{N}=4$ Super Yang-Mills theory in $d=4$, this is the theory at large $N$ and strong coupling.}
the correlator~\eqref{eun} has a particularly simple structure
\begin{align}\label{lte}
    G(\tau)  = G_T(\tau) + G_{[\phi\phi]}(\tau)\, .
\end{align}
$G_T(\tau)$ is the contribution from OPEs involving multiple stress tensors, $i.e.$ operators of the schematic form $(T_{\mu \nu})^n$, with $T_{\mu \nu}$ denoting the stress tensor,  while $G_{[\phi\phi]}(\tau)$ is the contribution from double-trace operators formed from $\phi$, $i.e.$ operators of the schematic form $\phi (\p^2)^n \p_{i_1} \cdots \p_{i_l} \phi$. 
$ G_T(\tau)$ will be referred to as the stress-tensor sector of the correlator \cite{Karlsson:2019dbd}.
It will also be convenient to define 
\be
\label{eq:llt}
\LL_T (\tau) =  -\lim_{\Delta \to \infty} \frac{1}{\Delta} \log G_T(\tau)
\ee
and similarly $\LL_{[\phi\phi]}(\tau)$.

In \cite{Fitzpatrick:2019zqz} a scheme for computing the OPE coefficients of multi-stress tensors for holographic theories {in even spacetime dimensions} has been presented, which enables us to compute $G_T (\tau)$ explicitly as a series expansion in $\tau$.%
\footnote{See \cite{Li:2019tpf,Fitzpatrick:2020yjb,Karlsson:2022osn,Huang:2022vet} for generalisations  to other bulk theories as well as to external
operators with spin.}
The scheme is based on an ansatz that solves the bulk equations of motion order by order in the near-boundary expansion.
The resulting OPE coefficients agree, for example, with the ones determined by bootstrap and related techniques
\cite{Karlsson:2019dbd,Li:2019zba,Li:2020dqm,Karlsson:2020ghx,Dodelson:2022eiz},
as well as with those obtained by solving the Fourier transformed equation  order by order in the OPE
\cite{Dodelson:2022yvn,Dodelson:2023vrw,Esper:2023jeq,Parisini:2023nbd}\footnote{See e.g. \cite{Katz:2014rla,Manenti:2019wxs} for additional examples of the OPE analysis of finite temperature holographic correlators.}.
Other nontrivial checks include comparison with the Regge limit holographic data
\cite{Kulaxizi:2019tkd,Karlsson:2019qfi,Karlsson:2019txu,Parnachev:2020zbr}
and geodesics in asymptotically AdS spacetimes \cite{Parnachev:2020fna,Rodriguez-Gomez:2021pfh}.

By studying the OPE data of $n$-stress tensor exchanges in the regime of large $n$,  we find that, for finite $\De$, $G_T (\tau)$ contains a singularity precisely of the form~\eqref{Tcor1},\footnote{Since~\eqref{Tcor1} only applies to the large $\De$ limit, the exponents are also consistent.}
\be \label{Tcor3}
G_T (\tau) \propto {1 \ov (\tau - \tau_c^\pm)^{2 \De-{d \ov 2}}} , \quad \tau \to \tau_c^\pm = {\b \ov 2} \pm i {\tilde \b \ov 2}  = i t_c \ .
\ee
This, combined with general structure of the double-trace contribution $G_{[\phi \phi]} (\tau)$, as well as the geodesic analysis, gives a boundary picture that sheds light on various questions mentioned  earlier. 
Here we highlight the main elements: 

\ben 
\item 
Black hole singularities are encoded in the analytic structure of the stress tensor sector of thermal correlation functions. In particular, the bouncing singularity is present at  finite $\De$ and can be accessed by analytically continuing the stress tensor sector contribution $G_T (\tau)$ to complex values of $\tau$ without the need of going to a different sheet, which is needed to obtain 
$\hat \LL(t)$. 

Heuristically, we may interpret the black hole geometry as being obtained from the empty AdS by ``condensing'' multiple gravitons, which are roughly dual to multiple stress tensors on the boundary. It thus makes intuitive sense that the black hole singularities reflect the analytic behavior of the stress tensor sector. The stress-tensor sector of thermal correlators thus possesses a large degree of universality and can serve as a direct probe of the black hole structure.

\item We expect that the OPE of $G(\tau)$ is uniformly convergent for all $\De$ only for $|\tau| < {\b}/{2}$. 
This is the regime where we can take the $\De\to\infty$ limit inside the OPE and in
doing so neglect the contributions of double-trace operators. 
This is why for $|\tau| < {\b}/{2}$  the stress-tensor sector OPE fully reproduces the geodesic length~\cite{Fitzpatrick:2019zqz}.
At  leading order in the large $\De$ limit, 
\begin{align}\label{gsb}
    \lim_{\Delta\to\infty}G (\tau) =
    \lim_{\Delta \to \infty}  \bca
   G_T(\tau) &  \tau < \frac{\beta}{2}\,, \cr
   G_{[\phi\phi]}(\tau) & \tau > \frac{\beta}{2}\,, \cr
   G_T (\tau) + G_{[\phi\phi]}(\tau)  & \tau = \frac{\beta}{2}  + i t_L, \; t_L \in \mathbb{R} \,.
   \eca   
\end{align}
The branch point singularity observed in $\LL (t)$ at $t = -{i \b \ov 2}$ in the geodesic analysis can  be understood from the large $\De$ limit of $G (\tau)$. The appearance of new geodesic saddles at  this point is the  consequence of the ``sudden'' turn-on of the double trace contribution. In particular, the two terms in the last line of~\eqref{gsb} can be identified respectively with the contributions of the two complex geodesics in the gravity analysis. 

In addition, for $\tau >\beta/2$, the double-trace contribution ensures that the full correlator satisfies the  KMS condition \cite{Kubo:1957mj, Martin:1959jp}. As such, double traces are inherently linked with the periodicity in the temporal circle.

\item We observe numerically that as $\De$ is increased from a finite value to infinity, $\LL_T(\tau)$ computed using the OPE transitions between  a function which is regular at $\tau=\b/2$ and has a singularity  at \eqref{Tcor3} and a function whose radius of convergence is $\b/2$, where it develops a branch cut.
\item For $d=2$, $G(\tau)$ is known exactly from conformal symmetry, and equals the corresponding Virasoro vacuum block of the heavy-heavy-light-light correlator~\cite{Fitzpatrick:2015foa}.
All contributions come from the Virasoro descendants of the identity, which are the multi-stress operators,\footnote{We discuss this in detail in Appendix~\ref{s.kms2d}.} 
with no double-trace contributions for any $\tau$
\be
G (\tau)  = G_T (\tau), \quad d=2 \ .
\ee
Since $G(\tau)$ cannot have the bouncing singularities~\eqref{Tcor3}, neither can $G_T (\tau)$. This is related to the corresponding bulk geometry being regular. 

When the CFT is put on a circle, the double-trace contributions are needed for  thermal correlators to be periodic on the spatial circle. On the other hand, the corresponding bulk BTZ geometry develops an orbifold singularity. This suggests that the appearance of the black hole orbifold singularity is intrinsically linked with  non-trivial double-trace contributions. This is in stark contrast with the situation in $d \geq 3$, where the analytic behavior of multiple stress-tensor exchanges appears to  reflect bulk curvature singularities. 

\item The behavior~\eqref{Tcor3} does not directly say anything regarding the momentum space behavior~\eqref{Tcor2}. 
It is not clear whether~\eqref{Tcor2} survives to finite $\De$. If it does,  it is tempting to speculate a relation 
\be 
G(i \om_E)  \sim \int_{-\infty}^\infty d \tau \, e^{ i \om_E \tau} G_T (\tau) , \quad \om_E \to \pm \infty ,
\ee
although with current available information, it is not possible to be more precise.
\een 

In Section~\ref{sec:gff} we show that \eqref{gsb} also applies in 
the generalized free field (GFF) case
where the thermal correlator is given by the sum over  thermal images of the vacuum correlator, 
\be
\label{eq:GFF}
  G(\tau)^{(GFF)} = \langle \phi(\tau) \phi(0) \rangle_\beta^{(GFF)} = \sum_{m\in \mathbb{Z}} {1\over (\tau + m\, \beta)^{2\Delta}  }\,,
\ee
The sum over images ensures that the correlator satisfies the KMS 
condition \cite{Kubo:1957mj,Martin:1959jp}, $G(\tau)=G(\b-\tau)$.
Note that in \eqref{eq:GFF} the $m=0$ term is the contribution of the identity, while all other terms correspond to multi-trace contributions \cite{Iliesiu:2018fao,Alday:2020eua}\footnote{See also \cite{El-Showk:2011yvt,Fitzpatrick:2014vua,Komargodski:2012ek} for related earlier developments and
\cite{Parisini:2023nbd,Marchetto:2023xap} for  examples of recent work on  manifestations of KMS conditions in CFT.}.
However, there are also differences -- unlike in the holographic case,  $\LL(\tau)^{(GFF)}$ does not develop a 
branch point at $\tau=\b/2$.

The fact that double-trace operators do not contribute to holographic correlators for $\tau<\beta/2$  in the large-$\Delta$ limit was discussed in \cite{Fitzpatrick:2019zqz}
(see also \cite{Rodriguez-Gomez:2021pfh}).
Another situation  where the boundary correlator only receives contributions from the stress-tensor sector appeared in~\cite{Parnachev:2020fna}, where a particular near-lightcone limit was considered. 
In this limit, the  correlators receive contributions only from the leading twist multi-stress tensor operators and can be related to  spacelike one-sided geodesics.

The rest of this paper is organised as follows. In Section~\ref{sec:SC} we review the relation between spacelike geodesics in eternal Schwarzschild-AdS black holes and thermal correlation functions. For definiteness, we focus on $d=4$. In particular, we analyse the singularity associated with the bouncing geodesic. In Section~\ref{s.bhope} we then analyse the OPE coefficients associated with $n$-stress-tensor exchanges in $d=4$. By analysing the large-$n$ behavior, we find~\eqref{Tcor3}. In Section~\ref{sec:inter}, we use the OPE analysis of Section~\ref{s.bhope} to give a boundary interpretation of the results of Section~\ref{sec:SC}.  In particular, we argue that the bouncing singularities originate from the singular behaviour found at finite $\Delta$ in the OPE analysis of the stress-tensor sector.
We discuss the results in other dimensions and the generalization to the boundary CFT on a sphere,  possible resolutions of the black hole singularities from $\apr$ and $G_N$ corrections, and various future perspectives in Section~\ref{sec:Disc}.

Some more technical details are presented in the appendices. In Appendix~\ref{a.cbd} we discuss the general structure of thermal two-point functions. In Appendix~\ref{app:eomcoefz} we state the partial differential equation that we solve to determine the holographic OPE data, while in Appendix~\ref{a.sumz} we discuss the validity of approximating the OPE series expansion by an integral. We also present some additional arguments to support the main claims of this paper. The analysis of the thermal correlator in two-dimensions is performed in detail in Appendix~\ref{s.kms2d}. In Appendix~\ref{app:DeltaTau} and Appendix~\ref{app:XCorr} we analyse the subleading terms in the correlator near the bouncing singularity. In the main part of the paper, we focus only on the $d=4$ case, but we show in Appendix~\ref{a.ope6d} that the stress-tensor sector has a singularity at the complexified time $t_c$ which corresponds to the bouncing geodesic  in other dimensions as well (we explicitly checked $d=6$ and $d=8$). Finally, in Appendix~\ref{app:LTanal} we discuss in detail the lowest-twist contributions to the OPE.


\section{Black hole singularity from geodesics}
\label{sec:SC}

The main object of interest in this paper are thermal two-point correlation functions of identical operators
\begin{equation}\label{eq:DefCorrFun}
        G(t,\vec{x})=\expval{\phi(t,\vec{x})\phi(0,\vec{0})}_\beta\,,
\end{equation}
where $\beta = T^{-1}$ denotes the inverse temperature. In field theories with  holographic duals  these correlators can be calculated using the Green's function of the scalar field propagating in the
asymptotically  Anti-de Sitter black hole. When the mass, $m$, associated with the bulk scalar field, or equivalently,  the conformal dimension  of the dual field theory operator, $\Delta$,  is large, then the correlation function can be approximated by summing over classical saddles of the relevant path integral \cite{Louko:2000tp}
\begin{align}
\label{eq:Saddles}
	G(t, \vec{x}) \sim \sum_{\rm saddles}e^{-\Delta \,L}\,.
\end{align}
These saddles correspond to the geodesics in the black-hole background that connect the boundary points at which the operators are inserted and $L$ denotes the regularised proper length of the geodesic.

We will focus on the case where the bulk spacetime is a black brane in five dimensions%
\footnote{Examples in higher  dimensions are discussed in Appendix~\ref{a.ope6d}.}
with the metric 
\begin{equation}
    \label{eq:d=4BlackBranemet}
     ds^2~=~-r^2\,f(r)\,dt^2+\frac{dr^2}{r^2\,f(r)}+r^2\,d\vec{x}^2\,,
\end{equation}
where $\vec{x}=(x,y,z)$, so that  $d\vec{x}^2$  denotes the flat metric on $\mathbb{R}^3$, and 
\begin{align}
\label{eq:d=4fFunction}
    f(r) \equiv 1 -\frac{\mu}{r^4}\,.
\end{align}
Note that in most of this paper we set the radius of AdS to unity. Near the spacetime boundary, $r\to \infty$, the metric \eqref{eq:d=4BlackBranemet} reduces to that of the Poincar\'e patch of  AdS$_5$. As such, the conformal boundary is just four-dimensional Minkowski space $\mathbb{R}^{1,3}$. The parameter $\mu$ is related to the inverse temperature $\beta$ through $\mu=(\pi/\beta)^4$. For the remainder of this section, we set $\mu =1$, which means that $\beta = T^{-1}= \pi$ and the location of the black-hole horizon is given by $r_0 = 1$. The curvature singularity is at $r=0$.

\begin{figure}[t]
    \centering
    \includegraphics[scale=1]{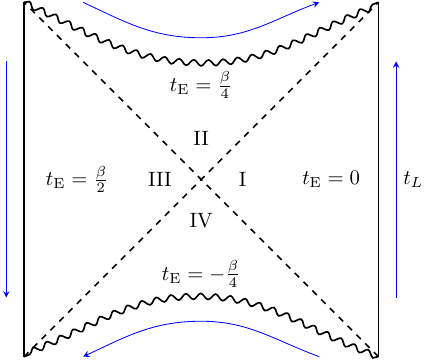}
    \caption{The Penrose diagram for the Lorentzian section of the maximally extended black-hole geometry in AdS$_5$. The spacetime separates into four regions with different constant values for the imaginary part of the time-coordinate.
    The blue arrows depict the direction of Lorentzian time, $t_L$, in each region.}
    \label{fig:Penrose0}
\end{figure}

We are interested in the maximally extended spacetime which can be described by using complexified Schwarzschild coordinates. The time coordinate $t$ then has a real and imaginary part, which we denote as\footnote{Note that we use a different sign convention for the imaginary part compared to \cite{Fidkowski:2003nf, Festuccia:2005pi}.}
\begin{align}
\label{eq:ComplexTime}
    t = t_{\rm L} - i\, t_{
    \rm E}\,, \qquad t_{\rm L} \in \mathbb{R}\,,\quad  0\leq t_{
    \rm E} < \beta\,. 
\end{align}
The Lorentzian section of this spacetime can be divided into four wedges in which the imaginary part takes  different constant values, summarised in 
Figure~\ref{fig:Penrose0}. 
Regions I and III describe two spacelike separated regions outside the  horizons. We choose the imaginary part of the time coordinate in Regions I and III to be given by $t_{ \rm E} = 0$ and $t_{ \rm E} = {\beta}/2$ respectively. Region II describes the interior of the black hole with $t_{ \rm E} = {\beta}/4$, while Region IV is the white hole region, where  $t_{ \rm E} = - {\beta}/4$. In essence, crossing a horizon corresponds to shifting the imaginary part of the time coordinate by  $\beta/4$.

\begin{figure}[t]
    \centering
    \includegraphics[width=1\textwidth]{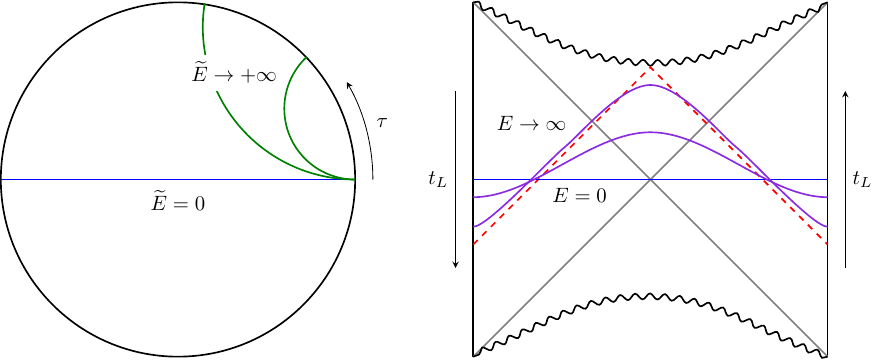}
    \caption{Diagrams for the Euclidean section (left) and the Lorentzian section (right) of the complexified black-hole spacetime.
    In both, we plot spacelike geodesics: The $\et =  E = 0$  geodesic (blue) can be drawn in both sections.
    On the left, the green curves depict geodesics with increasing  $\et$, which correspond to taking the limit $\tau \to 0$. 
    On the right, in purple, we plot real geodesics that probe the black-hole interior. As $E = - i \et \to \infty$, the geodesics become light-like signalling singular behaviour -- these are the Bouncing geodesics \cite{Fidkowski:2003nf, Festuccia:2005pi}.}
    \label{fig:Penrose}
\end{figure}

In what follows we  review the analysis of the geodesic approximation to 
the correlator $G(t_L - i\beta/2, \vec{x})$, which can be interpreted as a two-sided correlator with the operators in \eqref{eq:DefCorrFun} being inserted at different asymptotic regions of the complexified spacetime \eqref{eq:d=4BlackBranemet}. As shown in \cite{Fidkowski:2003nf, Festuccia:2005pi}, real spacelike geodesics that connect the two asymptotic regions probe the interior of the black hole, see Figure~\ref{fig:Penrose}. As they probe deeper into the interior, the geodesics  become more and more light-like with their proper length vanishing. This singular behavior is incompatible with the general properties of thermal correlation functions, which shows that such ``bouncing geodesics'' cannot contribute to the path integral.
However, in the next section we show that the  diverging behaviour related to the bouncing geodesics is still encoded in the stress-tensor sector of the OPE.

\subsection{Geodesics in the Euclidean section}

We begin by analysing geodesics in the Euclidean section of the black-brane spacetime
\begin{align}
\label{eq:EuclidMetric}
     ds^2~=~r^2\,f(r)\,d\tau^2~+~\frac{dr^2}{r^2\,f(r)}~+~r^2\,d\vec{x}^2\,,
\end{align}
where $f(r)$ is defined in (\refeq{eq:d=4fFunction}) and
$\tau \sim \tau + \beta$ is periodically identified.%
\footnote{The coordinate $\tau$ is related to the usual time coordinate through Wick rotation, $\tau = i\,t = t_{\rm E}+ i\,t_{\rm L}$, which explains our choice of sign for the imaginary part in \eqref{eq:ComplexTime}. In this subsection, we take $\tau \in \mathbb{R}$, while we will extend it to full complex space in subsequent sections.}
We can use the symmetries of the metric to reduce the problem to the motion in a one-dimensional effective potential. In general, we can introduce the energy, $\et$, associated with the time-translation invariance, and linear momenta $P_i$, related to the $\mathbb{R}^3$ isometry. In the main text we limit ourselves to the case with $x=0$, so that all linear momenta are set to zero.\footnote{The case with non-vanishing spatial separation is considered in Appendix~\ref{app:XCorr}.}

Let the geodesic be parameterised by an affine parameter $s$ and denote the derivative with respect to this parameter with a dot, for example $\dot{\tau}(s)$.
The energy of a geodesic is given by
\begin{align}
    \et = r^2\,f(r)\,\dot \tau\,,
\end{align}
defined in such a way that $\dot \tau >0$ for $\et>0$.
The condition that the geodesic is everywhere spacelike can be rearranged into 
\begin{align}
\label{eq:TurnEuclid}
    \dot r^2 = r^2\,f(r) - \et^2\,.
\end{align}
Geodesics in the Euclidean section are pictured on the left of Figure~\ref{fig:Penrose}.
They start at $r= \infty$ and probe the space up to a minimal value, $r_t$, which we call the turning point, before returning to the asymptotic boundary.
The turning point is given by the largest real root at which \eqref{eq:TurnEuclid} vanishes
\begin{align}
\label{eq:EuclideanTurning}
    r_t^2 = \frac12\left(\et^2+ \sqrt{\et^4+4}\right)\,.
\end{align}

The time difference between the endpoints of a geodesic  is given by \begin{align}
\label{eq:TauEuclid}
    \tau &\equiv \tau_f-\tau_i = 2 \int_{r_t}^{\infty}\frac{\et\,dr}{r^2\,f(r)\,\sqrt{r^2\,f(r)- \et^2}}\nonumber\\
    &= \frac12 \log\left(\frac{\et^2 +2\,\et+2}{\sqrt{\et^4+4}}\right)- \frac{i}{2}\log\left(\frac{\et^2 +2\,i\,\et-2}{\sqrt{\et^4+4}}\right)\,.
\end{align}
Let us focus on the behaviour at high energies, $\et \to \infty$. 
In this limit, the turning point goes toward the asymptotic boundary -- such geodesics probe only the asymptotic region of space.
To see how the time difference scales with the energy, we expand \eqref{eq:TauEuclid}
\begin{align}
\label{eq:EofTauExp}
    \tau = \frac{2}{\et}-  \frac{8}{5\,\et^5}+ \coo{\et^{-9}}\,, \quad \Longrightarrow \quad \et = \frac{2}{\tau }-\frac{\tau ^3}{10}+ \coo{\tau^{7}}\,,
\end{align}
meaning that as the energy is increased, the time difference goes to 0.

The regularised proper length of the geodesic is given by
\begin{align}
\label{eq:ProperLengthEuclid}
    L &= 2\lim_{r_{\rm max}\to \infty}\left[\int_{r_t}^{r_{\rm max}}\frac{dr}{\sqrt{r^2\,f(r) -\et^2}}- \log r_{\rm max}\right]= \frac12\,\log\left(\frac{16}{\et^4+4}\right)\,,
\end{align}
where $r_{\rm max}$ is the UV cut-off length \cite{Festuccia:2005pi}. By expressing the energy as a function of $\tau$ we find a logarithmic divergence at the origin
\begin{align}
\label{eq:OneSidedProperLength}
    L = 2\,\log\tau - \frac{\pi^4}{40}\left(\frac{\tau}{\beta}\right)^4 - \frac{11\,\pi^8}{14400}\left(\frac{\tau}{\beta}\right)^8+ \coo{\tau^{12}}\,,
\end{align}
where we have reinstated the appropriate units, using $\beta/\pi = 1$. When $\Delta \gg 1$, the sum over geodesics gives an approximation to the thermal correlation function.
And by exponentiating \eqref{eq:OneSidedProperLength} we get the contribution from the spacelike geodesic connecting two points on the Euclidean time circle
\begin{align}
    \label{eq:NearCoincidentCorrelator}
   e^{-\Delta\,L} =\frac{1}{\tau^{2\Delta}}\left[1+
    \frac{\Delta\,  \pi^4}{40}\,\left(\frac{\tau}{\beta}\right)^4
    +\frac{\left(9 \Delta ^2+22 \Delta \right) \pi
   ^8}{28800}\,\left(\frac{\tau}{\beta}\right)^8+
   \coo{\tau^{12}}\right]\,.
\end{align}
In the next section we will show that this result is completely reproduced by only the stress-energy sector in the large-$\Delta$ limit.

\subsection{Contributions from different saddles} \label{sec:geo}

The expressions~\eqref{eq:TauEuclid} and~\eqref{eq:ProperLengthEuclid} can be analytically continued to complex $\widetilde E$ to obtain geodesics in the real section of Schwarzschild spacetime. In particular, to obtain two-sided correlation functions, we need to consider 
\be \label{tLd}
\tau = {\pi \ov 2} + i t_L , \quad t_L \in \mathbb{R} \ .
\ee 
For this purpose, expanding~\eqref{eq:TauEuclid} around $\widetilde E =0_+$, we find 
\be \label{deo}
\tau - {\pi \ov 2} =  - {\widetilde E^3 \ov 6} + O(\widetilde E^5), \qquad L = {\log[2]} - {\widetilde E^4 \ov 8} + O(\widetilde E^8)\,,
\ee
which implies that $L (\tau)$ has a branch point singularity of the form $(\tau-{\pi \ov 2})^{4 \ov 3}$ at $\tau = {\pi \ov 2}$. 
Solving $\widetilde E$ in terms of $\tau$ we find three branches of solutions 
\be 
\widetilde E_0 = (6r)^{1 \ov 3}\,  e^{i \th \ov 3} 
 , \quad \widetilde E_1 = \widetilde E_0 \,e^{-{2 \pi i \ov 3}} , \quad 
\widetilde E_2 = \widetilde E_0\, e^{-{4 \pi i \ov 3}}, \quad {\pi \ov 2} - \tau \equiv r \,e^{i \th}  \ .
\label{eq:branches}
\ee
For ${\pi \ov 2} - \tau  > 0$ (i.e. $\th =0$), $\widetilde E_0$ branch should be chosen. To analytically continue  $L (\tau)$ to
$\tau$ given by~\eqref{tLd}, we decrease $\th$ from $0$ to $-{\pi \ov 2}$ (for definiteness take $t_L > 0$). A careful analysis~\cite{Fidkowski:2003nf}\footnote{See Section 3.4 of the paper. While the discussion in Section 3.4  was phrased as a model, it can be justified using Section 4.4 of~\cite{Festuccia:2005pi}.}
shows that in the large $\De$ limit:
\begin{enumerate}
    \item For  $\th  \in (-{\pi \ov 8},0]$, $\widetilde E_0$ is the only saddle contributing to $G(\tau)$. 
    \item For $\th \in (- {\pi \ov 2}, -{\pi \ov 8} ]$, $\widetilde E_0$ is the dominant saddle, but now $\widetilde E_1$ also contributes as a subdominant saddle, $i.e.$  
\be \label{enq}
G (\tau) \sim e^{- \De L (\widetilde E_0)} + e^{- \De L (\widetilde E_1)} \ .
\ee
    \item  At $\th = -{\pi \ov 2}$, $i.e.$ for two-sided geodesics, the two terms in~\eqref{enq}  have the same norm and thus contribute equally. Both $\widetilde E_0$ and $\widetilde E_1$ are complex, corresponding to complex geodesics in the black hole spacetime.\footnote{See \cite{Chapman:2022mqd,Aalsma:2022eru} for recent discussion of complexified geodesics in the de Sitter context.}
    \item  For  $\th \in [-{7 \ov 8}\pi, -{\pi \ov 2})$, we still have~\eqref{enq}, but now $\widetilde E_0$ contribution is subdominant. 
    \item For $\th \in [-\pi, -{7 \ov 8}\pi)$ only $\widetilde E_1$ contributes. In particular, for $\th = -\pi$, $i.e.$ $\tau > {\b}/{2}$, we have $\widetilde E_1$ becomes real, negative, and 
\be 
-{1 \ov \De} \log G (\tau) =   L (\widetilde E_1 (\tau)) =  L (\widetilde E_0 (\b-\tau)) =-{1 \ov \De} \log G (\b-\tau), \quad \tau \in \le({\b \ov 2}, \b \ri)  \ .
\ee 
This shows that $G(\tau)$ in the large $\Delta$ limit satisfies the KMS condition. 

\item The saddle corresponding to $\widetilde E_2$ does not lie on the steepest descent contour of  
the integral to obtain $G(\tau)$~\cite{Fidkowski:2003nf,Festuccia:2005pi}, and 
never contributes to the correlator.

\end{enumerate}

\subsection{Bouncing geodesics}
\label{ssec:bounce}

The $\widetilde E_2$ branch in \eqref{eq:branches}, which never contributes to the correlator, corresponds to the two-sided real geodesics presented in the right plot of Figure~\ref{fig:Penrose}. More explicitly, for $\th = - {\pi \ov 2}$ we have 
\begin{equation}
    \widetilde E_2  = i\,E\,,\qquad E \in \mathbb{R}_+ \, ,
\end{equation}
The turning point is then given by%
\footnote{The analytic structure of the turning point as a function of the energy,  $r_t(\et)$,  is discussed in \cite{Festuccia:2005pi}. This function has branch cuts in the complex plane that can be associated to quasinormal modes of the black hole. Then one can define $r_t(\et)$ in the full complex $\et$ plane by starting from \eqref{eq:TurnEuclid} for $\et \in \mathbb{R}$ and analytically continue through the origin. Physically this can be thought of as following the geodesics in the Euclidean section as the real part of the energy is decreased from  infinity to 0. At this point the geodesic crosses the cap or equivalently traverses the double-sided Lorentzian section (see the two geodesics in blue in Figure~\ref{fig:Penrose}). One then increases the imaginary part of the energy, $E$, causing the geodesics to probe the region behind the horizon.}
\begin{align}
\label{eq:P=0Turning}
    r_t^2 = \frac12\left(\sqrt{E ^4+4}-E ^2\right)\,.
\end{align}
One can see that the turning point is inside the horizon for $E ^2>0$. In fact, as $E \to \infty$, the turning point approaches the origin as $r_t \sim 1/E$, meaning that in this regime the geodesics probe the region near the singularity, see right of Figure~\ref{fig:Penrose}.

The time difference and the proper length of the two-sided geodesics corresponding to $\widetilde E_2$ can be obtained by taking $\et = i\,E$ in \eqref{eq:TauEuclid} and \eqref{eq:ProperLengthEuclid}
\begin{subequations}
\label{eq:LorentzianIntegrals}
    \begin{align}
    \label{eq:deltatRes1}
         \tau &=  \frac{\pi}{2}\left(1+i\right) +\frac{1}{2}\log\left(\frac{E ^2 -2i\,E  -2}{\sqrt{4+ E ^4}}\right)- \frac{i}{2}\log\left(\frac{E ^2 +2\,E  +2}{\sqrt{4+ E ^4}}\right)\,,\\
        \label{eq:cLRes1}
        \hat \LL & \equiv L (\widetilde E_2)  = \frac12\log{\left(\frac{16}{E ^4+4}\right)}\,.
    \end{align}
\end{subequations}
which matches the known results \cite{Fidkowski:2003nf, Festuccia:2005pi}. We have also identified $\hat \LL$, which was introduced around~\eqref{ghas}, as the geodesic distance associated with $\tilde E_2$. 
We keep the constant time shift explicit---the real part corresponds to the shift of $\tau = \beta/2$ which comes from the spacelike geodesic crossing two horizons as it goes from the I to the III patch in the Lorentzian section of the spacetime.

Denote the  time shift observed above as
\begin{align}
\label{eq:CriticalTauDef}
    \tau_c \equiv \frac{\pi}{2}(1+i) = \frac{\pi}{\sqrt{2}}\,e^{\frac{i\,\pi}{4}}\,,
\end{align}
and expanding \eqref{eq:deltatRes1} in large $E $ gives
\begin{align}
\label{eq:LargeEdeltat}
\tau=  \tau_c - \frac{2\,i}{E } + \frac{8\,i}{5\,E ^5}  + \coo{E ^{-9}}\,,
\end{align}
meaning that as $E \to \infty$, and  the spacelike geodesics become increasingly null-like, $\tau \to \tau_c$. Let
\begin{align}
    \label{eq:Deltatdef}
    \delta \tau \equiv \tau_c-\tau\,,
\end{align}
and perturbatively invert \eqref{eq:deltatRes1} to express $E $ as a function of $\delta \tau$
\begin{align}
    \label{eq:Eoft}
    E= \frac{2i}{\delta \tau} - \frac{i}{10}\,(\delta \tau)^3+ \coo{(\delta t)^{7}}\,.
\end{align}
One can then insert this into the expression for proper length and again find a logarithmic divergence, only now as $\tau \to \tau_c$
\begin{align}
    \label{eq:PropLenDeltat}
    \hat \cL = 2 \log\delta \tau + \frac{\pi^4}{160}\,\left(\frac{\delta \tau}{\tau_c}\right)^4+\coo{\delta \tau^{8}}\,,
\end{align}
where we have reinstated the units using $\pi^4 = \beta^4 = - 4\,\tau_c^4$. We note that the first correction to the logarithmic divergence appears at order $(\delta \tau)^4$.
We then find 
\begin{align}
\label{eq:GeoCOrrSmallDeltaTauExp}
   \hat G \equiv   e^{-\Delta \,\hat \cL}
    = \frac{1}{(\delta \tau)^{2\Delta}}\Bigg[1 -\frac{\Delta\,\pi^4}{160}\,\left(\frac{\delta \tau}{\tau_c}\right)^4 + \coo{\delta \tau^{8}}\Bigg]\,,
\end{align}
which exhibits the singular behavior~\eqref{Tcor1}.


\section{Black hole singularity from OPE}
\label{s.bhope}

In this section we holographically extract the OPE coefficients of the $n$-stress tensor contributions to the thermal correlator.
We then find the large-$n$ behaviour of these coefficients, which allows us to resum the stress-tensor sector of the correlator near the radius of convergence of the OPE.
In $d=4$, for finite $\Delta$ we find singularities in the complex $\tau$-plane, located at $\tau_c=\frac{\beta}{\sqrt{2}}e^{i\frac{\pi}{4}+ik\frac{\pi}{2}}$ for $k\in\mathbb{Z}$.
These correspond to the singularities associated with bouncing geodesics discussed in the previous section and are direct signatures of the black hole singularity. 
We further show that these bouncing singularities disappear in the large $\Delta$ limit, where we recover the branch point singularity at $\tau = \beta/2$. This is consistent with the geodesic analysis performed in the previous section.

\subsection{Stress-tensor sector of the correlation function}
\label{ss.holo}

Consider the scalar two-point function at finite temperature $T = \beta^{-1}$
\begin{equation}\label{eq:DefCorrFunagain}
        G(\tau,\vec x)=\expval{\phi(\tau,\vec x)\phi(0,0)}_\beta\,.
\end{equation}
Let the CFT be holographically dual to a  planar AdS-Schwarzschild black hole in $(d+1)$-dimensions and the scalar operators be dual to minimally coupled scalar fields in the bulk.
We again use the black hole metric in the Euclidean signature \eqref{eq:EuclidMetric}
    \begin{equation}\label{e.bhg}
         ds^2~=~r^2\,f(r)\,d\tau^2~+~\frac{dr^2}{r^2\,f(r)}~+~r^2\,d\vec{x}^2\,,
    \end{equation}
with  $\vec{x}=(x,y,z)$,  and $f(r)=1-\frac{\mu}{r^d}$.
In this section we keep the parameter \mbox{$\mu=(4\pi/d\,\beta)^d$} explicit. 
The equation of motion for the minimally coupled scalar
    \begin{equation}\label{e.weworig}
        (\Box-m^2)\phi=0\, ,\qquad m^2=\Delta(\Delta-d)
    \end{equation}
can be solved by a near-boundary expansion \cite{Fitzpatrick:2019zqz}.
We first perform a coordinate transformation $(\tau,\vec{x},r)\rightarrow(w,\rho,r)$
    \begin{equation}
    \rho^2=r^2\,\vec{x}^2\,,\qquad w^2=1+r^2(\tau^2+\vec{x}^2)    \,,
    \end{equation}
and write
    \begin{equation}\label{e.anz1}
        \phi(w,\rho,r)=\left(\frac{r}{w^2}\right)^\Delta \psi(w,\rho,r)\ ,
    \end{equation}
where $(r/w^2)^\Delta$ is the solution in the pure AdS space. 
The equation of motion \eqref{e.weworig} then reduces to a differential equation for $\psi(w,\rho,r)$, whose explicit form in arbitrary dimension $d$ is given in Appendix~\ref{app:eomcoefz}.
Using the standard AdS/CFT dictionary, the boundary correlator can be obtained by taking the limit
    \begin{equation}\label{eq:LimGen}
        G(\tau,\vec{x})
        =\frac{1}{(\tau^2+\vec{x}^2)^\Delta}\lim_{r\rightarrow\infty}\psi\,.
    \end{equation}
Determining the function $\psi$ using the bulk equation of motion determines the boundary correlation function.

There are two types of contributions to $\psi$ -- in terms of the CFT description, one corresponds to the stress-tensor sector, while the other contains the contributions of the double-trace operators.
We restrict ourselves to non-integer conformal dimensions, %
in which case the two sectors decouple%
\footnote{
For integer $\Delta$ the two sectors mix and situation becomes more subtle \cite{Fitzpatrick:2019zqz}.
}
    \begin{equation}
        \psi=\psi_{T}+\psi_{[\phi\phi]}\,,
    \end{equation}
where $\psi_{T}$ denotes the stress tensor sector, which also includes the contribution dual to the identity operator, while $\psi_{[\phi\phi]}$ denotes the double-trace contributions. We focus on the stress tensor sector. Its near-boundary expansion is given by \cite{Fitzpatrick:2019zqz}
    \begin{equation}
    \label{e.theansatz}
        \psi_{T}=1+\sum_{i=1}^\infty\sum_{j=0}^i\sum_{k=-i}^{\frac{d\,i}{2}-j}a_{j,k}^i\frac{\rho^{2j}w^{2k}}{r^{d\,i}}\ ,
    \end{equation}
where 1 corresponds to the contribution of the identity operator. By inserting this ansatz   into the equation of motion of the bulk scalar and expanding to arbitrary order in $1/r$, one is able to determine the coefficients $a_{j,k}^i$. Most importantly, through the dictionary \eqref{eq:LimGen}, this large-$r$ expansion on the bulk side systematically maps to the OPE of the boundary correlator, which for $\vec{x}=0$ reads%
\footnote{See Appendix \ref{a.cbd} for more details.}
    \begin{equation}\label{e.resumL}
G_T(\tau)=\frac{1}{\tau^{2\Delta}}\sum_{n=0}^\infty\Lambda_n\left(\frac{\tau}{\beta}\right)^{d\,n}\,,
    \end{equation}
    where the subscript in $G_T(\tau)$ denotes that this is only the 
stress-tensor sector of the full correlator, $G(\tau)$.%
\footnote{With an abuse of nomenclature, we will still refer to $G_T(\tau)$ as the correlator. The subscript should serve as a reminder that it is just the stress-tensor sector contribution.} Through this expression we are able to determine the stress-tensor contributions to $\Lambda_n$ using the near-boundary expansion of $\psi_{T}$.

It is important to emphasise that using this procedure one cannot determine the full correlation function but only the stress tensor sector. Namely, double trace operators are sensitive to the near-horizon behaviour, hence their OPE data cannot be determined from the near-boundary analysis that is central to this method \cite{Fitzpatrick:2019zqz}. We also want to stress that this method utilises two expansions in large-$r$: In \eqref{e.theansatz} one first uses such an expansion to solve the bulk equations of motion followed by a second $r\rightarrow\infty$ limit to obtain the correlation function \eqref{eq:LimGen}. This double-$r$ limit is crucial, but it makes the method conceptually less transparent. Nonetheless, we consider the OPE data obtained from this method to be exact to all orders in $\Delta$. While we are currently lacking a definite proof that this is the case, the results of this method were shown to be consistent with those obtained
by alternative techniques.

In what follows, we focus on the correlation functions where the operators are inserted at the same spatial points, with correlators at $\vec{x} \neq 0$ considered in Appendix~\ref{app:XCorr}.

\subsection{OPE coefficients and the KMS condition}
\label{ss:NCLim}

\begin{figure}
    \centering
    \includegraphics[width=\textwidth]{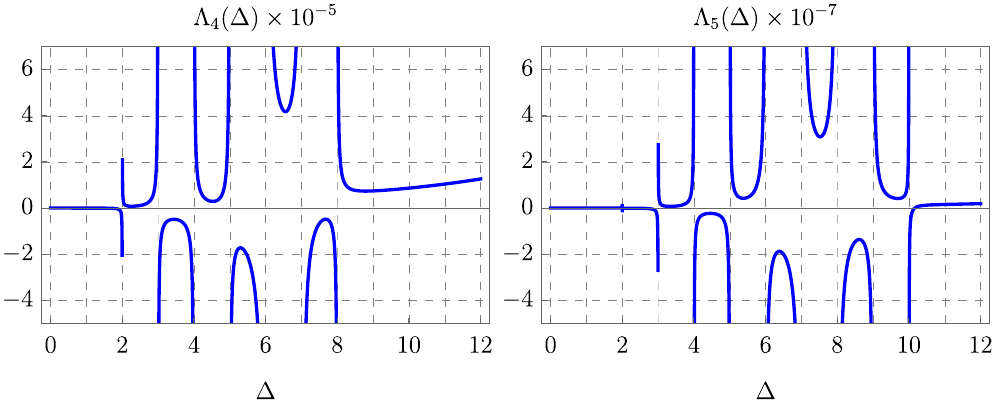}
    \vspace{-0.5cm}
    \caption{Plots for $\Lambda_4$ and $\Lambda_5$ as functions of $\Delta$. 
    These OPE coefficients have poles at $\Delta=2,3,\ldots,2n$ and are regular for $\Delta >2n$.}
    \label{fig:l4l5}
\end{figure}

Let us now set $d=4$. For small values of $n$, one can efficiently calculate the $\Lambda_n$ as explicit functions of $\Delta$. After that, calculating $\Lambda_n$ for general $\Delta$ becomes too time-demanding, so we first fix $\Delta$ to a certain value and only  then calculate $\Lambda_n$. Namely, as $n$ grows, these coefficients become more complicated functions of $\Delta$. For example, the first few terms in the small $\tau$ expansion of the correlator are given by
{\small{
\begin{equation}
\label{eq:NCCorr}
    G_T(\tau) \approx \frac{1}{\tau^{2\Delta}}\!\left[1 + \frac{\pi^4\,\Delta}{40}\left(\frac{\tau}{\beta}\right)^4  + \frac{\pi ^8 \,\Delta  \left(63 \Delta ^4-413 \Delta ^3+672 \Delta ^2-88 \Delta
   +144\right)}{201600 (\Delta -4) (\Delta -3) (\Delta -2)}\left(\frac{\tau}{\beta}\right)^8+ \ldots\right].
\end{equation}}}%
The expressions for $\Lambda_n$ with higher $n$ follow the same pattern as $\Lambda_2$ above (see Figure~\ref{fig:l4l5} for $\Lambda_4$ and $\Lambda_5$ as functions of $\Delta$) and 
can be schematically written as 
\begin{align}
\label{eq:LambdanStructure}
    \Lambda_n \sim \Delta\,(\Delta-1)\,\Gamma(\Delta-2n)\,h_n(\Delta)\,,
\end{align}
where $h_n(\Delta)$ is a polynomial function. 

\begin{figure}[t]
    \centering
    \includegraphics[width=\textwidth]{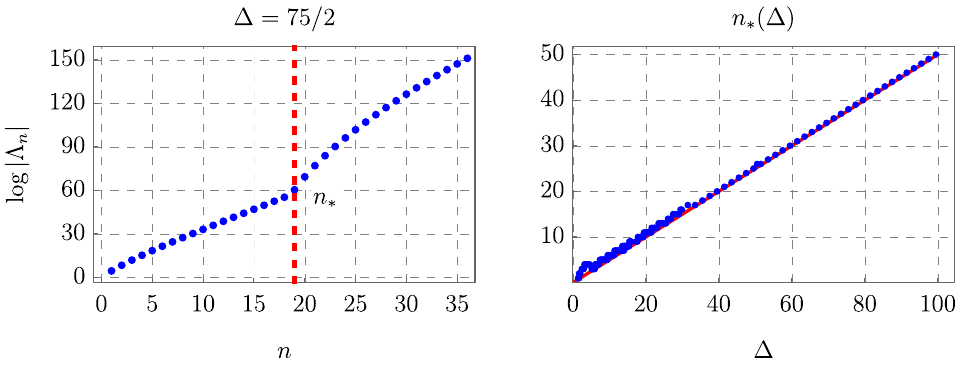}
    \vspace{-0.5cm}
    \caption{On the left we plot the values of $\log|\Lambda_n|$ for $\Delta = \frac{75}2$. We see that at $n_* = 19$, there is a change of behaviour of the OPE coefficients. 
    On the right, we plot the behaviour of $n_*$ as a function of $\Delta$ and find that $n_* = \Delta/2$ (red). }
    \label{fig:nstar}
\end{figure}

Fixing $n$ and varying $\Delta$, we can distinguish two regimes: $\Delta \leq 2n$, where we find poles at $\Delta=2,3,\ldots,2n$, and $\Delta >2n$, where the OPE coefficients have no poles in $\Delta$.
For fixed $\Delta$, the OPE coefficients in these two regimes behave differently, which can be seen in Figure~\ref{fig:nstar}. We see a significant change in the behaviour  at a particular value $n=n_*$.
Analysing the dependence of this cross-over point on the conformal dimension, we find that $n_* = \Delta/2$,
\footnote{In practice, we have taken $n_*$ to be the lowest integer greater than $\Delta/2$, which is why all points lie just above the $\Delta/2$ line on the right plot in Figure~\ref{fig:nstar}.}
which is exactly where we find the last pole in $\Lambda_n$.

That the cross-over happens at this precise value of $n$ should not be surprising. Namely, recall from \eqref{e.resumL} that the conformal dimension of the $(T_{\mu\nu})^n$ multi-stress tensor exchange in $d$ dimensions is $\Delta_{T^n} = n\,d$, while those for double-trace operators have $\De_m = 2 \De + 2m$, $m=0, 1, 2, \ldots$. They can mix when 
\mbox{$\De_{T^n} = \De_m$}, which are exactly the locations of the poles of $\De$  in $\Lam_n$ for a fixed $n$, and are 
only possible for $n \geq n_*$ with 
\begin{align}
    \label{eq:cross-over}
    n_* \equiv \frac{2 \Delta}{d} \stackrel{d=4}{=} {\De \ov 2} \,,\quad \Longleftrightarrow \quad \Delta_{T^n}^* \equiv n_* \,d = 2\,\Delta\, .
\end{align}
  This explains the different behavior of $\Lam_n$ for $n > n_*$ and $n < n_*$ observed in the left plot of Figure~\ref{fig:nstar}.
We can equivalently say that the cross-over point happens exactly at the point where double-trace operators start contributing;  for $n < n_*$, the series expansion for $G (\tau)$ has only contributions from $G_T (\tau)$.

\begin{figure}[t]
    \centering
   \includegraphics[scale=0.8]{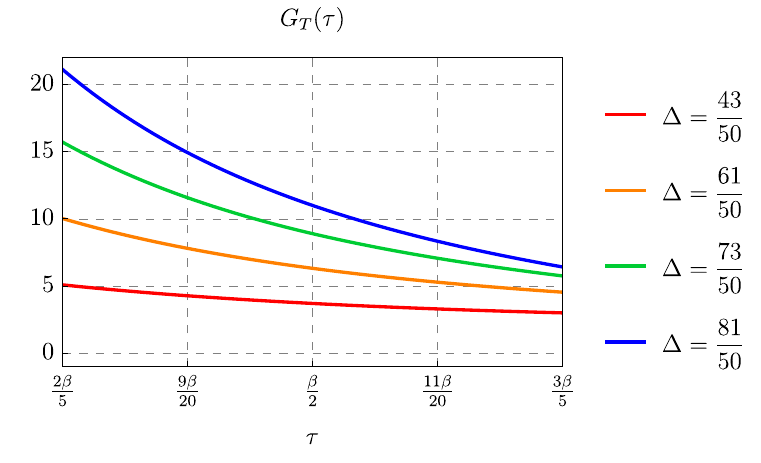}
    \caption{The value of stress-tensor contribution to the thermal correlator for different values of $\Delta$ near $\tau= \beta/2$. 
    We see that the stress-tensor sector is not symmetric around $\tau= \beta/2$ and thus does not satisfy the KMS condition, $G_T(\tau) = G_T(\beta-\tau)$.
 }
    \label{fig:noKMS}
\end{figure}

In our analysis, we have no control over the double-trace contributions. 
Therefore, for finite $\Delta$, $G_T(\tau)$ does not reproduce the full thermal correlation function. A simple check of the importance of the double-trace exchanges comes from the failure of the stress-tensor sector to satisfy the KMS condition.  In Figure~\ref{fig:noKMS}, we  plot numerically $G_T (\tau)$ for various $\De$, and see that $G_T(\tau)$ is not symmetric around $\tau= \beta/2$ and thus $G_T(\tau)\neq G_T(\beta-\tau)$. The stress-tensor sector contribution by itself does not satisfy the KMS condition and contains no knowledge about the periodicity of the~(Euclidean) time circle. Hence one role of the double-trace sector is to ensure that the full correlator satisfies the KMS condition.

\subsection{Asymptotic analysis of OPE coefficients for finite \texorpdfstring{$\De$}{Delta} and bouncing  singularities}

Let us now focus on the analysis of the coefficients $\Lambda_n$ for large values of $n$ and finite $\Delta$. We are thus interested in the regime where $n>n_*$ and both the stress-tensor and double-trace contributions are appearing in the full correlator. As already mentioned above, calculating $\Lambda_n$ as explicit functions  of $\Delta$ is currently out of our computational reach. Instead we first fix $\Delta$ to a finite number: For each value of $\Delta$ considered, we calculated $n \approx 50$ coefficients, which on a standard desktop computer takes around 10 days per $\Delta$.

\begin{figure}[t]
\includegraphics[scale=0.9]{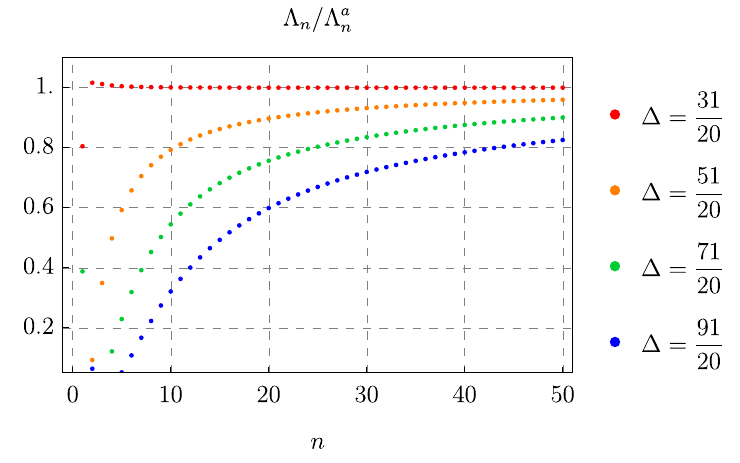}
	\centering
    \caption{Ratio of the explicit results for $\Lambda_n$ to the leading large-$n$ prediction for different values of $\Delta$ in $d=4$.}
    \label{fig:d4Ratio}
\end{figure}

We find that for large values of $n$, the  $\Lambda_n$ can be approximated by
(see Figure~\ref{fig:d4Ratio})
\begin{align}
\label{eq:Lambdand=4Ansatz}
   \Lambda_n^a = c(\Delta)\, \frac{n^{2\Delta -3}}{\left(\frac{1}{\sqrt{2}}\right)^{4n}\,e^{i\,\pi\,n}}\,.
\end{align}
One can be more precise and include $1/n$ corrections to \eqref{eq:Lambdand=4Ansatz}.  We analyse such terms and how we obtained the form for $\Lambda_n^a$ in detail in Appendix~\ref{app:DeltaTau}. The $1/n$ corrections become especially important at large $\Delta$, which can already be seen in Figure~\ref{fig:d4Ratio}, where the approach to the asymptotic form is slower for higher $\Delta$.

\begin{figure}[t]
    \centering
    \includegraphics[scale=0.8]{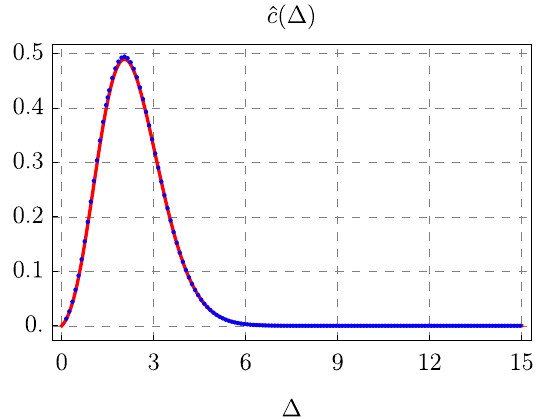}
    \caption{Numerical data for $\hat c(\Delta)$ (blue) compared with the function given in \eqref{eq:cdeltafun} (red).}
    \label{fig:c(Delta)}
\end{figure}

The coefficients $\Lambda_n^a$ contain a non-trivial function $c(\Delta)$, which has poles at $\Delta =2,3,\ldots$.%
\footnote{As mentioned earlier, at integer values of $\Delta$ the double trace operators can mix with the stress tensor sector. The addition of these double traces should cancel the divergence and make the correlation function finite \cite{Fitzpatrick:2019zqz}.}
This diverging behaviour can be isolated through
\begin{align}
\label{eq:cdeltafun}
    c(\Delta) \equiv \frac{\pi\,\Delta\,\left(\Delta-1\right)}{\sin\left(\pi\,\Delta\right)}\,\hat c(\Delta)\,, \qquad \qq{for}\Delta >0\,,
\end{align}
where the function $\hat c(\Delta)$ is free of poles.
It turns out (see Figure~\ref{fig:c(Delta)}), that this residual function can be approximated by\footnote{That is, for $\Delta\gtrsim 5/4$  and up to $\Delta \approx 15$, the ratio between the ``numerical'' values and  \eqref{eq:cdeltafun} is equal to 1 up to 2\%.} 
\begin{align}
\label{eq:chatfun}
    \hat c(\Delta)  \approx  \frac{\Delta}{\Gamma\left(2\,\Delta+ \frac{3}2\right)}\,\frac{4^{2\Delta}}{20}\,.
\end{align}

\begin{figure}[t]
\centering
\includegraphics[scale=1]{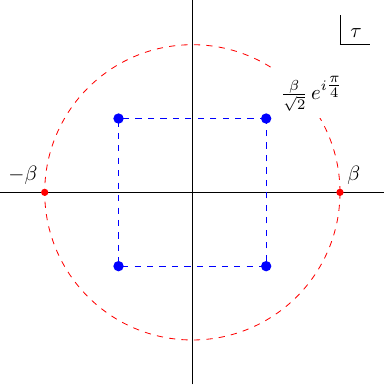}
\caption{The poles of the stress-tensor sector of the thermal correlator in four dimensions in the complex $\tau$ plane. The poles are located inside the circle of radius $\beta$.}
\label{fig:Polesd4}
\end{figure}

We now insert the asymptotic form of $\Lambda_n$ into the OPE of the correlator \eqref{e.resumL} and approximate the sum with an integral%
\footnote{We discuss the validity of this approximation in Appendix \ref{a.sumz}.}
\begin{align}
\label{e.sin1}
        G_T(\tau)&\approx\frac{1}{\tau^{2\Delta}}\int_0^\infty\Lambda_n^a\left(\frac{\tau}{\beta}\right)^{4n}d n
        %
        %
        =\frac{c(\Delta)\,\Gamma(2\Delta-2)}{\tau^{2\Delta}}\, \left[-\log\left(\frac{\tau^4}{\left(\frac{\beta}{\sqrt2}\right)^4\,e^{i\pi}}\right)\right]^{-(2\Delta-2)}\!.
\end{align}
The correlator will diverge whenever the argument of the logarithm is equal to 1, which is precisely at 
\begin{equation}\label{e.poles}
       \tau =  \tau_c\equiv\frac{\beta}{\sqrt{2}}e^{i\frac{\pi}{4}+ik\frac{\pi}{2}}\qq{for}k\in\mathbb{Z}\,.
\end{equation}
Let $\delta \tau \equiv \tau_c - \tau$. Near the critical values, the correlator takes the form
\begin{align}
\label{eq:CorrDivergenced4}
    G_T(\tau \approx \tau_c) \sim \frac{c(\Delta)\,\Gamma(2\Delta-2)}{4^{(2\Delta-2)}}\,\frac{1} {\tau_c^{2}}\frac{1}{\delta\tau^{2\Delta-2}}\,,
\end{align}
which is precisely of the form of the bouncing singularities~\eqref{Tcor1} (see Figure~\ref{fig:Polesd4}).
The four-fold rotational symmetry of the singularities~\eqref{e.poles} originates from the 
fact that multiplying by a fourth root of unity $\tau \to e^{i\frac{k\pi}{2}} \tau$
leaves all terms inside the sum in the OPE~\eqref{e.resumL} invariant.

At first sight, one can now take the large-$\Delta$ limit of the correlator near the poles. 
 To make a valid comparison, we first define
\begin{align}
\label{eq:LogOfCorr}
    \overline{\cL}_{T} \equiv -\frac{1}{\Delta}\,\log G_T(\tau)\,,
\end{align}
which, using \eqref{eq:CorrDivergenced4}, gives
\begin{align}
\label{eq:LeadingLogRes}
    \overline\cL_T \simeq -\frac{1}{\Delta}\log\left[\frac{c(\Delta)\,\Gamma(2\Delta-2)}{4^{(2\Delta-2)}}\,\frac{1} {\tau_c^{2}}\right]+ \frac{2\Delta-2}{\Delta}\log\delta\tau\,.
\end{align}
If then one inserts the expression \eqref{eq:cdeltafun} (including \eqref{eq:chatfun}) for $c(\Delta)$ then in the large-$\Delta$ limit all $\Delta$ dependence disappears%
\footnote{At integer values of $\Delta$, the prefactor $c(\Delta)$ diverges (see \eqref{eq:cdeltafun}). However, in these cases, one finds $\log(\Delta)/\Delta$ behaviour which vanishes as $\Delta\to \infty$.} and
\begin{align}
\label{eq:ProperLengthNearLim}
    \lim_{\Delta \to \infty}\overline \cL_T \simeq  2\log\delta\tau \,.
\end{align}
This appears to give the precise form of divergence expected from the bouncing geodesic \eqref{eq:PropLenDeltat}.

However, the order of limits for~\eqref{eq:ProperLengthNearLim} is different from that of~\eqref{eq:PropLenDeltat}.
To obtain (\refeq{eq:ProperLengthNearLim}), we first focused on the limit $\tau\to\tau_c$ and then took the large $\De$ limit. 
On the other hand, the bulk geodesic analysis corresponds to taking the large $\De$ limit first. 
That the order of limits is important can be seen from the structure of the $\Lambda_n$ as functions of $\Delta$, 
as indicated in Figure~\ref{fig:l4l5} and~\ref{fig:nstar}. To obtain~\eqref{eq:CorrDivergenced4}, 
we first fixed $\Delta$ and analyzed the asymptotic large $n$ behavior of $\Lam_n$, which is in the regime $n>n_* = \Delta/2$. 
However, the geodesic analysis corresponds to taking first $\Delta\to \infty$, where the cross-over point also goes to infinity, $n_* \to \infty$, and thus  we are always in the regime $n < n_*$.

\subsection{Asymptotic behavior in the large \texorpdfstring{$\De$}{Delta} limit} \label{s.abintldl}

To properly analyse the large $\Delta$ limit and make contact with the geodesic results we first need to take $\Delta \to \infty$ and only then determine the OPE coefficients $\Lambda_n$. We show in this subsection that the 
Euclidean geodesic results~\eqref{eq:TauEuclid} and~\eqref{eq:ProperLengthEuclid} are indeed recovered.

We begin by expanding the logarithm of \eqref{eq:NCCorr} in small $\tau$
\begin{align}
\label{eq:LOS}
    \overline{\cL}_T = 2\,\log\tau - \frac{\pi^4}{40}\left(\frac{\tau}{\beta}\right)^4 - \frac{\left(77 \Delta ^3-483 \Delta ^2+712 \Delta +72\right)\,\pi ^8  }{100800\,(\Delta -4) (\Delta -3) (\Delta -2)}\,\left(\frac{\tau}{\beta}\right)^8 + \ldots\,.
\end{align}
and compare this with the proper length of the geodesic in Euclidean space \eqref{eq:OneSidedProperLength}. 
We see that logarithmic and the $\tau^4$ terms match,%
\footnote{These terms are related to the exchange of identity and a single stress-tensor operator in the OPE and are fixed by symmetry.}
while the $\tau^8$ (and all higher order) terms differ.
This should not be surprising: The geodesic result can be meaningfully compared to the correlation function only in the  $\Delta\to \infty$ limit, in which case
\begin{align}\label{ldeB}
   \cL_T\equiv  \lim_{\Delta\to \infty}\overline{\cL}_T = 2\,\log\tau - \frac{\pi^4}{40}\left(\frac{\tau}{\beta}\right)^4 - \frac{11\,\pi^8}{14400}\,\left(\frac{\tau}{\beta}\right)^8 + \ldots\,,
\end{align}
is equal to \eqref{eq:OneSidedProperLength}. We checked the agreement explicitly up to order $(\tau/\beta)^{20}$.
This importantly shows that in the limit where the conformal dimension of the probe field is large, the stress tensor sector completely reproduces the geodesic result. 

When $\Delta \to \infty$, we can also analyse the asymptotic behaviour of the expansion coefficients in $\cL_T$. 
In particular, inserting \eqref{e.resumL} into \eqref{eq:LogOfCorr}, we obtain
\begin{align}
    \overline{\cL}_T = 2\log\tau - \sum_{n=1}^{\infty} \frac{\widetilde \Lambda_n}{\Delta}\left(\frac{\tau}{\beta}\right)^{4n}\,,
\end{align}
where $\widetilde \Lambda_n$ are appropriate combinations of $\Lambda_n$ and we have used the fact that $\Lambda_0=1$.
One can then take the limit $\Delta \to \infty$ as 
\begin{align}
\label{eq:cltexp}
    \cL_T= \lim_{\Delta \to \infty}\overline{\cL}_T = 2 \log \tau - \sum_{n=1}^{\infty} L_n\,\left(\frac{\tau}{\beta}\right)^{4n}\,,
\end{align}
where we have assumed that one can take the limit inside the series and thus
\begin{align}
    L_n \equiv \lim_{\Delta \to \infty}\frac{\widetilde \Lambda_n}{\Delta}\,.
\end{align}
\begin{figure}[t]
    \centering
    \includegraphics[scale=0.8]{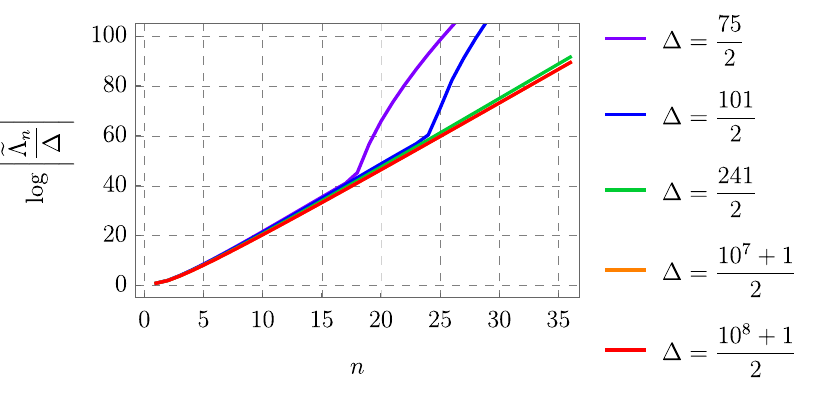}
    \vspace{-0.5cm}
    \caption{The values of $\log\left|\widetilde\Lambda_n/\Delta\right|$ for different $\Delta$. The discrete data is joined for a clearer presentation. When $\Delta$ is small, we observe a change of behaviour at $n = n_* = \Delta/2$. For large $\Delta$, this cross-over point disappears and the data converges to a fixed curve, analysed in \eqref{eq:Ansatz1}.}
    \label{fig:plogL}
\end{figure}

\begin{figure}[t]
    \centering
    \includegraphics[scale=0.8]{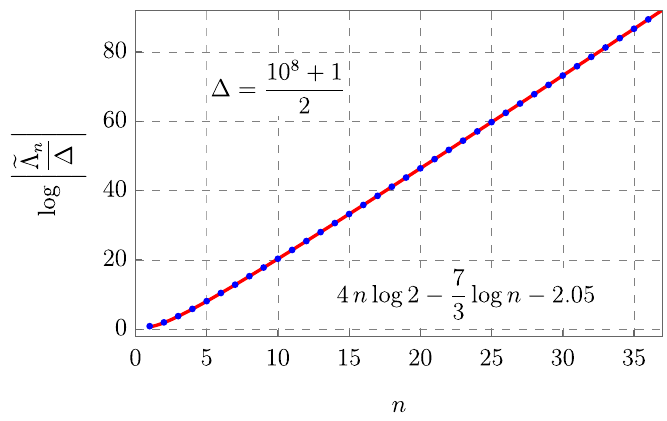}
    \vspace{-0.5cm}
    \caption{Comparing the holographic data for $\Delta = (10^8+1)/2$ to the ansatz \eqref{eq:Ansatz1} and the numerical data given by \eqref{eq:NumericalData}.}
    \label{fig:plogfit}
\end{figure}

The first few $L_n$ are given implicitly in \eqref{ldeB}. For higher values of $n$, we again analyse the data for large but fixed  $\Delta$. For some  $\Delta$, we plot the values of $\log\left|\widetilde\Lambda_n/\Delta\right|$ in Figure~\ref{fig:plogL}. For small enough  $\Delta$, we again observe the cross-over at $n=n_*$ to the asymptotic behaviour analysed in the previous subsection. For large $\Delta$, not only does this cross over get pushed to infinity, but the curve stabilises to an asymptotic value, which we assume is given by
\begin{align}
\frac{\widetilde \Lambda_n}{\Delta}  \xrightarrow[\Delta\to\infty]{} L_n + \coo{\frac{1}{\Delta}}\,.
\end{align}
In practice, we analyse  $\Delta = (10^8+1)/2$ and estimate $L_n$ up to $1/\Delta$ corrections. We find that at large values of $n$, one can make an ansatz 
\begin{align}
\label{eq:Ansatz1}
    L_n = c\,n^b\,a^n\left(1+ \coo{\frac1n}\right)\,,
\end{align}
where $a$, $b$, and $c$ are all constants. 
Then for $n\gg 1$
\begin{align}
    \log L_n = \log c + b \,\log n+ n\,\log a + \coo{\frac1n}\,,
\end{align}
which can be directly fitted to the $\widetilde \Lambda_n/\Delta$ data for $\Delta = (10^8+1)/2$ and we find%
\footnote{We used the data from $n=10$ to $n=36$ and used the ansatz \eqref{eq:Ansatz1} with up to $1/n^{12}$ corrections.}
(see also Figure~\ref{fig:plogfit})
\begin{align}
\label{eq:NumericalData}
   & \log a = 4\,\log 2 \pm 10^{-6}\,,&&  b = -\frac73  \pm 10^{-4} && \log c = -2.050 \pm 10^{-3}\,,
\end{align}
In what follows we neglect the uncertainties which can be traced back to both $1/\Delta$ and $1/n$ effects. 
We insert these values into the expansion \eqref{eq:cltexp} and approximate the sum by an integral%
\footnote{We use the function \begin{align*}
    E_n(x) \equiv \int_1^{\infty}t^{-n}\,e^{-t\,x}\,dt\,.
\end{align*}}
\begin{align}
    \cL_T &\approx 2 \log\tau - c\,\sum_{n=1}^{\infty}n^{-\frac{7}{3}}\,2^{4n}\,\left(\frac{\tau}{\beta}\right)^{4n}\approx 2\log\tau -c \int_1^\infty\,dn\,n^{-\frac73}\,\left(\frac{2\,\tau}{\beta}\right)^{4n}\nonumber\\*
   &= 2\log \tau - E_{\frac73}\left(-\log\left(\frac{2\tau}{\beta}\right)^4\right)\,.
    \label{eq:ExponentialIntegal}
\end{align}
This function has a branch cut at all points at which the argument of the logarithm is equal to 1, which is 
\begin{align}
    \tau_{b} = \frac{\beta}{2}\,e^{\frac{i\pi\,k}{2}}\,,\qquad k \in \mathbb{Z}\,, 
\end{align}
which suggests four singularities in the complex-$\tau$ plane.
Focusing on $\tau =\beta/2$, we expand \eqref{eq:ExponentialIntegal} around this point and find
\begin{align}
\label{eq:sing_bovertwo}
    \cL_T \approx  - c\,\Gamma\left(-\frac{4}{3}\right)\,\left(\frac{8}{\beta}\right)^{\frac{4}{3}}\,\left[\frac{\beta}{2}-\tau\right]^{\frac43} + \ldots \approx - 1.364 \le({\pi \ov 2}-\tau \ri)^{4 \ov 3} + \ldots \,,
\end{align}
where we kept only the leading non-analytic behavior at $\tau = \beta/2$, and in the second equality inserted the value of $c$ and taken $\b = \pi$. The leading non-analytic behavior at  $\tau = \beta/2$ of the Euclidean geodesic results~\eqref{eq:TauEuclid} and~\eqref{eq:ProperLengthEuclid} can be readily found from~\eqref{deo}, which gives 
\be 
L = - {6^{4 \ov 3} \ov 8} \le({\pi \ov 2}-\tau \ri)^{4 \ov 3} + \ldots \approx  - 1.363  \le({\pi \ov 2}-\tau \ri)^{4 \ov 3} + \ldots  \ .
\ee
We see that the asymptotic analysis reproduces the location, the exponent, and the prefactor of the branch point structure observed from bulk geodesics.

What if we take the large $\Delta$ limit directly in the correlator, instead of using its logarithm? One can perform the same analysis as above for a large value of $\Delta$ and obtain
\begin{align}
\log\Lambda_n \sim n \log \Delta - \log \Gamma(n+1)  + n\,\log\left(\frac{\pi^4}{40}\right)+\coo{\frac{1}{\Delta}}\,,
\end{align}
where the numerical value of $\log\left(\frac{\pi^4}{40}\right) \approx 0.89004$ gets reproduced by fitting up to $\approx 10^{-6}$ for $\Delta = (10^8+1)/2$.
Exponentiating the above result yields
\begin{align}
    \Lambda_n = \frac{1}{n!}\,\left(\frac{\Delta\,\pi^4}{40}\right)^n\left(1+ \coo{\frac{1}{\Delta}}\right)\,.
    \label{eq:StarOPE}
\end{align}
Inserting this into the OPE and assuming that one can trust these OPE coefficients up to $n \to \infty$ yields
\begin{align}
    \label{eq:expres}
     G_T(\tau) 
     \approx  \frac{1}{\tau^{2\Delta}}\sum_{n=0}^{\infty}\,\frac{1}{n!}\left(\frac{\pi^4\,\Delta\,\tau^4}{40\,\beta^4}\right)^n 
     \approx \frac{1}{\tau^{2\Delta}}\exp\left[\frac{\pi^4\,\Delta\,\tau^4}{40\,\beta^4}\right]\,,
\end{align}
where we neglected $1/\Delta$ corrections. This result matches  \cite{Rodriguez-Gomez:2021pfh}, where starting from \eqref{eq:NearCoincidentCorrelator}, a double-scaling limit is taken with  $\Delta \to \infty$ while $\tau/\beta \to 0$ such that $\Delta\tau^4/\beta^4$ is held fixed.
The significance of this double-scaling limit is that it can be obtained from the exponentiation of the single stress tensor exchange.
Indeed, we see that the argument of the exponent is precisely the first non-logarithmic term in the proper length in \eqref{eq:OneSidedProperLength}.
Therefore, taking the large $\Delta$ limit at the level of the correlator neglects all the higher order terms in \eqref{ldeB}, since the information about these terms is encoded in the $1/\Delta$ corrections in \eqref{eq:StarOPE}, which are discarded.
Since these higher order terms are crucial for the analytic structure of the proper length, we are unable to see the branch cut when taking the $\Delta\to\infty$ limit at the level of the correlation function.
As seen above, one needs to analyse the proper length in order to see the branch cut.

\subsection{Summary} \label{sec:sum}

Here we summarize the main results obtained in this section: 

\ben 
\item For a finite $\De$, as indicated by Figure~\ref{fig:noKMS}, $G_T (\tau)$ does not satisfy the KMS condition on its own. 
 
The double-trace contribution is thus also needed such that the KMS condition is satisfied by the full correlator. 

\item \label{item:2}
For a finite $\De$, $G_T (\tau)$ has singular behavior~\eqref{eq:CorrDivergenced4} at~\eqref{e.poles}. 
 
Full thermal correlation functions cannot have such singular behavior, and thus~\eqref{eq:CorrDivergenced4} should be cancelled by $G_{[\phi\phi]} (\tau)$. That is, $G_{[\phi\phi]} (\tau)$ must have the same singular behavior, but with an opposite sign. 

\item\label{item:3}
 In the $\De \to \infty$ limit, the small $\tau$ expansion of $\LL_T (\tau)$ (defined in~\eqref{ldeB}) precisely recovers the small $\tau$ expansion of $L (\tau)$ obtained from gravity. 

While we have only checked the agreement between $\LL_T (\tau)$ and $L (\tau)$ in this limit up to $\tau^{20}$, we will assume below that the agreement in fact holds to all orders in the small $\tau$ expansion. 

\item There is a crossover behavior in $\Lambda_n$
at $n_*=\Delta/2$. As a result, in the large $\De$ limit, the singularity of $G_T (\tau)$ at $\tau_c$ obtained at a finite $\De$ is not directly seen, instead we find a branch point singularity in $\LL_T (\tau)$ at $\tau = \b/{ 2}$.

A possible scenario is that as $\De$, and therefore $n_*$, increases the size of the region around $\tau=\tau_c$ where 
$G_T(\tau)$ diverges shrinks and finally vanishes when $\Delta\to \infty$.
At the same time, as $\De$ increases, the behavior of $\LL_T (\tau)$ near $\tau=\beta/2$ gets closer and closer to the singular behavior observed in the geodesic analysis \eqref{eq:sing_bovertwo}.
This would be another signature of the
noncommutativity of limits $\tau \to \tau_c$ and $\De\to\infty$.

\een


\section{Boundary interpretation of the gravity results}\label{sec:inter}

We now use the results obtained from the OPE analysis of Section~\ref{s.bhope} to interpret the gravity results reviewed in Section~\ref{sec:SC}. While the singular behavior~\eqref{eq:CorrDivergenced4} has the same form  as that in~\eqref{Tcor1}, 
the connection is not immediate, as~\eqref{eq:CorrDivergenced4} and~\eqref{Tcor1} appear in different quantities and in different regimes (one in $G_T$ at finite $\De$, while the other in $\hat G$ at infinite $\De$).  
What is more, as discussed at the end of last section, the singular behavior~\eqref{eq:CorrDivergenced4} disappears in the large $\Delta$ limit, which obscures its connection to the bouncing singularities and  black hole singularities. 
 
In this section, we argue that the singular behavior~\eqref{eq:CorrDivergenced4} can indeed be identified as the boundary origin 
of~\eqref{Tcor1} and thus the black hole singularities. 
Below we will restore $\b$, and when we say ``for all $\tau$'' it always refers to ${\rm Re} \, \tau \in (0, \b)$.

\subsection{Role of the double-trace contributions in the large \texorpdfstring{$\De$}{Delta} limit}\label{s.dt}

We first discuss the role of the double-trace contributions in the large $\De$ limit. In Section~\ref{s.bhope} we found that in the $\De \to \infty$ limit, the small $\tau$ expansion of $\LL_T (\tau)$  precisely recovers that of $L (\tau)$ obtained from 
the geodesic analysis (i.e. item~\ref{item:3} of Section~\ref{sec:sum}). 
Furthermore, we have shown that in the large $\Delta$ limit, $\LL_T (\tau)$  has a branch point singularity of the form $({\b \ov 2}-\tau)^{4 \ov 3}$ at $\tau ={\b}/{2}$. 
That is, for large $\De$, $\LL_T (\tau)$ (and $L(\tau)$) has a radius of convergence $\b/2$, and 
the double-trace contribution $G_{[\phi \phi]} (\tau)$ can be neglected in the large $\De$ limit for $\tau < {\b}/{2}$. 

That $G_{[\phi \phi]} (\tau)$ can be neglected for small $\tau$ has a simple explanation from the OPE structure.  The double-trace contribution $G_{[\phi\phi]} (\tau)$ has a small $\tau$ OPE of the form 
\begin{align}\label{dtope}
G_{[\phi\phi]}(\tau) = \frac{1}{\beta^{2\Delta}}\,\sum_{k=0}^{\infty}\,D_k\left(\frac{\tau}{\beta}\right)^{2\,k}
\end{align}
where $D_k$ are some constants. The appearance of even powers in~\eqref{dtope} comes from the fact that the double-trace operators $\phi (\p^2)^n \p_{i_1} \cdots \p_{i_l} \phi$
have dimensions $2 \De + 2n +l$ and only the ones with even $l$ can have nonzero thermal expectation values in~\eqref{eun}. 

Comparing~\eqref{dtope} with the analogous expression
for the stress tensor sector~\eqref{e.resumL} gives 
\begin{align}
    \frac{G_{[\phi\phi]}(\tau)}{G_T(\tau)} \sim \left(\frac{\tau}{\beta}\right)^{2\Delta}\,\left(1 + \coo{\tau^2}\right), 
\end{align}
with the full $G (\tau)$ having the structure 
\be 
G(\tau) = {1 \ov \tau^{2 \De}} \le[\sum_{n=0}^\infty\Lambda_n\left(\frac{\tau}{\beta}\right)^{d\,n}\ 
+\ \sum_{k=0}^{\infty}\,D_k\left(\frac{\tau}{\beta}\right)^{2\,k + 2 \De} \ri] \ .
\ee
We may  conclude from the above equation that, in the large $\De$ limit, the double-trace contribution is always suppressed for all $\tau$. This, however, assumes that we can take the large $\De$ limit {\it inside} the sum, which 
requires  $G(\tau)$ to be uniformly convergent for all $\Delta$ for all $\tau$. This is incorrect, since as mentioned earlier, $\LL_T (\tau)$ develops a branch point singularity at $\tau = {\b}/ { 2}$ in the large $\De$ limit, and only has the radius of convergence equal to $\b \ov 2$.\footnote{Even if the series $G_T (\tau)$ remains convergent in the large $\De$ limit for all $\tau$, it is possible that the full series $G (\tau)$ may not be uniformly convergent for all $\De$ for all $\tau$. We will see an example of this in Section~\ref{sec:gff}.}
Thus for $|\tau| \geq {\b}/ { 2}$, we need to sum  the series for $G (\tau)$ first for finite $\De$ and then take the large $\Delta$ limit.

For $|\tau| \geq {\b}/ { 2}$, we expect $G_{[\phi\phi]}(\tau)$ will be needed. After all, as mentioned earlier, the double-trace contributions are needed for the full correlator to satisfy the KMS condition. Now comparing with the discussion of Section~\ref{sec:geo}, we can identify 
\be \label{ll0}
\LL_T (\tau) = L (\widetilde E_0 (\tau))
\ee
where $L$ on the RHS is the geodesic distance obtained from gravity. From~\eqref{enq}, we can further identify 
\be \label{ll00}
\LL_{[\phi\phi]} (\tau) =  L (\widetilde E_1 (\tau))
\ee
which becomes important for $\th \leq -{\pi \ov 2}$, with $\th$ defined as ${\b \ov 2} - \tau = r e^{i \th}$. 
Furthermore, the relative dominance between $\widetilde E_0$ and $\widetilde E_1$ saddles discussed there 
can now be written in terms of the boundary language as 
\begin{align}
\label{eq:RealCorr}
    \lim_{\Delta\to\infty}G (\tau) =
    \lim_{\Delta \to \infty}  \bca
   G_T(\tau) &  \tau < \frac{\beta}{2} \cr
   G_{[\phi\phi]}(\tau) & \tau > \frac{\beta}{2} \cr
   G_T (\tau) + G_{[\phi\phi]}(\tau)  & \tau = \frac{\beta}{2}  + i t_L, \; t_L \in \mathbb{R} 
   \eca      \ .
\end{align}
In particular, $G_T (\tau)$ and $G_{[\phi\phi]}(\tau)$ correspond respectively  to the contributions from the two complex geodesics 
on the gravity side. 

One can ask why the uniform convergence stops at $\tau = \beta/2$. If we consider a theory with time-reversal symmetry, $\tau \to - \tau$, that satisfies the KMS condition, $G(\tau) = G(\beta-\tau) $, then the branch point at which the OPE sum and the $\Delta \to \infty$ limit cannot be exchanged  can be in principle at any $0\leq \beta^*\leq \beta/2$. It would be interesting to  explore this direction further. 

\subsection{Boundary interpretation of the bouncing singularities}\label{s.biotbs}

We now turn to the boundary interpretation of the bouncing singularities~\eqref{Tcor1}.

The identification~\eqref{ll0} is consistent with our numerical observation that the singular behavior~\eqref{eq:CorrDivergenced4}  is not directly seen in the large $\De$ limit, as $\widetilde E_0$ corresponds to a complex geodesic for $\tau = {\b}/ { 2} + i t_L$, and is regular at $\tau_c = {\b \ov 2} (1+i)$. 
The same statement applies to $G_{[\phi \phi]} (\tau)$. As we discussed in item~\eqref{item:2} of Section~\ref{sec:sum},
at a finite $\De$, $G_{[\phi \phi]} (\tau)$ should also have a bouncing singularity at $\tau_c$, but the identification~\eqref{ll00} implies that the singularity is no longer there in $\LL_{[\phi \phi]} (\tau)$. 
Nevertheless, we would like to argue that the bouncing geodesic singularities~\eqref{Tcor1} do originate from~\eqref{eq:CorrDivergenced4}. 

More explicitly, after obtaining $G_T (\tau)$ from the OPE of multiple stress tensors, we can expand $\cL_T (\tau)$ near $\tau = {\b}/ { 2}$ as 
\be \label{loq1}
\LL_T (\tau) \sim  \le({\b \ov 2} - \tau\ri)^{4 \ov 3} + \ldots \equiv f_T (y)\,, \quad y \equiv \le({\b \ov 2} - \tau\ri)^{1 \ov 3}  \,,
\ee
where we neglect the constant prefactor and terms that are regular at $\tau = \beta/2$. 
Since on the gravity side $\tilde E_1 (\tau)$ and $\tilde E_2 (\tau)$ are related to $\tilde E_0 (\tau)$ 
by  phase multiplications~\eqref{eq:branches}, we can then write 
\be \label{loq2}
\LL_{[\phi \phi]} (\tau) = f_T \le(e^{-{2 \pi i \ov 3}} y \ri), \quad \hat \LL (\tau)  = f_T \le(e^{-{4 \pi i \ov 3}} y \ri) \ .
\ee
That is,  through the branch point singularity developed by $\LL_T (\tau)$ at $\tau = {\b}/ { 2}$, both $\LL_{[\phi \phi]} (\tau)$ coming from $G_{[\phi \phi]} (\tau)$, and $\hat \LL (\tau)$~\eqref{eq:PropLenDeltat} corresponding to the bouncing geodesic are fully determined from $\LL_T (\tau)$.

While the singular behavior~\eqref{eq:CorrDivergenced4} of 
$G_T (\tau)$ and $G_{[\phi \phi]} (\tau)$ have seemingly both disappeared in themselves in the large $\De$ limit, the bouncing singularities are not lost, but are transferred to $\hat \LL (\tau)$.  They are just not as manifest. 
This phenomenon is quite remarkable. As emphasized in Section~\ref{s.bhope}, the bouncing singularities of $G_T (\tau)$ at finite $\De$ are controlled by the asymptotic large $n$ behavior in the regime $n > n_*$, while  $\LL_T (\tau)$ is given by $\Lam_n$  in the regime $n< n_*$ after first taking $n_* \to \infty$. 
Nevertheless, $\LL_T (\tau)$ does contain the information regarding the bouncing singularities, albeit indirectly.

At finite $\De$, both $G_T (\tau)$ and $G_{[\phi \phi]} (\tau)$ have bouncing singularities, then why do we say that they originate from the stress tensor sector, not from the double-trace sector? We already saw that in the large $\De$ limit, both 
$\LL_{[\phi \phi]} (\tau)$ and $\hat \LL (\tau)$ are determined from $\LL_T (\tau)$. This is not an accident; in fact, $G_{[\phi \phi]} (\tau)$ is also determined by $G_T (\tau)$ for any $\De$. 

To make the point clear, recall the OPE computation of a vacuum four-point function 
\be\label{o4}
\vev{VVWW} = \sum_O C_{VVO} C_{WWO} \vev{OO}  ,
\ee
where the four-point function is fully determined from the OPE data and vacuum two-point functions. 
In contrast, in the OPE computation~\eqref{eun} of $G(\tau)$, the right hand side of the equation involves the thermal expectation values of double-trace operators 
\be 
\tvev{\phi (\p^2)^n \p_{i_1} \cdots \p_{i_l} \phi}_\b
\ee
whose values require knowledge of thermal two-point functions of $\phi$. In other words, both sides of~\eqref{eun} depend on $G(\tau)$, and it should be viewed as an equation {\it to solve} for $G(\tau)$, rather than as an expression to calculate the left hand side as is the case for~\eqref{o4}.  The inputs that we need to solve~\eqref{eun} are $G_T (\tau)$, the OPE coefficients 
of $\phi (\p^2)^n \p_{i_1} \cdots \p_{i_l} \phi$, as well as imposing that $G(\tau)$  obeys the KMS condition and be analytic in ${\rm Re} \, \tau \in (0, \b)$.  Thus we can say $G(\tau)$ and $G_{[\phi \phi]} (\tau)$ are determined from $G_T (\tau)$. 

Finally, we comment that $G_T (\tau)$ is closely connected to the bulk geometry, and thus enjoys some level of universality. 
The OPE coefficients for $k$-stress tensor exchanges correspond in the bulk to couplings of the bulk dual $\Phi$ of $\phi$ 
to multiple gravitons, $i.e.$ couplings of the form $\Phi \Phi h^k$, where $h$ schematically denotes the bulk graviton. The thermal one-point function $v_n$ for $k$ stress tensors is given schematically by $(\langle T_{\mu \nu} \rangle_\beta)^k$, $i.e.$ $k$-th power of the boundary stress tensor in the black hole geometry.

\subsection{A speculation on the momentum space behavior and bouncing singularities} 

Now suppose the behavior~\eqref{Tcor2} is also present at finite $\De$. Then given~\eqref{eq:CorrDivergenced4}, 
it is tempting to speculate that 
\be \label{ejp}
G(i \om_E) \sim \int_{-\infty}^\infty d \tau \, e^{i \om_E \tau} \, G_T (\tau)  \ .
\ee
We use $\sim$ in the above equation to indicate that while (under some assumptions) the right hand side gives the correct qualitative behavior 
for $G (i \om_E)$ in the limit $\om_E \to \pm \infty$, at the moment we do not have enough information to specify a precise relation. As $\om_E \to +\infty$, we assume the behavior of $G_T (\tau)$ at infinity is such that we can close the contour in the upper half complex $\tau$-plane. Then the integral will mainly receive contributions from the neighborhood of $\tau_c^+ = {\b \ov 2} + i {\widetilde \b \ov 2}$, leading to 
\be 
G(i \om_E) \sim e^{i \om_E \tau_c} \om_E^{2 \De-3} \propto  (i \om_E)^{2 \De-3} e^{i (i \om_E) t_c^+} , \quad \om_E \to +\infty \ .
\ee
Similarly, for $\om_E \to - \infty$, we close the contour in the lower half complex $\tau$-plane, with the main contribution coming from integration around $\tau_c^- = {\b \ov 2} - i {\widetilde \b \ov 2}$, leading to 
\be 
G(i \om_E) \sim  (i \om_E)^{2 \De-3} e^{i (i \om_E) t_c^-} , \quad \om_E \to -\infty \ .
\ee

\subsection{Double-trace contributions in the GFF example} \label{sec:gff}

Now consider the following example 
\be
\label{eq:GFF1}
G(\tau)^{(GFF)}=   \langle \phi(\tau) \phi(0) \rangle_\beta^{(GFF)} = \sum_{m\in \mathbb{Z}} {1\over (\tau + m\, \beta)^{2\Delta}  }\,,
\ee
which is obtained by taking the vacuum Euclidean two-point function of $\phi$ and adding images in the $\tau$ direction such that it has periodicity $\b$. The $m=0$ term is the contribution of the identity while all other terms correspond to multi-trace contributions~\cite{Iliesiu:2018fao,Alday:2020eua}, as expanding $m \neq 0$ terms in $\tau$ we find a power series of the form~\eqref{dtope}.  
It will also be convenient to define
\be\label{eq:Lgff}
   \LL(\tau)^{(GFF)} = -\lim_{\De\to\infty} {1\over\De} \log G(\tau)^{(GFF)}
\ee
It is easy to compute $\LL(\tau)^{(GFF)} $: it is given by
\be
   \LL(\tau)^{(GFF)} = 2 \log \tau, \; \tau<{\b\over 2}, \qquad 
    \LL(\tau)^{(GFF)} = 2 \log (\beta-\tau), \; \tau>{\b\over 2}
\ee
and has a cusp at $\tau=\b/2$.

We will see momentarily that the series in $\tau$ 
which defines ${1\over\De} \log G(\tau)^{(GFF)}$ is uniformly convergent for all $\De$ only for $\tau < {\b}/ { 2}$.
Moreover, the GFF example satisfies ~\eqref{eq:RealCorr},
which also applies in holography.
It will be convenient to write
\be
\label{eq:logg}
  -{1\over\De} \log G(\tau)^{(GFF)} = 2 \log \tau - {1\over\De} 
          \log \left(1 +\sum_{m\neq0} {\tau^{2\De}\over (\tau+m \b)^{2\De}} \right)
\ee
where we have separated the $m=0$ term (the identity operator contribution). 
We can now expand the second term for small $\tau$, 
\begin{equation}
  -{1\over\De} \log G(\tau)^{(GFF)}
    = 2 \log \tau+ \lim_{n \to \infty} \LL^{(n)}(\tau),\qquad
   \LL^{(n)}(\tau)=  \sum_{k=0}^{n}c_k(\Delta)\,\le({\tau \ov \b} \ri)^{2\Delta +2 k}  
\end{equation}
where we have written the infinite series as a limit. 
Taking the large $\De$ limit inside the series amounts to the order of limits 
\begin{equation}\label{wbq}
    \lim_{n \to \infty}\,\lim_{\Delta\to \infty}\LL^{(n)}(\tau) =0,
    \qquad \tau <\beta\,,
\end{equation}
in which case  all
terms vanish.
However,  the limits $n\to \infty$ and $\Delta \to \infty$ may not be exchangeable.
For the other order of limits, we should take the large $\De$ limit directly in~\eqref{eq:logg}, which gives 
\begin{equation}\label{ejhw}
   \lim_{\Delta\to \infty}\,\lim_{n\to \infty}  \LL^{(n)}(\tau) 
    =\begin{cases}
    0, & \tau<\frac{\beta}{2} \\
    -2 \log \tau + 2 \log (\beta-\tau) , &  \tau>\frac{\beta}{2}
  \end{cases} \ .
\end{equation}
By comparing~\eqref{wbq} and~\eqref{ejhw}  we see that the limits can only be exchanged when $\tau<\beta/2$. %
From the Moore-Osgood theorem, it follows that the limit $n\to \infty$~(the OPE) cannot be uniformly convergent in $\Delta$ for $\tau >\beta/2$.
This example nicely illustrates the role of double traces: when $\tau$ is small, only the stress-tensor sector (identity) contributes in the large-$\Delta$ limit and the double traces can be safely ignored. 
When $\tau >\beta/2$, the double trace contribution dominates. 

Furthermore, for $\tau = {\b \ov 2} + i t_L, t_L \in \mathbb{R}$ 
at large $\De$ terms with $m=0$ and $m=-1$ both contribute
\be 
G^{(GFF)}\left(\tau={\b \ov 2} + i t_L\right) \approx \frac{1}{\tau^{2\Delta}} + \frac{1}{(\tau- \beta )^{2\Delta}}  \ .
\ee
We thus see the GFF example~\eqref{eq:GFF1} has exactly the same behavior as~\eqref{eq:RealCorr} except that $G_T (\tau)$ only has contribution from the identity and thus is always trivially convergent for all $\tau$ even in the large $\De$ limit. 


\section{Discussion}\label{sec:Disc} 

We showed that the stress tensor sector of the thermal two-point function of  scalar operators exhibits  bouncing singularities at finite $\De$.  We  argued that this singular behavior matches the one observed in~\cite{Fidkowski:2003nf} from the bulk geodesic analysis, and can be interpreted as a boundary reflection of  black hole singularities. Since contributions from multi-stress tensor exchanges do not depend on the specific details of the operators, they encode the universal features of  black hole singularities as probed by generic bulk fields. 
Furthermore, we expect that two-point functions of general ``light'' operators in generic ``heavy'' states  exhibit  thermal 
behavior at  leading order in the  $1/N$ expansion. Hence our conclusion can also be used to explain the universality of black hole singularities in single-sided black holes formed from gravitational collapse. 

Our results also help elucidate the role of double-trace operator contributions, and in particular connect them to geodesic results in the limit of  large conformal dimensions. 

Below we first discuss  results in other dimensions, then give a boundary interpretation of the gravity results for the boundary on a sphere,  and finally offer some future perspectives. 

\subsection{Other spacetime dimensions}

In the main text, we have focused on $d=4$ for definiteness. Here we discuss the results in  other dimensions. 

For $d=2$, which is discussed in detail in Appendix~\ref{s.kms2d}, Euclidean function $G(\tau, x)$~(including the spatial dependence) is known exactly from conformal symmetry, and can be shown to solely come from  the Virasoro descendants of the identity, which are the multi-stress operators. The stress tensor sector  satisfies the KMS condition by itself, and does not have any unphysical singularity. 
The corresponding bulk geometry is described by the AdS Rindler, and there is no singularity behind the Rindler horizon.  
 
The $d=4$ discussion of the main text can be straightforwardly generalized to $d=6$ and $d=8$. The geodesic analysis of~\cite{Fidkowski:2003nf} yields the bouncing singularities for general $d$ at 
\begin{equation}\label{tauc0}
   \tau_c  = {\beta \over 2} \pm i {\beta \over 2}  \ {\cos {\pi\over d} \over \sin {\pi\over d}} = \pm i {\beta e^{\mp {i \pi\over d}} \over 2 \sin {\pi\over d}}\ . 
\end{equation}
As discussed in Appendix~\ref{a.ope6d},  these singularities are exactly reproduced from the asymptotic analysis of the stress tensor OPE in $d=6$ and $d=8$, with the behavior near the singularities 
\be 
G_T (\tau) \propto {1 \ov (\de \tau)^{2 \De -{d \ov 2}}} , \quad \de \tau = \tau_c - \tau  \ .
\ee
The $-{d \ov 2}$ term in the exponent $2 \De - {d \ov 2}$ is somewhat curious and it would be interesting to understand its meaning further.

Note that for $d=6$, $|\tau_c| = \b$, while for $d=8$, $|\tau_c| > \b$, which again highlights that the stress tensor sector contribution $G_T (\tau)$ does not obey the KMS condition. It is also curious to note 
that for both $d=6$ and $d=8$ there are additional singularities other than~\eqref{tauc0} (or their reflection in the left $\tau$ plane) at 
\bea
&& d=6: \qquad \tau =  \b , \\
&& d=8: \quad \tau =  {\beta \ov 2}  \frac{e^{\pm i\frac{\pi}{8}}}{\sin\frac{\pi}{8}}\,,
\eea
These singularities have ${\rm Re} \, \tau > {\b}/ { 2}$. 
It would be interesting to see whether they play a role similar to the
bouncing singularities.  

\subsection{Boundary theory on a sphere} 

Consider the boundary CFT on a sphere $S^{d-1}$ of radius $R$ for $d >2$. 
The decomposition~\eqref{lte} still applies but the structure of each term becomes significantly more complicated. For example, now the descendants of the multiple stress tensor operators can also contribute, and there is a new dimensionless parameter ${\b \ov R}$.
Nevertheless, we expect that the boundary theory interpretation we gave in this paper should still apply qualitatively. 
That is, there should be singularities in the stress tensor sector contribution $G_T (\tau)$ at a general finite $\De$, which in the large $\De$ limit, ``become'' the bouncing singularities seen in the geodesic analysis. Below we give a boundary interpretation of the gravity analysis given in Section 3.3 of~\cite{Fidkowski:2003nf}. 

\begin{figure}
\centering
\includegraphics[scale=1]{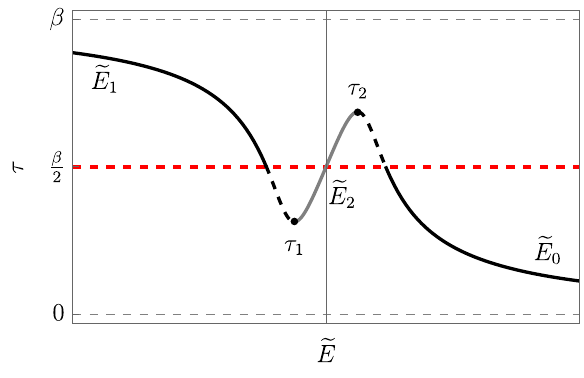}
\caption{$\tau$ as function of $\widetilde E$ obtained from the Euclidean geodesic analysis for the boundary theory on a sphere.
}
\label{fig:sphere}
\end{figure}

Figure~\ref{fig:sphere} gives Euclidean time separation $\tau$ as a function of $\widetilde E$ obtained from bulk Euclidean geodesics. In contrast with~\eqref{deo}, at $\tau = {\b}/ { 2}$, there are three different real solutions of $\widetilde E$, labelled as $\widetilde E_0, \widetilde E_1, \widetilde E_2$  in the figure.  
Contribution from the $\widetilde E_0$  branch, which is the unique real solution for sufficiently small $\tau$, should again be identified with the large $\De$ limit of $\LL_T$. Contribution from the $ \widetilde E_1$ branch should be identified with the double trace contributions $\LL_{[\phi \phi]}$, while the middle branch $\widetilde E_2$ (in gray) does not contribute  and gives rise to the bouncing geodesics when extended to complex $\tau = {\b \ov 2} + i t_L$. $\widetilde E_0, \widetilde E_1$ again correspond to complex geodesics for such complex $\tau$. The geodesic saddle $\widetilde E_1$ corresponding to the double-trace piece starts appearing at some value $\tau_1 < {\b}/ { 2}$ (denoted with a dot in Figure~\ref{fig:sphere}), but is subdominant for $\tau < {\b}/ { 2}$. Its contribution becomes the same as that from $\widetilde E_0$ at $\tau = {\b}/ { 2}$. For $\tau > {\b}/ { 2}$,  $\widetilde E_1$ (i.e. the double-trace piece) dominates. We see that the structure is identical to~\eqref{eq:RealCorr}. 

In the large $\De$ limit, $\LL_T (\tau)$ is now analytic at $\b/2$, but develops a square root branch point singularity at some value $\tau_2  > {\b}/ { 2}$ (see Figure~\ref{fig:sphere}). 
Near $\tau = \tau_2$, the geodesic distance  $\hat \LL (\tau)$ corresponding to $\widetilde E_2$  branch can be obtained from $\LL_T$ as follows, 
\be \label{eya}
\LL_T (\tau) = f_T (y), \quad  y \equiv (\tau_2 - \tau)^{1 \ov 2}, \quad \hat \LL (\tau) = f_T (-y) \ .
\ee
Analytically continuing $\hat \LL (\tau)$ to complex $\tau = {\b \ov 2} + i t_L$ should yield the bouncing singularities. 
In this case,  the double-trace piece $\LL_{[\phi \phi]}$ is still determined from $\LL_T (\tau)$ through a more involved procedure. After finding $\hat \LL (\tau)$ from $\LL_T (\tau)$ at $\tau_2$ using~\eqref{eya}, we  continue $\hat \LL (\tau)$ to $\tau_1$, where $\hat \LL (\tau)$ should encounter another square root branch point singularity.  At $\tau_1$ we can obtain $\LL_{[\phi\phi]} (\tau)$ from $\hat \LL (\tau)$ using a parallel procedure as~\eqref{eya}. 

\subsection{Future perspectives} 

Here we comment on some immediate extensions of our results and other future directions:

\ben 
\item {\bf Subleading corrections and spatial dependence.} 

One can try to go a step further and analyze subleading corrections and spatial dependence to the leading singular behavior we found in $G_T (\tau)$ -- we discuss corrections in $\delta\tau$ in Appendix~\ref{app:DeltaTau} and the corrections in small $x$ in Appendix~\ref{app:XCorr}.
Our current analysis is not enough for drawing further conclusions.  
We also analyzed  the contributions of lowest-twist operators in the stress-tensor sector. We summarise our early analysis in Appendix~\ref{app:LTanal}.
Interestingly, we  find that the leading twist OPE coefficients grow quicker than the  $\Lambda_n$ coefficients 
(which contain contributions from the $[T_{\mu\nu}]^n$ operators of all twist)
at large $n$ and furthermore do not exactly reproduce the locations of the bouncing singularities. 
This shows that the leading-twist exchanges are not the dominant exchanges in the stress-tensor sector and that the subleading twists are crucial to reproduce the singular behavior. 

\item {\bf Leading twist OPE.} 

It is not surprising that the resummation of the leading twist contributions 
does not lead to the correct bouncing singularities.
The kinematical regime where the leading twist contributions are dominant
is the near-lightcone regime, where $x^-\to0$ with $x^- (x^+)^3/\b^4$ fixed.
Clearly, this regime is very different from the $\vec x=0$ kinematics discussed in 
this paper.
The near-lightcone limit has its advantages -- the OPE coefficients are universal \cite{Fitzpatrick:2019zqz}
and, furthermore, can be computed using CFT bootstrap \cite{Karlsson:2019dbd}.
In addition, the geodesic analysis in this limit takes a simple form \cite{Parnachev:2020fna}.
One may hope that generalization of the results of the present paper to the near-lightcone regime
may provide a useful way of computing analytic observables which can probe   black hole singularities.

\item {\bf Nature of singularities of BTZ black holes?}

For $d=2$, when the spatial direction is a circle, the bulk system is described by a  BTZ black hole, which has an orbifold singularity. The singularity is invisible to the kind of geodesic analysis reviewed in Section~\ref{sec:SC}, and thus to the kind of boundary correlation functions discussed in this paper. The singularity is, however, visible from correlation function of the form 
\be \label{owm}
\tvev{\Phi (X) \phi (0,0)}_\b
\ee
when the bulk point $X$ is taken to the black hole singularity~\cite{HamKab06b} (here $\Phi (X)$ is the bulk field dual to boundary operator $\phi$). $\Phi (X)$ can be expressed in terms of boundary operators using bulk reconstruction, and thus~\eqref{owm} can be written in terms of boundary correlation functions. It would be interesting to see whether it is possible to understand the nature of the BTZ singularities in terms of features of the stress tensor or double-trace contributions to the thermal correlation functions. 

\item {\bf OPE structure for the boundary CFT on a sphere.} 

It would be desirable to use the stress tensor sector for the CFT on a sphere 
for $ d> 2$ to check explicitly whether the bouncing singularities can be recovered
and whether the radius of convergence of the stress-tensor sector
at large $\De$ is larger than $\b/2$.
For this purpose one can use the results of~\cite{Fitzpatrick:2019zqz}
for a spherical black hole geometry and repeat our analysis.

\item {\bf Effects on the black hole singularities from $\apr$ or $G_N$ corrections.}

Connecting   black hole singularities to the singularities of the stress tensor sector of boundary thermal correlators opens 
up new avenues for understanding effects on the black hole singularities from $\apr$ or $G_N$ corrections, and their possible 
resolutions at finite $\apr$ or $G_N$. 

We will use the example of ${\cal N} =4$ super-Yang-Mills theory with gauge group $SU(N)$ as an illustration, for which $\apr$ and $G_N$ translates into 
$1/\sqrt{\lam}$ and $1/N^2$, where $\lam$ is the 't Hooft coupling. 

Consider first a finite, but large $\lam$ in the large $N$ limit, so that we can still talk about 
bulk gravity in terms of spacetime geometry. In this case, there are two important changes:
(i) The coefficients $\Lam_n$ in~\eqref{e.resumL} become $\lam$-dependent \cite{Fitzpatrick:2020yjb}. (ii) 
At finite $\lam$, single-trace operators other than the stress tensor, which can include light operators dual to other supergravity fields and operators corresponding to stringy modes, may develop nonzero thermal expectation values, and can 
contribute to the exchanges in~\eqref{eun}. More explicitly, suppose the 
thermal expectation value of a single-trace operator $O$ is nonzero at finite $\lam$, and $O$ appears in the OPE 
expansion of $\phi (\tau)  \phi (0)$, then in~\eqref{lte} we should add contributions from the exchanges of $O^n$ for $n \in \mathbb{N}$. Denote such a contribution as $G_O (\tau)$, and we should sum over all possible such $G_O (\tau)$. 

(i) arises from $\lam$-dependence of the OPE coefficients as well as the $\lam$-dependence of the thermal expectation value of the stress tensor. These corrections could change the location of the bouncing singularities we saw in $G_T (\tau)$, and can in principle be studied perturbatively in $1/\sqrt{\lam}$ expansion. 
(ii) will make the connection of the singularities in $G_T$ to spacetime singularities discussed in Section~\ref{sec:inter}  more intricate. For example, in the large $\De$ limit,  both $G_T (\tau)$ and $G_O (\tau)$ may contribute in the small $\tau$ regime, with the contribution from $G_O$ interpreted as smearing of geodesics from interacting with stringy excitations. 

At finite $N$ (i.e. finite $G_N$ in the bulk), equation~\eqref{lte} no longer applies. Neither the notion of multiple stress tensor nor double-trace operators make exact sense. But in the large $N$ limit, it should be possible to study perturbative $1/N$ corrections to $G_T (\tau)$, and could lead to insights into the nature of the black hole singularities under perturbative $G_N$ corrections. 

\item {\bf Black hole in the presence of matter fields.} 

In this paper we considered the simplest Schwarzschild black hole. It is interesting to consider solutions with matter fields turned on which can deform the black hole interior as well as the region near the black hole singularities (see e.g.~\cite{Frenkel:2020ysx} for an example). On the boundary side, this corresponds to turning on marginal or relevant deformations. It would be interesting to see whether the stress energy sector 
still captures the bouncing singularities. 

\item {\bf Possible boundary origin of the BKL behavior.} 

In the presence of perturbations, there can be intricate chaotic dynamics in the approach to the black hole singularities~\cite{BKL70,Mis69}, called the BKL  or mixmaster behavior. Recently, such chaotic behavior has been obtained inside a four dimensional asymptotically AdS planar black hole~\cite{DeCHat23}. The connection between the black hole singularities and multiple stress tensor exchange on the boundary should be helpful to understand the boundary origin of the chaotic behavior.
\een


\section*{Acknowledgements}

We would like to thank I. Araya, S. Collier, M. Dodelson, C. Esper, S. Hartnoll, L. Iliesiu, D. Jafferis, R. Karlsson, D. Kolchmeyer, M. Kulaxizi, A. Levine, J.~Maldacena, K. Skenderis, S. Solodukhin, J. Sorce, S. Zhiboedov for discussions. The work of N\v{C} and AP is supported in part by the Science Foundation Ireland under the grant agreement 22/EPSRC/3832. SV was supported in part by the
Irish Research Council Government of Ireland Postgraduate Fellowship under project
award number GOIPG/2023/3661. HL is supported by the Office of High Energy Physics of U.S. Department of Energy under grant Contract Number  DE-SC0012567 and DE-SC0020360 (MIT contract \# 578218).
N\v{C}, AP and SV thank Mainz Institute for Theoretical Physics, where
part of this work was completed, for hospitality.
AP also thanks Harvard University, where
part of this work was completed, for hospitality.


\appendix

\section{Structure of the correlation functions}\label{a.cbd}

In this appendix we discuss the stress-tensor sector of the thermal two-point function of scalar operators \eqref{eq:DefCorrFunagain}. The contribution of the $n$-stress tensor $(T_{\mu\nu})^n$ to the correlator is, up to a constant, the $(T_{\mu\nu})^n$  conformal partial wave\footnote{Here, we are referring to conformal partial waves as defined in \cite{Dolan:2000ut,Dolan:2003hv}. Note that in modern CFT literature this term is sometimes used in a different context.} (CPW) of the heavy-heavy-light-light (HHLL) correlator on $\mathbb{R}\times S^{d-1}_R$ \cite{Karlsson:2021duj}. Light and heavy refer to how the conformal dimensions of the inserted operators scale with the central charge, $C_T$, of the CFT. The conformal dimension of the light operator, $\Delta$, does not scale with the central charge, while for the heavy operator it does,  $\Delta_H \sim \coo{C_T}$.

Starting in flat space, the HHLL four-point function can be expanded in the corresponding CPWs $\widehat{W}_{\Delta',J'}$ as
    \begin{equation}\label{e.cordef}
G(z,\overline z) \equiv \expval{\mathcal{O}_H(0)\phi(z,\overline{z})\phi(1)\mathcal{O}_H(\infty)}=\sum_{\Delta',J'}c_{\Delta',J'}\widehat{W}_{\Delta',J'}\ ,
    \end{equation}
where the sum runs over all primaries and $c_{\Delta',J'}$ is a combination of OPE coefficients.
The flat space CPW are given by \cite{Dolan:2000ut,Dolan:2003hv}%
\footnote{Compared to \cite{Dolan:2000ut} we omit the factor $(-2)^{-J}$ in the conformal waves.}
{\allowdisplaybreaks{\small{
    \begin{align}
        d=2\quad\,&\widehat{W}_{\Delta',J'}=\frac{1}{(Z\Zb)^\Delta}\left(k_{\Delta'+J'}(Z)k_{\Delta'-J'}(\Zb)+k_{\Delta'-J'}(Z)k_{\Delta'+J'}(\Zb)\right)\label{e.2db}\\
        d=4\quad\,&\widehat{W}_{\Delta',J'}=\frac{1}{(Z\Zb)^\Delta}\frac{Z\Zb}{Z-\Zb}\left(k_{\Delta'+J'}(Z)k_{\Delta'-J'-2}(\Zb)-k_{\Delta'-J'-2}(Z)k_{\Delta'+J'}(\Zb)\right)\label{e.4db}\\   
        d=6\quad\,&\widehat{W}_{\Delta',J'}=\frac{1}{(Z\overline{Z})^\Delta}\Big(\mathcal{F}_{0,0}(Z,\overline{Z})-\frac{J'+3}{J'+1}\mathcal{F}_{-1,1}(Z,\overline{Z})\nonumber\\
        &\hspace{1.5cm}-\frac{(\Delta'-4)(\Delta'+J')^2}{16(\Delta'-2)(\Delta'+J'+1)(\Delta'+J'-1)}\mathcal{F}_{1,1}(Z,\overline{Z})\label{e.6db}\\
        &\hspace{1.5cm}+\frac{(\Delta'-4)(J'+3)}{(\Delta'-2)(J'+1)}\frac{(\Delta'-J'-4)^2}{16(\Delta'-J'-5)(\Delta'-J'-3)}\mathcal{F}_{0,2}(Z,\overline{Z})\Big)\ ,\nonumber
    \end{align}}}}where
{\allowdisplaybreaks{\small{
    \begin{align}
    &k_\eta(\xi)=\xi^{\frac{\eta}{2}}\!\!\!\!\phantom{F}_2F_1(\tfrac{\eta}{2},\tfrac{\eta}{2},\eta,\xi)\,,\\
    &\mathcal{F}_{n,m}(Z,\overline{Z})=\frac{(Z\overline{Z})^{\frac12 (\Delta'-J')}}{(\overline{Z}-Z)^3}(\mathbb{F}_{n,m}(Z,\overline{Z})-\mathbb{F}_{n,m}(\overline{Z},Z))\,,\\
    &\mathbb{F}_{n,m}(Z,\overline{Z})=\overline{Z}^{J'+n+3}Z^m\!\!\!\!\phantom{F}_2F_1\Big(\frac{\Delta'+J'}{2}+n,\frac{\Delta'+J'}{2}+n;\Delta'+J'+2n;\overline{Z}\Big)\nonumber\\
    &\hspace{2.2cm}\times\!\!\!\!\phantom{F}_2F_1\Big(\frac{\Delta'-J'}{2}-3+m,\frac{\Delta'-J'}{2}-3+m;\Delta'-J'-6+2m;Z\Big)\ .
    \end{align}
}}}%
We will use that in the t-channel the relation between $(Z,\Zb)$ and the cross-ratios $(z,\overline{z})$ is \mbox{$(Z,\Zb)=(1-z,1-\overline{z})$}.

We now map the flat space to $\mathbb{R}\times S^{d-1}_R$ using 
    \begin{equation}\label{e.cylmap}
            z=1-Z=e^{-\frac{\tau}{R}-i\frac{x}{R}}\,, \qquad \overline{z} = 1- \overline{Z} = e^{-\frac{\tau}{R}+i\frac{\vec{x}}{R}}\,, 
    \end{equation}
where $R$ is the sphere radius and $\vec{x}$ schematically denotes all coordinates besides $\tau$.
This transformation introduces an overall prefactor in conformal partial waves
    \begin{equation}\label{e.cylcpw}
        W_{\Delta',J'}=R^{-2\Delta}(z\overline{z})^\frac{\Delta}{2}\widehat{W}_{\Delta',J'}=R^{-2\Delta}\left((1-Z)(1-\overline{Z})\right)^\frac{\Delta}{2}\widehat{W}_{\Delta',J'}\,.
    \end{equation}
The $W_{\Delta',J'}$ are the CPW on $\mathbb{R}\times S^{d-1}_R$, where the two light operators are inserted at 0 and $(\tau,\vec{x})$ while the heavy operators sit at $-\infty$ and $+\infty$.
Via the operator-state correspondence, the correlator on $\mathbb{R}\times S^{d-1}_R$ can be seen as a two-point function in a heavy state $\ev{\phi(\tau,x)\phi(0,0)}{\mathcal{O}_H}$. 

Let us now focus on the exchanges of $n$-stress tensors, which have conformal dimension $\Delta'=d\,n$ and spin $J'=0,2,4,\ldots,2n$.
Since these thermalize in heavy states \cite{Karlsson:2021duj}, their expansion in CPW \eqref{e.cylcpw} precisely matches the corresponding expansion in terms of thermal conformal blocks \cite{Iliesiu:2018fao}.
We are interested in the OPE limit\footnote{In this limit we effectively work on $\mathbb{R}^d$. It is however important to distinguish this from the original flat space-time on which correlator \eqref{e.cordef} was formulated.} $z,\overline{z}\rightarrow1$ (or equivalently $Z,\Zb\rightarrow0$), which in $\mathbb{R}\times S^{d-1}_R$ corresponds to $\tau,\abs{\vec{x}}\ll R$.

First consider the case with $x=\abs{\vec{x}}=0$.  In the regime $\tau\ll R$,
the corresponding CPWs simplify to 
    \begin{align}
    d=2\qquad&W_{2n,J'}\approx2\tau^{-2\Delta}\left(\frac{\tau}{R}\right)^{2n}\label{e.assu}\\
    d=4\qquad&W_{4n,J'}\approx(1+J')\tau^{-2\Delta}\left(\frac{\tau}{R}\right)^{4n}\label{e.for4d}\\
    d=6\qquad&W_{6n,J'}\approx\frac16(2+J')(3+J')\tau^{-2\Delta}\left(\frac{\tau}{R}\right)^{6n}\ .   
    \end{align}
Each $n$-stress tensor exchange is multiplied by a factor of $\mu^n$, where%
\footnote{In holographic theories, $\mu$ is also the mass parameter appearing in the metric of the AdS-Schwarzschild black hole, see for example \eqref{eq:d=4fFunction}.
For a black brane in $d+1$ dimensions, the relation between $\mu$ and the (inverse) Hawking temperature is 
    \begin{equation*}
        \mu=\left(\frac{4\pi R}{d\,\beta}\right)^d\ ,
    \end{equation*}
    which scales with $R/\beta$ exactly as in \eqref{eq:Mutemp}.}
\begin{align}
\label{eq:Mutemp}
    \mu \propto \frac{\Delta_H}{C_T}
    \propto \varepsilon\,\frac{R^d}{C_T}
    \propto \left(\frac{R}{\beta}\right)^d\,.
\end{align}
In the above, we used that the energy density $\varepsilon$ is proportional to $C_T\,T^{d}$ \cite{El-Showk:2011yvt, Iliesiu:2018fao}.
It is convenient to isolate these dimensionful factors in the OPE coefficients and define
\begin{equation}
        c_{d\,n,J'} \equiv  \lambda_{n,J'}\left(\frac{R}{\beta}\right)^{d\,n}\,,
\end{equation}
as in this case all dependence on the sphere radius $R$ completely cancels out in the stress-tensor sector of the thermal correlator%
\footnote{The subscript in $G_T(\tau)$ denotes that this is the stress-tensor sector and \emph{not} the full correlator.}
{\allowdisplaybreaks{\small{
    \begin{align}
        d=2\quad\,&G_T(\tau)=\frac{1}{\tau^{2\Delta}}\left(1+2\lambda_{1,2}\left(\frac{\tau}{\beta}\right)^2+\sum_{n=2}^\infty\left[\sum_{J'}\,2\lambda_{n,J'}\right]\left(\frac{\tau}{\beta}\right)^{2n}\right)\,,\\
        d=4\quad\,&G_T(\tau)=\frac{1}{\tau^{2\Delta}}\left(1+3\lambda_{1,2}\left(\frac{\tau}{\beta}\right)^4+\sum_{n=2}^\infty\left[ \sum_{J'}\,\lambda_{n,J'}(1+J')\right]\left(\frac{\tau}{\beta}\right)^{4n}\right)\,,\label{e.lebara}\\
        d=6\quad\,&G_T(\tau)=\frac{1}{\tau^{2\Delta}}\left(1+\frac{10}{3}\lambda_{1,2}\left(\frac{\tau}{\beta}\right)^6+\sum_{n=2}^\infty\left[ \sum_{J'}\,\lambda_{n,J'}\frac{(2+J')(3+J')}{6}\right]\left(\frac{\tau}{\beta}\right)^{6n}\right)\,,\label{e.6dcb}
    \end{align}}}}%
where the sums over $J'$ run over $J'=0,2,\ldots 2\,n$.
Finally, it is useful to introduce summed coefficients $\Lambda_n$ through
    \begin{equation}\label{e.topeL}
        G_T(\tau)=\frac{1}{\tau^{2\Delta}}\sum_{n=0}^\infty\Lambda_n\left(\frac{\tau}{\beta}\right)^{d\,n}\,,
    \end{equation}
which we use in the main text.

Now let $x$ be  small, but non-vanishing, such that \mbox{$x\ll\tau\ll R$}.
For concreteness, we focus only on $d=4$.
Mapping the flat space CPW \eqref{e.4db} to $\mathbb{R}\times S^{d-1}_R$, taking the OPE limit, and then expanding in $x\ll\tau$, we get the stress-tensor sector of the correlator as
    \begin{equation}\label{e.extendedg}
    \begin{split}
        G_T(\tau,x)&=\frac{1}{\tau^{2\Delta}}\left(1-\frac{\Delta\,x^2}{\tau^2}\right) \sum_{n=0}^\infty \left[\Lambda_n+\left(2\,n\,\Lambda_n +\wlo\right)\frac{x^2}{\tau^2}\right]\left(\frac{\tau}{\beta}\right)^{4\,n}\!\!+\coo{x^4},
    \end{split}
    \end{equation}
with $\Lambda_n$ defined in \eqref{e.topeL} and
    \begin{equation}\label{e.dufamzeposledna}
        \wlo=-\frac{1}{6}\sum_{J'=0,2,\ldots,2n}\!\!\!J'\,(1+J')\,(2+J')\,\lambda_{n,J'}\ .
    \end{equation}
Note that $n=0$ corresponds to the identity contribution and thus $\widetilde \Lambda^{(1)}_0=0$, while at  $n=1$ the only non-vanishing contribution comes from the stress tensor with $J'=2$.

To make the notation for $x\neq0$ more transparent, we generalise equation\ \eqref{e.topeL} to
    \begin{equation}\label{e.exnenula}
        G_T(\tau,x)=\frac{1}{\tau^{2\Delta}}\sum_{n=0}^\infty\left[\Lambda^{(0)}_n+\frac{x^2}{\tau^2}\,\Lambda^{(1)}_n+\coo{\frac{x^4}{\tau^4}}\right]\left(\frac{\tau}{\beta}\right)^{dn}\ ,
    \end{equation}
where $\Lambda^{(0)}_n\equiv\Lambda_n$ and $\Lambda^{(1)}_n\equiv(2n-\Delta)\Lambda_n+\wlo$. In the same way one can define $\Lambda_n^{(m)}$ for any $m$. 
Note that the holographic method we used to determine the stress-tensor contributions extracts $\lambda_{n,J'}$ \cite{Fitzpatrick:2019zqz}.
This means that in principle one can use this method  to obtain the coefficients $\Lambda^{(m)}_n$ to arbitrary high orders in $m$ and $n$.

Let us conclude this appendix by comparing the above analysis with
the expansion of the correlator using the thermal conformal blocks formulated  on $\mathbb{S}_\beta^1\times\mathbb{R}^{d-1}$  \cite{Iliesiu:2018fao}
    \begin{equation}\label{e.fgege}
        G_T(\tau,x)=\frac{1}{\abs{\tau^2+x^2}^\Delta}\sum_{n=0}^\infty\sum_{J'=0,2,\ldots,2n}\!\!\!\hat{\lambda}_{n,J'}\frac{\abs{\tau^2+x^2}^\frac{dn}{2}}{\beta^{dn}}C_{J'}^{\left(\frac{d+2}{2}\right)}\left(\frac{\tau}{\sqrt{\tau^2+x^2}}\right)\ ,
    \end{equation}
where $C_{J'}^{(\nu)}(\eta)$ are Gegenbauer polynomials and $\hat{\lambda}_{n,J'}$ are dimensionless coefficients. Imposing $x=0$ and restricting to $d>2$, \eqref{e.fgege} simplifies to
    \begin{equation}\label{e.fgeg2}
        G_T(\tau)=\frac{1}{\tau^{2\Delta}}\sum_{n=0}^\infty\sum_{J'=0,2,\ldots,2n}\!\!\!\hat{\lambda}_{n,J'}\smqty(d-3+J \\ J)\left(\frac{\tau}{\beta}\right)^{dn}\ .
    \end{equation}
This can now be compared with the equations \eqref{e.lebara} and \eqref{e.6dcb}. For $d=4$ one finds $\hat{\lambda}_{n,J'}=\lambda_{n,J'}$, while in $d=6$ there is a conventional difference by a factor $(J+1)$ in the coefficients $\lambda_{n,J'}$ and $\hat{\lambda}_{n,J'}$.


    \section{Bulk equation of motion}
    \label{app:eomcoefz}
    In this appendix we give the explicit form of the equation of motion
    \begin{equation}\label{e.wew}
        (\Box-m^2)\phi=0\, ,\qquad m^2=\Delta(\Delta-d)
    \end{equation}
in the background of ($d+1$)-dimensional  planar Euclidean Schwarzschild-AdS black hole after the coordinate  transformation 
    \begin{equation}
    \rho^2=r^2\vec{x}^2\qq{and}w^2=1+r^2(\tau^2+\vec{x}^2). 
    \end{equation}
    Introducing the following ansatz for the scalar bulk field
    \begin{equation}\label{e.anz1again}
        \phi(w,\rho,r)=\left(\frac{r}{w^2}\right)^\Delta \psi(w,\rho,r)\ ,
    \end{equation}
where $(r/w^2)^\Delta$ is the solution in the pure AdS space, ``reduces'' the Klein-Gordon equation \eqref{e.wew} to  \cite{Fitzpatrick:2019zqz}
{\small{
    \begin{equation}\label{e.theequation}
    \left(\partial_r^2\!+\!C_1\partial_w^2\!+\!C_2\partial_\rho^2\!+\!C_3\partial_r\partial_w\!+\!C_4\partial_r\partial_\rho\!+\!C_5\partial_w\partial_\rho\!+\!C_6\partial_r\!+\!C_7\partial_w\!+\!C_8\partial_\rho\!+\!C_9\right)\psi=0\ ,
    \end{equation}}}with
the coefficients $C_i$ given by
{\allowdisplaybreaks{\small{   
    \begin{align}
        C_1&=\frac{f\left(\rho^2+(w^2-1)^2f\right)+w^2-\rho^2-1}{r^2w^2f^2}\,,\\
        C_2&=\frac{1+f\rho^2}{r^2f}\,,\\
        C_3&=\frac{2}{rw}(w^2-1)\,,\\
        C_4&=\frac{2\rho}{r}\,,\\
        C_5&=\frac{2\rho}{r^2wf}\left(1+(w^2-1)f\right)\,,\\
        C_6&=\frac{1}{f}\dv{f}{r}+\frac{w^2(10-4\Delta)+8\Delta}{2rw^2}+\frac{d-4}{r}\,,\\
        C_7&=\left(\frac{1}{rwf}\dv{f}{r}-\frac{w^2(2\Delta-5)-4\Delta-1}{r^2w^3}\right)\left(w^2-1\right)\nonumber\\
        &\phantom{=}+\frac{3w^2-\rho^2(1+4\Delta)}{r^2w^3f}+\frac{1+\rho^2+4\left(1-w^2+\rho^2\right)\Delta}{r^2w^3f^2}+(d-4)\frac{C_5}{2\rho}\,,\\
        C_8&=\frac{2(w^2-2\rho^2\Delta)+\rho^2\left(w^2(5-2\Delta)+4\Delta\right)f}{r^2w^2\rho f}+\frac{\rho}{rf}\dv{f}{r}+(d-4)\frac{C_2}{\rho}\,,\\
        C_9&=\frac{\Delta}{w^2}\Bigg(\frac{(w^2-2)^2\Delta+4(1+w^2-w^4)}{r^2w^2}+\frac{4\rho^2(\Delta+1)-w^4(\Delta-4)-6w^2}{r^2w^2f}\nonumber\\
        &\phantom{=}+\frac{2w^2(1+2\Delta)-4(1+\rho^2)(1+\Delta)}{r^2w^2f^2}-\frac{w^2-2}{rf}\dv{f}{r}-(d-4)\frac{(w^2-2)(f-1)}{r^2f}\Bigg)\ ,
    \end{align}}}}where
$f=1-\frac{\mu}{r^d}$.


\section{Validity of approximating the OPE by an integral}\label{a.sumz}

In the main text we have approximated the OPE sums by an integral, such as in \eqref{e.sin1}. In this appendix we now discuss the validity of this approximation.

We want to resum the OPE \eqref{e.resumL} where the coefficients are given by their asymptotic form \eqref{eq:Lambdand=4Ansatz}. Assume that this asymptotic form is a good approximation to the actual OPE coefficients after a certain value $n = n_*$ and that one can neglect terms with $n<n_*$. The first assumption comes from our analysis of OPE data. We can justify the second assumption by noting that 
near the critical points $\tau \approx \tau_c$, which  are at the radius of convergence of the stress-tensor OPE, all terms in the $\tau$ expansion contribute. 
One sees from \eqref{eq:Lambdand=4Ansatz} that for large enough $\Delta$ the OPE coefficients are increasing with $n$ and thus close enough to the critical points, the large-$n$ terms will be the most important. All in all, to calculate the stress-tensor contribution to the correlator, one has to evaluate a sum of the type
    \begin{equation}\label{e.vzorovasume}
        \sum_{n=n_*}^\infty n^{a\Delta+b}y^{d n}=y^{d\,n_*}\,\Phi\left(y^d,-a\Delta-b,n_*\right)\ ,
    \end{equation}
where  $\Phi$ is the Hurwitz-Lerch transcendent, $\abs{y}\leq1$, and $a>0$ and $b$ are constants. We want to compute this sum in the limit $y\rightarrow1$, which corresponds to the correlator near $\tau \approx \tau_c$. For $\Delta>-\frac{a+b}{2}$, one can expand
    \begin{equation}
    \begin{split}
        &y^{d\,n_*}\,\Phi(y^d,-a\Delta-b,n_*)=\\
        &\hspace{1.3cm}\Gamma(1+a\Delta+b)(-\log y^d)^{-(1+a\Delta+b)}+\sum_{k=0}^\infty\zeta(-a\Delta-b-k,n_*)\frac{\left(\log y^d\right)^k}{k!}\ ,
    \end{split}
    \end{equation}
where $\zeta$ is the generalised Riemann zeta function. 
The second term is regular at $y=1$ for all $k$, so it gives a subleading contribution to the correlator near the critical point.
Therefore, near $\tau \approx \tau_c$ we can approximate the sum \eqref{e.vzorovasume} by
    \begin{equation}
    \sum_{n=n_*}^\infty n^{a\Delta+b}y^{dn}\approx\Gamma(1+a\Delta+b)(-\log y^d)^{-(1+a\Delta+b)}=\int_0^\infty n^{a\Delta+b}y^{dn}\,dn\ .        
    \end{equation}
This justifies the exchange of the sum for an integral from 0 to $\infty$.


	\section{The KMS pole and OPE in \texorpdfstring{$d=2$}{d=2} }\label{s.kms2d}

In this appendix we show how the KMS pole emerges from the OPE in $d=2$. The finite temperature two-point function is known in a closed form and  the KMS pole can be seen explicitly without the use of the OPE.
Nevertheless, the recovery of the KMS pole from the OPE can serve as a guideline for the analysis in higher dimensional cases, where the thermal two-point functions are not known exactly.

\subsection{KMS pole}

Consider a scalar two-point function at finite temperature $T=\beta^{-1}$ 
    \begin{equation}\label{e.def}
        G(\tau,x)=\expval{\phi(\tau,x)\phi(0,0)}_\beta\ .
    \end{equation}
This can be rewritten, using the periodicity of the trace, as
    \begin{equation}
    \begin{split}
    \label{e.kmsc}
G(\tau,x)=\frac{1}{Z}\Tr e^{-\beta H}\phi(\tau,x)\phi(0,0)
=\frac{1}{Z}\Tr e^{-\beta H}\phi(\beta,0)\phi(\tau,x)=G(\beta-\tau,-x)\ ,
    \end{split}
    \end{equation}
which is the KMS condition \cite{Kubo:1957mj,Martin:1959jp}. Let us consider the case where $x=0$. For a unit-normalised scalar with scaling dimension $\Delta$, the small $\tau$ behaviour is
    \begin{equation}
G(\tau)\xrightarrow{\tau\rightarrow0}\frac{1}{\abs{\tau}^{2\Delta}}\ .
    \end{equation}
From \eqref{e.kmsc}, it then follows that $G$ also has a \textit{KMS pole} at
    \begin{equation}\label{e.pole}
      G(\tau) \xrightarrow{\tau\rightarrow\beta}\frac{1}{\abs{\beta-\tau}^{2\Delta}}\ .
    \end{equation}
This analysis is valid for any quantum field theory at finite temperature. 
When dealing with CFTs one has the additional tool of the OPE with a non-vanishing radius of convergence, which is typically determined by the first singular point encountered in the complex plane.
Thus, if there are no singularities closer to the origin, the KMS pole should be encoded in the asymptotic behaviour of the OPE.

\subsection{OPE analysis}

Let us consider a CFT in two dimensions. 
Since $S_\beta\times\mathbb{R}$ is conformally equivalent to $\mathbb{R}^2$, the two-point correlation function  is known in a closed form \cite{Iliesiu:2018fao}, and in the case of two identical scalar operators it is given by
    \begin{equation}\label{e.expl}
G(\tau, x)=\left[\frac{\beta}{\pi}\sinh\left(\frac{\pi(x-i\tau)}{\beta}\right)\right]^{-\Delta}\left[\frac{\beta}{\pi} \sinh\left(\frac{\pi(x+i\tau)}{\beta}\right)\right]^{-\Delta}\ ,
    \end{equation}
where we have chosen the normalisation such that in the zero-temperature limit we recover unit norm, $i.e.$ $\lim_{\beta\rightarrow\infty}G(\tau,x)=(\tau^2+x^2)^{-\Delta}$.

\begin{figure}
\includegraphics[width=0.8\textwidth]{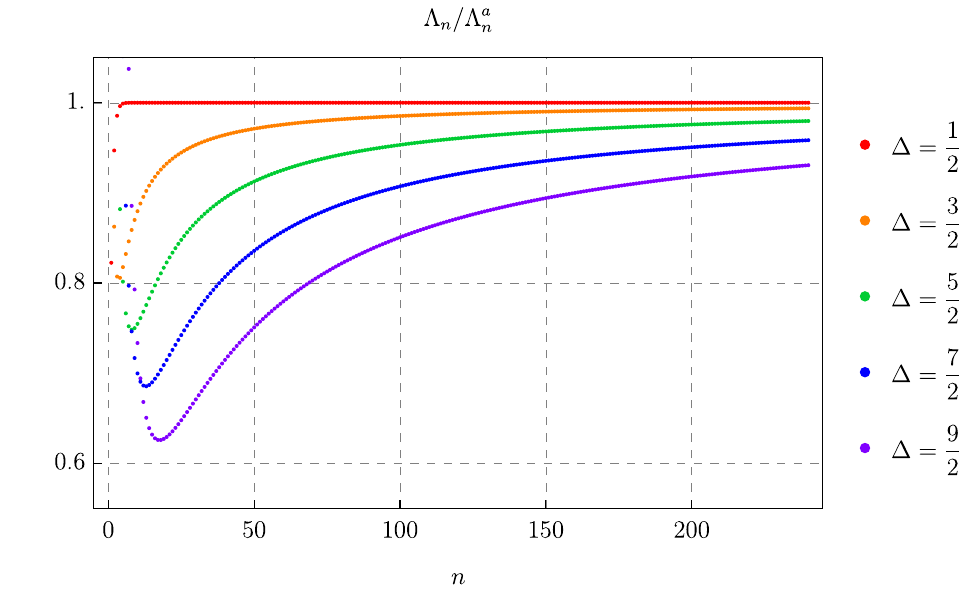}
	\centering
    \caption{Ratio of the explicit results for $\Lambda_n$ to the leading large-$n$ prediction $\Lambda_n^a$, given in \eqref{eq:AsyFormd2}, for different values of $\Delta$ in $d=2$.}
    \label{fig.Lambdand=2}
\end{figure}

Let us first set $x=0$. In a two-dimensional CFT, only the Virasoro vacuum module contributes to the thermal correlator. In other words,  the only non-zero contribution comes from the multi-stress tensors and thus the full correlator can be expanded as
    \begin{equation}\label{e.sum}
G(\tau)=\frac{1}{\tau^{2\Delta}}\sum_n\Lambda_n\left(\frac{\tau}{\beta}\right)^{2n}\ .
    \end{equation}
The coefficients $\Lambda_n$ can simply be read off from the  expansion of the correlator \eqref{e.expl} near $\tau =0$. We are interested in the behaviour of the OPE near its convergence radius, where we expect all terms in the expansion to be of similar magnitude. As such, we are interested in the behaviour of $\Lambda_n$ for large values of $n$. One finds that for large enough $n$ these can be written in a $1/n$ expansion (see figure~\ref{fig.Lambdand=2})
\begin{equation}\label{eq:LambdaNd=2}
    \begin{split}
    \Lambda_n &= \frac{2^{2\Delta}}{\Gamma(2\Delta)}\,n^{2\Delta-1}\sum_{k=0}^\infty\frac{c_k(\Delta)}{n^k}
    = \frac{2^{2\Delta}}{\Gamma(2\Delta)}\,n^{2\Delta-1}\left(1+ \frac{c_1(\Delta)}{n} + \ldots  \right)\,,
    \end{split}
\end{equation}
where we choose the overall prefactor in such a way that $c_0(\Delta) = 1$, while all remaining  $c_k(\Delta)$  can be arbitrary functions of $\Delta$.
We see that as $n \to \infty$ the coefficients tend to 
\begin{align}
    \label{eq:AsyFormd2}
    \Lambda_n^a &= \frac{2^{2\Delta}}{\Gamma(2\Delta)}\,n^{2\Delta-1}\,,
\end{align}
which determines the leading order behaviour of the correlator.
For $\tau$ near the radius of convergence of the OPE \eqref{e.sum}, we can approximate the sum by an integral
\begin{equation}\label{eq.Corr2dLead}
\begin{split}
    G(\tau) &\approx  \frac{1}{\tau^{2\Delta}}\int_0^\infty \Lambda_n^a\,\left(\frac{\tau}{\beta}\right)^{2n} \,d n
    =\left[-\frac{\tau}{2}\log\left(\frac{\tau^2}{\beta^2}\right)\right]^{-2\Delta}\,,
\end{split}
\end{equation}
where we have taken \eqref{eq:AsyFormd2} for the OPE coefficients.
We encounter a singularity when the argument of the logarithm is equal to 1, which happens at $\tau = \pm \beta$ where we find
    \begin{equation}
        G(\tau) \xrightarrow{\tau\rightarrow\pm \beta}\frac{1}{|\beta-\tau|^{2\Delta}}\,,
    \end{equation}
which are exactly the KMS poles \eqref{e.pole}.
This explicitly shows that the asymptotic OPE analysis reproduces the first non-trivial poles of the full correlator \eqref{e.expl} in the complex-$\tau$ plane.
However, as may be expected, the OPE analysis does not contain any information about the higher order poles  at $|\tau| > \beta $.

\paragraph{Subleading analysis:}

\begin{figure}[t]
    \centering
    \includegraphics[width=\textwidth]{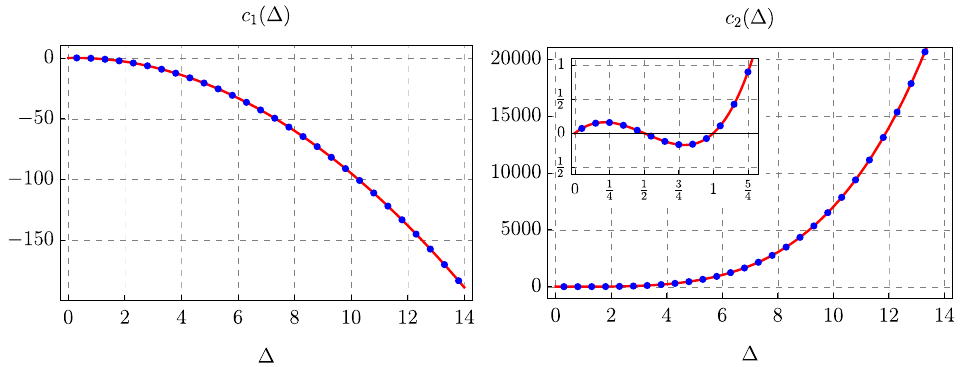}
    \vspace*{-0.6cm}
    \caption{The comparison between $c_1(\Delta)$ (left) and $c_2(\Delta)$ (right) as obtained from expansion of the exact correlator (red curve) and the results from the direct analysis of $\Lambda_n$ for large $n$ (blue markers).}
    \label{fig:c1c22d}
\end{figure}

The above analysis shows that KMS pole is already contained in $\Lambda_n^a$, the leading behaviour of the OPE coefficients at large $n$. 
Let us now analyse the $1/n$ corrections in \eqref{eq:LambdaNd=2}.
The coefficients $c_k(\Delta)$ are determined by carefully analysing $\Lambda_n$ as a function of $\Delta$,\footnote{We discuss how to determine the form of \eqref{eq:LambdaNd=2} and the values $c_k(\Delta)$ from $\Lambda_n$ in more detail in Appendix~\ref{app:DeltaTau}, where we analyse  the correlator in $d=4$.} see blue markers in Figure\ \ref{fig:c1c22d}. 

Alternatively, one can follow a different approach to obtain coefficients $c_k(\Delta)$:
Insert the full expansion \eqref{eq:LambdaNd=2} into the sum \eqref{e.sum} and again approximate it with an integral
\begin{equation}\label{eq.Gd2SubDeltaTauInt}
    \begin{split}
    G(\tau) =& \frac{2^{2\Delta}}{\Gamma(2\Delta)\, \tau^{2\Delta}}\int_0^\infty \!n^{2\Delta-1}\left(\frac{\tau}{\beta}\right)^{2n}\left(1+ \frac{c_1(\Delta)}{n} +\frac{c_2(\Delta)}{n^2}+ \ldots  \right)d n\\
    =& \left[-\tau\log\left(\frac{|\tau|}{\beta}\right)\right]^{-2\Delta}\,\Bigg[1+ \frac{2\,c_1(\Delta)}{2\Delta-1}\,\left(-\log\left(\frac{|\tau|}{\beta}\right)\right) \\
    &\hspace{2.4cm}+ \frac{4\,c_2(\Delta)}{(2\Delta-1)(2\Delta-2)}\,\left(-\log\left(\frac{|\tau|}{\beta}\right)^2\right)+ \ldots\Bigg]\,.
    \end{split}
\end{equation}
Expanding this result around%
\footnote{One can equally expand around $\tau = - \beta$, but we will focus on this pole without the loss of generality.}
$\tau = \beta$ then gives
\begin{align}
    \label{eq:Gtaud2Exp}
    G(\tau) &\xrightarrow{\tau\rightarrow\beta}\frac{1}{(\beta-\tau)^{2\Delta}}\Bigg[1 + \frac{1}{\beta}\left(\Delta + \frac{2\,c_1(\Delta)}{2\Delta-1}\right) (\beta -\tau)\\
    &+ \frac{1}{\beta^2}\left(\frac{\Delta(6 \Delta +7)}{12}+ \frac{(2\Delta+1)c_1(\Delta)}{2\Delta-1}+ \frac{2\,c_2(\Delta)}{(2\Delta-1)(\Delta-1)}\right) (\beta -\tau)^2 + \ldots \Bigg]\nonumber\,.
\end{align}
We thus see that the $1/n$ corrections in \eqref{eq:LambdaNd=2} translate to $(\beta-\tau)$ corrections to the correlator near the KMS pole.

We can now use the exact form of the correlator \eqref{e.expl} to  determine the coefficients $c_k(\Delta)$.
Expand \eqref{e.expl} around $\tau = \beta$ to first subleading order
\begin{align}
    \label{eq.G2dExpSubTau}
    G(\tau) = \frac{1}{(\beta - \tau)^{2\Delta}}\left(1+ \frac{\Delta\,\pi^2}{3\,\beta^2}\,(\beta - \tau)^2+ \coo{(\beta-\tau)^3}\right)\,.
\end{align}
This expression and \eqref{eq:Gtaud2Exp} should match, which leads to
\begin{align}
    \label{eq:2dCkPred}
    c_1(\Delta) =\Delta\,\left(\frac12-\Delta\right)\,,\qquad
    c_2(\Delta) = \frac{1}{24}\Delta\,(\Delta-1)(2\Delta-1)(6\Delta + 4\,\pi^2 -1)\,,
\end{align}
which can be continued to arbitrary $k$.
One can compare these expressions with the direct data obtained from analysing the asymptotic form of $\Lambda_n$ and find perfect agreement, see Figure~\ref{fig:c1c22d}.

\paragraph{Nonzero $x$:}
Let us now set $x \neq 0$, so that $x\ll \tau$. 
Essentially, we  work in the limit where $x$ is the smallest length scale in the expression.
Expanding the exact correlator \eqref{e.expl} in small $x$ gives to leading order
\begin{align}
\label{eq:GtxExp}
    G(\tau, x) = \left[\frac{\beta}{\pi}\sin\left(\frac{\pi\,\tau}{\beta}\right)\right]^{-2\Delta} + \Delta\,x^2\,\left[\frac{\beta}{\pi}\sin\left(\frac{\pi\,\tau}{\beta}\right)\right]^{-2(\Delta+1)}+ \coo{x^4}\,,
\end{align}
where each term contains non-trivial poles at $\tau = m\,\beta$, with $m\in \mathbb{Z}$.
We expand each term in the above series individually around $\tau = \beta$, focusing only on the leading behaviour 
\begin{align}
\label{eq:d2SmallxPoles}
    G(\tau, x) \approx \left(\beta - \tau\right)^{-2\Delta}\left[ 1 + \frac{1}{(\beta-\tau)^2}\,\frac{\Delta\,x^2}{\beta^2}\right]\,.
\end{align}
Naively, one would think  that the $x^2$ term  is more divergent than the leading order term.
However, this expansion is only valid if $x$ is the smallest parameter in the expression, \textit{i.e.}  $x^2\ll (\beta -\tau)^2$, in which case all corrections to the leading divergent behaviour in \eqref{eq:d2SmallxPoles} are small. 

Similar to the analysis of the subleading $(\beta-\tau)$ contributions we would like to reproduce this result using the OPE analysis. 
For that we take \eqref{eq:GtxExp} and expand each term individually in a series in $\tau$
\begin{align}
\label{eq:GtauxExp2}
    G(\tau, x) = \frac{1}{\tau^{2\Delta}}\left[\sum_{n=0}^\infty \Lambda_n^{(0)}\,\left(\frac{\tau}{\beta}\right)^{2n} + \frac{\Delta \,x^2}{\beta^2} \sum_{n=0}^\infty \Lambda_n^{(1)}\,\left(\frac{\tau}{\beta}\right)^{2n} + \coo{x^4} \right]\,.
\end{align}
Again, we can make an ansatz for asymptotic behaviour of the OPE coefficients
\begin{align}
    \Lambda_n^{(\alpha)} = c_0^{(\alpha)}(\Delta)\, n^{a\, \Delta +b}\,,
\end{align}
where we ignore all $1/n$ corrections, since we are only interested in the leading behaviour at each order in $x$. We find 
\begin{align}
    \Lambda_n^{(0)} \approx \frac{2^{2\Delta}}{\Gamma(2\Delta)}\,n^{2\Delta-1}\,,\qquad \Lambda_n^{(1)} \approx \frac{2^{2(\Delta+1)}}{\Gamma(2(\Delta+1))}\,n^{2\Delta+1}\,.
\end{align}
Inserting these values into \eqref{eq:GtauxExp2} and replacing the sum with the integral leads to 
\begin{align}
    G(\tau,x) &\approx \frac{1}{\tau^{2\Delta}}\int_0^\infty\left[\frac{2^{2\Delta}}{\Gamma(2\Delta)}\,n^{2\Delta-1}+\frac{\Delta\,x^2}{\beta^2} \,\frac{2^{2(\Delta+1)}}{\Gamma(2(\Delta+1))}\,n^{2\Delta+1}\right]\left(\frac{\tau}{\beta}\right)^{2n}d n\,\nonumber\\*
    &= \frac{1}{\tau^{2\Delta}}
\left[\left(-\log\frac{|\tau|}{\beta}\right)^{-2\Delta} +\frac{\Delta\,x^2}{\beta^2} \,\left(-\log\frac{|\tau|}{\beta}\right)^{-2(\Delta+1)}\right]\,.
\end{align}
By expanding this near $\tau=\beta$ we exactly reproduce \eqref{eq:d2SmallxPoles} -- the asymptotic OPE analysis reproduces the KMS poles of the exact two-point correlation function for $x\neq0$.

Let us conclude this appendix by noting that in the main text and in the appendices that follow  we repeat this analysis for thermal correlators in higher dimensions. 
In these cases we can still extract $c_k(\Delta)$ (respectively $\Lambda_n^{(1)}$) by careful analysis of the OPE coefficients. However, since a closed form expression of the thermal correlator is not known, to cross-check these results we compare them with the geodesic analysis at subleading $\delta\tau$ (respectively $x\neq0$).


\section{Analysis at subleading \texorpdfstring{$\delta\tau$}{delta tau} in \texorpdfstring{$d=4$}{d=4}}
\label{app:DeltaTau}

In this appendix we analyse the contributions to $\mathcal{L}_T(\tau)$ subleading in $\delta \tau$ and how these arise as $1/n$ corrections to the asymptotic form of the OPE coefficients. This analysis is valid for any finite $\Delta$.
When $\Delta\rightarrow\infty$ the situation becomes more subtle. As we argued in Section\ \ref{s.abintldl}, in this limit, the cross-over point also goes to infinity, $n^*= \Delta/2\to \infty$, which  leads to disappearance of the bouncing singularity in $\mathcal{L}_T(\tau)$.  However, as we saw in \eqref{eq:ProperLengthNearLim},  taking $\Delta$ large \emph{after} expanding the logarithm of the correlator near the bouncing singularity leads to the behaviour expected from the bouncing geodesic. Below we examine if this match persists beyond the leading order singularity.
We also detail the procedure that we used in the main text to determine the asymptotic form of the OPE coefficients $\Lambda_n$.

The main object of interest are the OPE coefficients $\Lambda_n$, as defined in \eqref{e.resumL}. In particular, we are interested in their behaviour when $n$ is large. 
Ideally, one would find the exact expressions of these OPE coefficients as functions of $\Delta$, however, in practice, finding such expressions for large $n$ is computationally too expensive. As already mentioned in the main part of this paper, it is  more efficient to first fix $\Delta$ and then calculate $\Lambda_n$ for that specific value. In this way, we are able to calculate $\Lambda_n$ up to $n \approx 50$ in about 5 days on a standard desktop machine.

\begin{figure}[t]
     \centering
    \includegraphics[width=\textwidth]{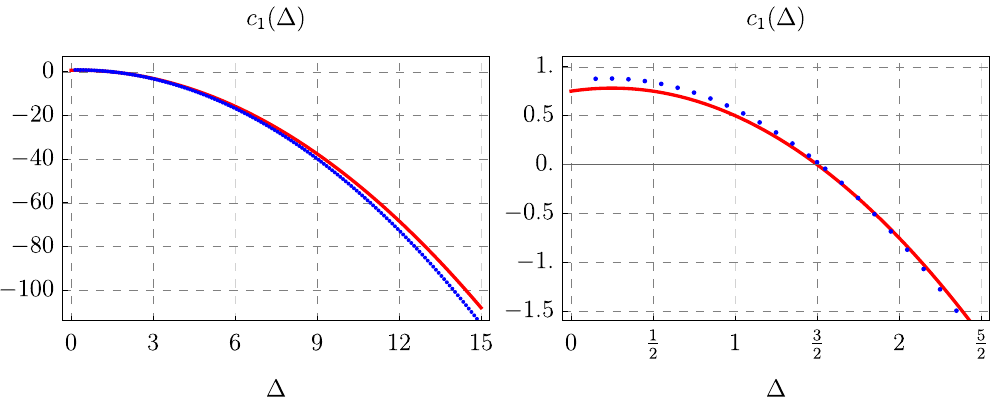}
    \vspace*{-0.7cm}
    \caption{The numerical values for $c_1(\Delta)$ (blue) compared with the data predicted by the geodesic analysis (red). On the right, the close-up shows that even at small $\Delta$ the geodesic results seems to match the data to a relatively high degree. This suggests that even in the full correlator, there is no term linear in $\delta\tau=\tau_c-\tau$.}
    \label{fig:c1s}
\end{figure}

\begin{figure}[t]
    \centering
    \includegraphics[width=\textwidth]{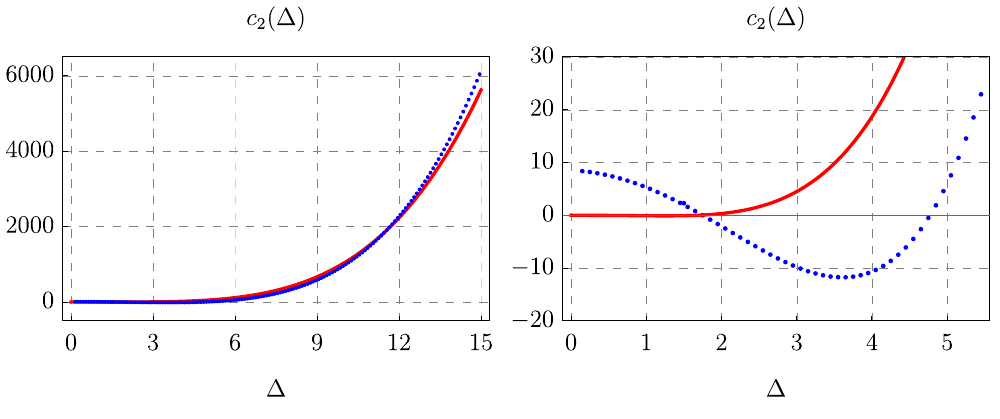}
        \vspace*{-0.7cm}
    \caption{The numerical values for $c_2(\Delta)$ (blue) compared with the data predicted by the geodesic analysis (red). On the right, the close-up shows that at small $\Delta$ the geodesic prediction disagrees with the numerical data. This is expected, since there is no reason for the geodesic result to be applicable outside the regime where $\Delta\gg 1$.}
    \label{fig:c2s}
\end{figure}

We find that for large enough $n$ the OPE coefficients can be described by
\begin{equation}\label{eq:rybana}
    \Lambda_n = c(\Delta)\frac{n^{2\Delta-3}}{\left(\frac{1}{\sqrt{2}}\right)^{4n}e^{i\pi \,n}}\,\sum_{k=0}^\infty\frac{c_k(\Delta)}{n^k}\ ,
\end{equation}
where we choose $c_0(\Delta)=1$, so that this expression is in accord with the dominant contribution \eqref{eq:Lambdand=4Ansatz} used in the main text. 
Let us here briefly explain how we obtained this expression. 
First, one draws inspiration from the two-dimensional analysis \eqref{eq:LambdaNd=2} and considers an ansatz for the dominant contribution $\Lambda_n^a = e^{-i\pi\,n} \widetilde k^n\,c(\Delta)n^{a\,\Delta+b}$. The first factor comes from the observed oscillating sign of $\Lambda_n$, while $\widetilde k$, $a$, and $b$ are constants and $c(\Delta)$ is a function which all need to be determined by our analysis. The values of the three constants can be determined by analysing different ratios of $\Lambda_n$ as $n$ and $\Delta$ are varied. The expressions for $c(\Delta)$ and $c_k(\Delta)$ are then obtained numerically by analysing $\Lambda_n$  as $n$ is varied for fixed values of $\Delta$. One first assumes the $1/n$ expansion \eqref{eq:rybana} and compares it with the values of $\Lambda_n$ as a function of $n$. This allows us to read off the coefficients $c_k(\Delta)$. This procedure gives, for example, the values given in the blue markers in Figure~\ref{fig:c(Delta)}, Figure~\ref{fig:c1s}, and Figure~\ref{fig:c2s}. 

Let us note that the accuracy of our analysis is limited by the number of $\Lambda_n$ we can calculate: The larger the $n$, the more accurate the asymptotic form \eqref{eq:rybana} will be. A higher $n$ also means that we can include, in practice, a higher number of $1/n$ corrections, which in turn allow for a more accurate values of $c_k(\Delta)$. An estimate of the value of $n$ necessary for an accurate determination of $\Lambda_n$ for a given $\Delta$ is given at the end of this appendix.

The $1/n$ terms in \eqref{eq:rybana} are mapped to the ($\tau_c - \tau$) corrections of the correlator. To see this, we insert the full $1/n$ expansion of $\Lambda_n$ into \eqref{e.resumL} and replace the sum with an integral
\begin{align}
    \label{eq:CorrFullSum}
    G_T(\tau)&\approx
    \frac{1}{\tau^{2\Delta}}\int_0^\infty\Lambda_n\left(\frac{\tau}{\beta}\right)^{4n}d n =
    \frac{c(\Delta)}{\tau^{2\Delta}}\int_0^\infty n^{2\Delta-3}\left(\frac{\tau}{\tau_c}\right)^{4n}\sum_{k=0}^\infty\frac{c_k(\Delta)}{n^k}\,d n\nonumber\\
    &= \frac{c(\Delta)}{\tau^{2\Delta}}\sum_{k=0}^\infty c_k(\Delta)\,\Gamma(2\Delta-2-k)\,\left[-\log\left(\frac{\tau^4}{\tau_c^4}\right)\right]^{-(2\Delta-2-k)}\,,
\end{align}
where $\tau_c$ is schematically one of the critical points defined in \eqref{e.poles}.
We see that near $\tau\approx \tau_c$ we get a sum of diverging terms with a pole of order $2\Delta-2-k$: The higher the $k$, the milder the singularity.

In four dimensions, we lack an exact expression for the correlator (or the stress-tensor sector of the correlator), so we cannot determine $c_k(\Delta)$ in a similar manner as in the two-dimensional case. 
However, we can take the large-$\Delta$ limit and compare these results with the expectation from the geodesic analysis. In particular, we know that the correlator obtained from the bouncing geodesic expanded around the lightcone singularity receives the first correction to the leading result at fourth order \eqref{eq:GeoCOrrSmallDeltaTauExp}.
Therefore, we  insert \eqref{eq:CorrFullSum} into \eqref{eq:LogOfCorr} and  expand in  $\delta\tau = \tau_c-\tau$
\begin{align}
    \label{eq:LogCorrNearPoleExp}
     \cL_T \approx -\frac{1}{\Delta}\log\left[\frac{c(\Delta)\,\Gamma(2\Delta-2)}{4^{(2\Delta-2)}}\,\frac{1} {\tau_c^{2}}\right]+ \frac{2\Delta-2}{\Delta}\log\delta\tau+\sum_{k=1}^{\infty}\gamma_k(\Delta)\left(\frac{\delta\tau}{\tau_c}\right)^k\,,
\end{align}
where $\gamma_k(\Delta)$ are non-trivial combinations of $c_k(\Delta)$.
Let us check if the $\Delta\to \infty$ limit reproduces the geodesic result \eqref{eq:PropLenDeltat}, which means that 
\begin{equation}\label{e.geodude}
    \gamma_1(\Delta) = 0\,, \qquad  \gamma_2(\Delta) = 0\,,\qquad  \gamma_3(\Delta) = 0\,,\qquad \gamma_4(\Delta) = \frac{\pi^4}{160}\,,\quad \ldots\,,
\end{equation}
up to $1/\Delta$ corrections. We only focus on $\gamma_1(\Delta)$ and $\gamma_2(\Delta)$. The former is given by
\begin{align}
\label{eq:Gamma1}
    \gamma_1(\Delta) = 1+ \frac{1}{\Delta}+ \frac{4\,c_1(\Delta)}{\Delta(2\Delta-3)}\,,
\end{align}
which can be expanded in large-$\Delta$ to first few orders
\begin{align}
\label{eq:Gamma1Exp}
    \gamma_1(\Delta) = 1 + \frac{1}{\Delta} + \frac{2\,c_1(\Delta)}{\Delta^2} + \frac{3\,c_1(\Delta)}{\Delta^3}+\coo{\Delta^{-4}}\,.
\end{align}
For this expression to be consistent with the bouncing geodesic, the leading behaviour of $c_1(\Delta)$ has to be given by 
\begin{equation}
\label{eq:c1exp}
    c_1(\Delta) \sim -\frac{\Delta^2}{2} + \ldots\,,
\end{equation}
which cancels out the leading order term in \eqref{eq:Gamma1Exp}. However, we can be slightly bolder and assume that $\gamma_1(\Delta) =0$ for all $\Delta$, which ensures absence of a linear term in the proper length and the correlator. In this case, we can solve \eqref{eq:Gamma1} directly and find
\begin{align}
\label{eq:c1}
    c_1(\Delta) &= -\frac{1}{4} (\Delta +1) (2 \Delta -3)\,,
\end{align}
whose large $\Delta$ behaviour agrees with \eqref{eq:c1exp}. In Figure~\ref{fig:c1s}, we plot the numerical data (in blue) against this prediction (in red). We find good agreement not only at large $\Delta$, but for all values which strongly suggests that $c_1(\Delta)$ is given by \eqref{eq:c1}.
The expression for $\gamma_2(\Delta)$ is more complicated
\begin{subequations}
    \begin{align}
        \gamma_2(\Delta) &= -\frac{7}{12} - \frac{5}{12\,\Delta} - \frac{2\,c_1(\Delta)}{\Delta(2\Delta-3)}+ \frac{8\,c_1(\Delta)^2}{\Delta(2\Delta-3)^2}- \frac{8\,c_2(\Delta)}{\Delta(6-7\Delta+2\Delta^2)}\label{eq:Gamma2}\\
        & \approx -\frac{7}{12} - \frac{5}{12\,\Delta} - \frac{c_1(\Delta)}{\Delta^2} -\frac{4\,c_2(\Delta)- 2c_1(\Delta)^2 +\frac32\,c_1(\Delta)}{\Delta^3}+ \coo{\Delta^{-4}}\,.
        \label{eq:Gamma2exp}
    \end{align}
\end{subequations}
Above we have not yet inserted the value of $c_1(\Delta)$ because it will help us illustrate an important point as to why higher $c_k(\Delta)$ are inaccessible. As seen in \eqref{eq:c1exp}, even  in the most agnostic estimate $c_1(\Delta)$ scales at most quadratically at large values of $\Delta$. Therefore, the $c_1(\Delta)^2$ factor in the fourth term will contribute at order $\Delta$ in \eqref{eq:Gamma2exp}.%
\footnote{We see from \eqref{eq:Gamma2} that $c_1(\Delta)$ appears at most quadratically in $\gamma_2$ and thus no higher order term in the $1/\Delta$ expansion will contribute at order $\Delta$ when the value of $c_1(\Delta)$ is inserted.} Such term is incompatible with a smooth large-$\Delta$ limit, independent of whether the geodesic is the bouncing geodesic or a non-singular complex geodesic.
It therefore needs to be cancelled by the leading order term in $c_2(\Delta)$, which in this case gives
\begin{align}
    c_2(\Delta) \sim \frac{\Delta^4}{8}+ \ldots.
\end{align}
It needs to be stressed that this behaviour does not contain any information about whether the correlator is given by the bouncing geodesics or by the combination of two complex geodesics. This is contained in the subleading behaviour of $c_2(\Delta)$, which can be determined uniquely only if one knows the subleading behaviour of $c_1(\Delta)$.
To illustrate this, let us introduce an expansion
    \begin{equation}
        c_1(\Delta) = c_1^{(2)}\,\Delta^2 + c_1^{(1)}\,\Delta + c_1^{(0)} + \ldots\,,\quad 
        c_2(\Delta) = c_2^{(4)}\,\Delta^4 + c_2^{(3)}\,\Delta^3 + c_2^{(2)}\,\Delta^2 + \ldots\,,
    \end{equation}
and insert it into \eqref{eq:Gamma2exp}
\begin{align}
    \gamma_2(\Delta) &= 2\Delta\left[\left(c_1^{(2)}\right)^2-2\,c_2^{(4)}\right] \\*
    &\quad + \frac{1}{12}\left[-7 - 12\,c_1^{(2)}+ 48 \,c_1^{(1)}\,c_1^{(2)}+ 72\,\left(c_1^{(2)}- 48 c_2^{(3)}- 168 c_2^{(4)}\right)^2\right] + \coo{\frac1{\Delta}}\,.\nonumber
\end{align}
The term that scales with $\Delta$ can be determined as explained above. But the term that does not scale with $\Delta$, and gives a finite contribution in the large-$\Delta$ limit, depends on the subleading terms of both $c_1(\Delta)$ and $c_2(\Delta)$. Since we believe that $c_1(\Delta)$ can be determined exactly \eqref{eq:c1}, one can determine $c_2^{(3)}$ and write
\begin{align}
    c_2(\Delta) = \frac{\Delta^4}{8}- \frac{5\,\Delta^3}{24} + \coo{\Delta^2}\,,
\end{align}
but other, lower order terms are out of reach, since they are not constrained by the geodesic result. We can compare this result with the data obtained from solving the bulk equations of motion, which is pictured in Figure~\ref{fig:c2s}. We see that while the asymptotic behaviours match, we cannot say anything about the behaviour of $c_2(\Delta)$ for small values of $\Delta$. In fact, one can directly solve \eqref{eq:Gamma2} and determine $c_2(\Delta)$ using \eqref{eq:c1} in the process. Even in this case one find significant deviation from the data obtained from the holographic calculation.

Of course, the  geodesic analysis is expected to hold only at large values of the conformal dimensions. We observed this already in Section~\ref{ss:NCLim}, where the geodesic analysis failed to reproduce all $1/\Delta$ corrections at higher orders of the $\tau =0$ OPE. However, $c_1(\Delta)$ predicted from the analysis of the bouncing geodesic is in good agreement with the holographic data for all values of $\Delta$. This suggests that the linear correction to the singularity, $\gamma_1(\Delta)$, vanishes for all conformal dimensions and not just in the large-$\Delta$ limit. This is generically not true for all other $\gamma_k(\Delta)$, starting with $\gamma_2(\Delta)$, and suggest that for finite values of $\Delta$, the contribution to the correlator from the stress-tensor sector near the bouncing singularity goes as
    \begin{align}
    \label{eq:CorrNearPoleExp}
    G_T(\tau \approx \tau_c) \sim \frac{c(\Delta)\,\Gamma(2\Delta-2)}{2^{2(2\Delta-2)}}\,\frac{1} {\tau_c^{2}}\frac{1}{\left(\tau_c-\tau\right)^{2\Delta-2}}\left[1+ \gamma_2(\Delta)\left(\frac{\tau_c-\tau}{\tau_c}\right)^2+\ldots\right]\,,
\end{align}
where $\gamma_2(\Delta)$ is some non-trivial function of $\Delta$.

Higher order $c_k(\Delta)$ cannot be uniquely determined because we are unable to determine $c_2(\Delta)$ to all orders -- the reasoning is the same as with the subleading order in $c_2(\Delta)$ only being accessible if we know $c_1(\Delta)$ to subleading order.
One is only  able to determine that the leading behaviour at large $\Delta$ is
\begin{align}
\label{eq:ckLargeDelta}
    c_k(\Delta)\sim \frac{(-1)^k\,\Delta^{2k}}{2^{k}\,k!}\,.
\end{align}
One can show that this behaviour is compatible with the holographic data obtained from solving the bulk equations of motion.
However, as already discussed above, \eqref{eq:ckLargeDelta} does not  contain any information about the specific geodesic that contributes to the correlator -- this information is encoded in the subleading terms which we are unable to access with out current precision.

The large-$\Delta$ behaviour \eqref{eq:ckLargeDelta} serves as a useful rough estimate for the amount of terms in the $1/n$ expansion of $\Lambda_n$ that give a considerable contribution at a fixed value of $\Delta$. Let us assume that we can determine the $\Lambda_n$ coefficients up to some number $n_{\rm max}$. At this point, the $1/n$ terms in \eqref{eq:rybana} scale roughly as $\Delta^2/n_{\rm max}$. For $n_{\rm max} \lesssim \Delta^2$ the apparent $1/n$ expansion will not look convergent, since successive terms in the expansion will increase. In other words, the analysis can only be trusted for $\Delta\lesssim\sqrt{n_{\rm max}}$. In practice, we are able to reach $n_{\rm max} \approx 50$, which would suggest that we can fully trust the results up to $\Delta \approx 7$. It would thus be important if the method of obtaining $\Lambda_n$ could be optimised so that $n_{\rm max}$ would be increased.

The main conceptual goal of this appendix was to check whether taking the large-$\Delta$ limit of the stress-tensor sector \textit{after} we expanded the
correlator near the bouncing singularity  reproduces the results predicted by the bouncing geodesic. In particular, we focused on the subleading behaviour in $\delta \tau$. While we found no obvious disagreement, we have also not found any conclusive evidence that confirms a relation. Therefore, a more thorough analysis is needed to establish a definite connection between these two results.


\section{Analysis at non-zero \texorpdfstring{$x$}{x} in \texorpdfstring{$d=4$}{d=4}}
\label{app:XCorr}

In this appendix we consider thermal correlation functions of scalar operators where the operators are inserted at a finite spatial distance, $x\neq 0$. We begin by discussing spacelike geodesics connecting the two insertion points and the effect of non-zero spatial distance on the bouncing geodesics. We then perform the OPE analysis for the stress-tensor sector at $x\neq 0$, expand the logarithm of the correlator near the bouncing singularity and \textit{then} take the large-$\Delta$ limit. We find that due to a slower convergence, even the leading correction to the bouncing singularity at non-vanishing $x$ cannot be conclusively matched between the geodesic and stress-tensor OPE results. We conclude that a more detailed analysis is needed to make a definite statement. 

\subsection{Semi-classical analysis}
\label{ss:SCApp}

In contrast to the main part of the text, we work here directly in the Lorentzian signature and consider spacelike geodesics  in \eqref{eq:d=4BlackBranemet}. Let us parameterise the geodesics as $(t(s), r(s), x(s))$, where $s\in \mathbb{R}$ is the affine parameter. We can introduce conserved charges 
\begin{align}
\label{eq:ConservedQuantities}
    E~ \equiv~ r^2\,f(r) \, \dot t\,, \qquad P_i ~\equiv~ r^2\,\dot x_i\,,
\end{align}
where the dot denotes the derivative with respect to the affine parameter, which reduce finding the geodesics to a one-dimensional problem. We refer to $P_i$ as the momentum and $E$ as the (imaginary) energy, though we will omit the imaginary adjective in this appendix. The signs in \eqref{eq:ConservedQuantities} are chosen in such a way that in Patch I of the complexified spacetime (Figure~\ref{fig:Penrose}) the time increases for $E>0$. Finally, we use the isometry of $\mathbb{R}^3$ and rotate the spacetime such that only one component of the momentum is non-vanishing, for example $P_1 \equiv P$. 
Generalising to having all momenta non-trivial is straightforward.

\begin{figure}
        \centering
        \includegraphics[width = 0.7\textwidth]{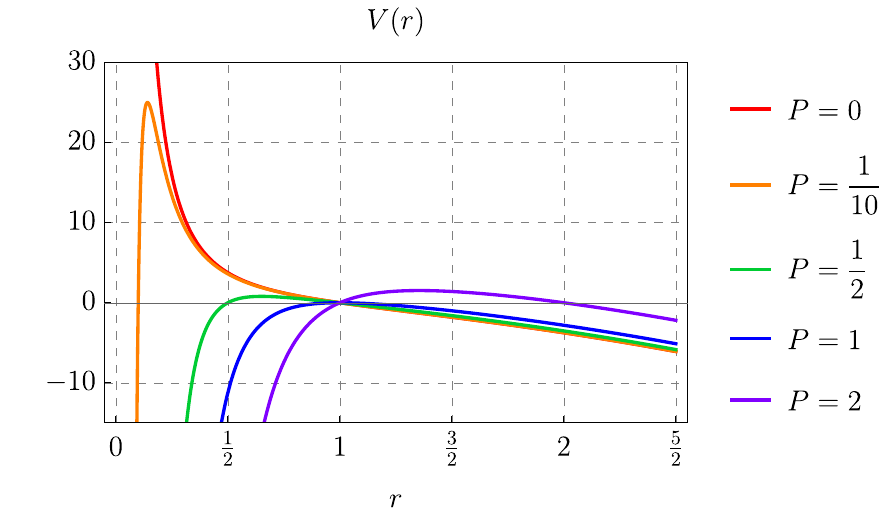}
        \caption{The potential $V(r)$ for different values of momentum $P$. When $P>1$, the turning points are always outside the horizon and such geodesics do not probe the singularity. When $0<P<1$, the potential has a maximum inside the horizon. In our analysis, we consider the case where $P$ (and thus $x$) is the smallest parameter of the problem and analyse perturbative corrections to the leading bouncing singularity due to non-vanishing spatial separation. }
        \label{fig:Potential1}
    \end{figure}

Spacelike geodesics in \eqref{eq:d=4BlackBranemet} have to satisfy the local constraint
\begin{align}
    \label{eq:SpacelikeConstraint}
 -r^2\,f(r) \, \dot t^2 + \frac{\dot r^2}{r^2\,f(r)} + r^2 \, \dot x^2 = 1\,,
\end{align}
which can be rewritten, using the expressions for the conserved charges, as
\begin{align}
    \label{eq:SpacelikeScattering}
    \dot r^2 -r^2\, f(r)\left(1 -\frac{P^2}{r^2}\right)  = E^2\,.
\end{align}
This effectively reduces the problem to classical scattering of a point particle in the potential (see Figure~\ref{fig:Potential1})
\begin{align}
\label{eq:ScatteringPotential}
    V(r) = -r^2\, f(r)\left(1 -\frac{P^2}{r^2}\right)\,.
\end{align}
The nature of the geodesics depends on the values of the conserved charges, which can  be seen from the turning point determined by the solution of $E^2 = V(r)$.
When $P=0$, the potential is monotonically decreasing and goes to positive infinity at $r\to 0$. As $E\to \pm\infty$, we get close to the singularity -- these are the bouncing geodesics discussed in the main text and in \cite{Fidkowski:2003nf, Festuccia:2005pi}. When $P\neq 0$, we have to distinguish between $P>1$ and $P<1$ regimes, as was already pointed out in \cite{Fidkowski:2003nf}. For $P>1$, there exist one real turning point, which is always outside the horizon. Such geodesics never probe the region near the singularity and will not be of interest in our analysis. For $P<1$, we can see that the behaviour near the origin changes -- the potential now reaches a maximal value inside the horizon before going to negative infinity. While this drastically alters the strict $|E|\to\infty$ limit, the behaviour does not change if we consider the regime where $P\,E \ll 1$. Namely, this regime is where we can think of spatial displacement $x$ (which is, as we will show, linearly related to the momentum $P$) as the smallest scale in the problem and treat it as an expansion parameter.%
\footnote{Another justification as to why this is a valid expansion comes from the WKB analysis of the correlator \cite{Festuccia:2005pi}. There it was shown that the Fourier conjugate to the spatial distance $x$ is actually $Q= -i\,P$. For purely imaginary momentum,  the potential is monotonically decreasing as a function of $r$. In that case the behaviour of bouncing geodesics does not change as we increase $x$.} As such, we can compare the small $x$ expansion on the geodesic side with the same expansion in the stress-tensor OPE \eqref{e.exnenula} -- in analogy of what was done in the two-dimensional example discussed in Appendix~\ref{s.kms2d}.

We want to calculate its regularised proper length and express it in terms of time and position displacements. The latter are given by
\begin{subequations}
\label{eq:Displacements}
    \begin{align}
        \label{eq:deltat}
           t &\equiv t_f - t_i = 2\int_{r_t}^{\infty}\frac{E\,dr}{r^2\,f(r)\sqrt{E^2 +r^2\, f(r) \left(1-\frac{P^2}{r^2}\right)}}\,,
        \\
        \label{eq:deltax}
        x &= x_f - x_i = 2\int_{r_t}^{\infty}\frac{P\,dr}{r^2\,\sqrt{E^2 + r^2\,f(r) \left(1-\frac{P^2}{r^2}\right)}}\,.
    \end{align} 
\end{subequations}
whereas the proper length integral is
\begin{align}
\label{eq:LengthInt}
    \cL = 2\int_{r_t}^{r_{\rm max}}\,\frac{dr}{\sqrt{E^2 + r^2\,f(r) \left(1-\frac{P^2}{r^2}\right)}}\,.
\end{align}
In the above, $r_t$ denotes the turning point of the geodesic, which is the largest real root of $E^2 = V(r)$, and we have used a cut-off parameter $r_{\rm max}$ that will help us to regularise the integral. 

For  $P\neq 0$ these integrals can be evaluated in terms of incomplete elliptic functions \cite{festucciathesis}. To express the solutions, we first define $q \equiv r^2$ and note that the turning point equation is a cubic equation in $q$. Let us denote the three, in general complex, solutions as $q_1$, $q_2$, and $q_3$, where $q_1$ corresponds to the largest real root and is identified with the turning point. Then one can show that the integrals \eqref{eq:Displacements} become
\begin{subequations}
\label{eq:EllipticalDisplacements}
    \begin{align}
         t &= \frac{E}{\sqrt{q_1(q_3-q_2)}}\left[\frac{(q_3-q_1)}{(q_1-1)(q_3-1)}\Pi\left(\frac{c}{a}, \phi, s\right)\right.\nonumber\\*
    &\left.\quad - \frac{(q_3-q_1)}{(q_1+1)(q_3+1)}\Pi\left(\frac{\bar c}{a}, \phi, s\right)+ \frac{2q_3^2}{(q_3-1)(q_3+1)}F\left(\phi, s\right)\right]\,,
        \\
         x &= \frac{2\,P}{\sqrt{q_1\left(q_3-q_2\right)}}\, F(\phi, s)\,,
    \end{align}
\end{subequations}
where $F(\phi,s)$ and $\Pi(z,\phi, s)$ are incomplete elliptic integrals of the first and third kind respectively, and we have used
\begin{gather}
\label{eq:Parameters}
    a = \frac{q_3-q_2}{q_1-q_2}\,,
    \qquad c = \frac{q_3-1}{q_1-1}\,,
    \qquad \bar c = \frac{q_3+1}{q_1+1}\,, 
    \quad s = \frac{q_3}{q_1\,a}\,,\qquad 
    %
    \phi = \arcsin{\sqrt{a}}\,.
\end{gather}
Finally, the proper length integral is given by
\begin{align}
    \label{eq:ProperLengthElliptic}
    \cL = \lim_{r_{\rm max}\to \infty}\Bigg\{2\sqrt{\frac{q_1}{q_3-q_2}}\,\left[s\,a\,F(\tilde \phi, s) + (1-s\,a) \Pi\left(\frac1{a},\tilde \phi, s\right)\right]- 2\,\log r_{\rm max} \Bigg\}\,,
\end{align}
where
\begin{align}
    \tilde \phi \equiv\arcsin\left( \sqrt{a\frac{r_{\rm max}^2- q_1}{r_{\rm max}^2-q_3}}\right)\,.
\end{align}
One can check that when $P=0$, these expressions reduce to \eqref{eq:LorentzianIntegrals}.

We now use these expressions to calculate the leading $x$ correction to the correlator near the bouncing singularity. Our method is to expand the expressions in $P$ and take the $E\to \infty$ limit order by order. For example, at first order in $P$, the turning point \eqref{eq:P=0Turning} is given by
\begin{align}
    q_1 &= \frac12\left(\sqrt{4 + E^4}- E^2\right) - P^2\,\frac{E^2}{\sqrt{4+ E^2}}\,,+ \coo{P^4}\,,
\end{align}
which, when taking $E\to \infty$ at each term separately, gives
\begin{align}
    q_1 &= \frac{1}{E^2}\Big[1- \varepsilon^2 - \coo{\varepsilon^4}\Big]\,.
\end{align}
In the above, we have used
\begin{align}
\label{eq:EpsilonParameter}
    \varepsilon \equiv P\,E\,,
\end{align}
which is a convenient expansion parameter in this double limit, since this calculation only gives sensible results if $ x$ (or $P$ in this instance) is the smallest parameter in the expansion. As such, $P \ll 1/E$ or equivalently $\varepsilon = P\,E \ll 1$, which is why this combination appears naturally in this limit.

One can then expand \eqref{eq:EllipticalDisplacements} in $\varepsilon$  and express the conserved charges as
\begin{align}
    \label{eq:EllipticalDisplacementsExpansion}
     E = \frac{2}{\delta t}\,, \qquad \varepsilon = \frac{2}{\pi}\,\frac{ x}{\delta t}\,,
\end{align}
where $\delta t = -i\delta \tau =  -t- i\tau_c$. The proper length integral expanded in $\varepsilon$ is
\begin{align}
\label{eq:cLExp}
    \cL \approx 2\log2 -2\log E + \frac{\pi}{2E^2}\,\varepsilon^2 \,,
\end{align}
which, when expressed in terms of position space coordinates, gives
\begin{align}
\label{eq:SummaryProperLength}
    \cL = 2 \log(\delta t) +\frac{ x^2}{2\pi}\,.
\end{align}
In a saddle point approximation, the contribution from such a geodesic to the correlator would be, to leading order in $x$
\begin{align}
\label{eq:SummaryCorrelator1}
    e^{-\Delta \,\cL} \sim \frac{1}{(\delta t)^{2\Delta}}\,\left(1-\frac{\Delta}{2\pi}\, x^2\right) = \frac{1}{(\delta t)^{2\Delta}}\,\left(1-\frac{\Delta\,\pi}{2\,\beta^2}\, x^2\right)\,,
\end{align}
where in the last expression we reinstated the inverse temperature. This is the result that we will compare to the OPE analysis.

\subsection{OPE analysis}
\label{ss:OPENonZeroX}

Let us now discuss the CFT side of the $x\neq0$ story. As discussed in Appendix~\ref{a.cbd}, for $0<x\ll\tau$ the decomposition of the correlator generalises to
        \begin{equation}\label{e.xxcoco}
        G(\tau,x)=\frac{1}{\tau^{2\Delta}}\sum_{n=0}^\infty\left[\Lambda^{(0)}_n+\frac{x^2}{\tau^2}\Lambda^{(1)}_n+\coo{\frac{x^4}{\tau^4}}\right]\left(\frac{\tau}{\beta}\right)^{4n}\ ,
    \end{equation}
where $\Lambda_n^{(0)}\equiv\Lambda_n$ is the contribution at $x=0$ that we examined in the main text. The coefficients $\Lambda^{(1)}$  can be further decomposed as
\begin{align}
\label{eq:Lambda1Decomp}
    \Lambda^{(1)}_n\equiv(2n-\Delta)\,\Lambda_n+\widetilde{\Lambda}^{(1)}_n\,,
\end{align}
where $\wlo$ are given by \eqref{e.dufamzeposledna} and can thus be determined using the same holographic calculation as $\Lambda_n^{(0)}$. Since the latter have already been analysed, we focus here on the asymptotic behaviour of $\wlo$. We find that this data is well described by 
\begin{align}
\label{eq:tLambda1Exp}
    \wlo  = c(\Delta)\, \frac{n^{2\Delta -2}}{\left(\frac{1}{\sqrt{2}}\right)^{4n}e^{i\pi\,n}}\sum_{k=0}^{\infty}\frac{d_k(\Delta)}{n^k}\,.
\end{align}
The most important difference compared to \eqref{eq:Lambdand=4Ansatz} is the power of the factor of $n$. As we will see, this has significant effects on the degree of the singularity at first non-trivial correction in $x$ near the bouncing singularity. The prefactor $c(\Delta)$ is chosen to be the same as in $\Lambda_n^{(0)}$, in which case $d_0(\Delta)$ is not necessarily equal to 1. Let us note that the convergence of the numerical data obtained from solving the bulk equation of motion to \eqref{eq:tLambda1Exp} is slower compared to the convergence in the $x=0$ case. We can trace this back to more prefactors in the expression for $\wlo$ \eqref{e.dufamzeposledna} compared to $\Lambda_n^{(0)}$  \eqref{e.lebara}. As such, the results in this appendix are less reliable. Nonetheless, one is able to extract some information even with $n\approx 50$ data points at each value of $\Delta$.

\begin{figure}
\centering
    \includegraphics[scale=0.8]{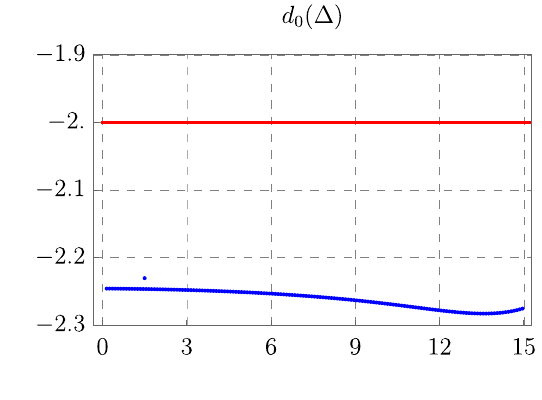}
    \vspace*{-0.8cm}
    \caption{The geodesic prediction for $d_0(\Delta)$ (red) compared to the numerical data obtained using the holographic method (blue). We see that there is significant mismatch between the two results. We relate this to the slower convergence of the OPE data to the asymptotic form.}
    \label{fig:new}
\end{figure}

Combining \eqref{eq:tLambda1Exp} with the expansion for $\Lambda^{(0)}_n$ gives
\begin{align}
    \Lambda_n^{(1)} =  c(\Delta)\, \frac{n^{2\Delta -2}}{\left(\frac{1}{\sqrt{2}}\,e^{\frac{i\pi}{4}}\right)^{4n}}\sum_{k=0}^{\infty}\frac{d_k(\Delta) + 2\,c_k(\Delta) - \Delta\,c_{k-1}(\Delta)}{n^k}\,,
\end{align}
where we define $c_{-1}(\Delta) = 0$. This, combined with the asymptotic form of $\Lambda^{(0)}_n$, can be inserted into the correlator expansion
{\allowdisplaybreaks{\small{
    \begin{align}
 G_T(\tau, x)&\approx \frac{c(\Delta)}{\tau^{2\Delta}}\int_0^{\infty}\!\!\!dn\left[n^{2\Delta-3} + \frac{x^2}{\tau^2}\,n^{2\Delta-2}\left(d_0(\Delta)+2 + \frac{d_1(\Delta)+ 2c_1(\Delta) -\Delta}{n}\right)\right]\left(\frac{\tau^4}{\tau_c^4}\right)^{n}\nonumber\\
 &= \frac{c(\Delta)\,\Gamma(2\Delta-2)}{\tau^{2\Delta}} \left[-\log\left(\frac{\tau^4}{\tau_c^4}\right)\right]^{2-2\Delta}\\
 &\quad\qquad  \times\Bigg\lbrace1 + \frac{x^2}{\tau^2}\bigg[
 \frac{(2\Delta-1)(d_0(\Delta)+2)}{-\log\left(\frac{\tau^4}{\tau_c^4}\right)}
 + d_1(\Delta)+ 2c_1(\Delta) -\Delta\bigg] \nonumber\Bigg\rbrace\label{eq:Gtx1}\,,
\end{align}
}}}%
where we have again replaced the sum with an integral, inserted 
$\tau_c$ using \eqref{e.poles}, and used that $c_0(\Delta)=1$. In this expression we work only at leading order in $n$ in the $x^0$ term while keeping the subleading contribution in the $x^2$ term. We will shortly show why the subleading correction is important. We can now expand the correlator around $\tau \approx \tau_c$, which gives the bouncing singularity together with several corrections
\begin{equation}
    G(\tau, x)\approx 
    \frac{c(\Delta)\,\Gamma(2\Delta-2)}{4^{2\Delta-2}}\,\frac{1} {\tau_c^{2}\,\delta\tau^{2\Delta-2}} \Bigg[1  + \frac{x^2}{\tau_c^2} \tilde \gamma_1(\Delta)+ \frac{x^2}{\tau_c\,\delta\tau}\,\frac{(2\Delta-1)(2+d_0(\Delta))}{4}\Bigg]\,,
\end{equation}
where we defined
\begin{align}
    \tilde \gamma_1(\Delta) \equiv d_1(\Delta) +\frac{-5 + 8\Delta+ 4 \Delta^2}{8}\,d_0(\Delta) + 2\,c_1(\Delta) - \frac54+ \Delta + \Delta^2\,.
\end{align}
To compare this analysis with the proper length of the bouncing geodesic we use \eqref{eq:LogOfCorr} and take the large-$\Delta$ limit
\begin{align}
\label{eq:Geodesicx2}
    \lim_{\Delta \to \infty}\cL_T &= 2 \log \delta\tau + \frac{x^2}{\tau_c^2}\Bigg[\frac{3-2\Delta}{4\Delta} - \frac{d_1(\Delta)}{\Delta}- \frac{2c_1(\Delta)}{\Delta}- \frac{3(2\Delta-1)d_0(\Delta)}{8\Delta}\Bigg]\nonumber\\
    & \quad +\frac{x^2}{\tau_c\,\delta\tau}\,\frac{(2\Delta-1)(2+d_0(\Delta))}{4\Delta} \,,
\end{align}
where it should be understood that one needs to take $\Delta\to \infty$ in all terms in this expression. We immediately notice that in contrast to \eqref{eq:SummaryProperLength}, the leading correction at $x^2$ comes at order $1/\delta\tau$. For the two expressions to be compatible, we would have to find
\begin{align}
\label{eq:d0value}
    d_0(\Delta) = - 2 + \coo{\frac{1}{\Delta}}\,.
\end{align}
We compare this prediction with the numerical data coming from solving the bulk equations of motion in Figure~\ref{fig:new}.
We note that in the range $0\lesssim \Delta \lesssim15$, the numerical data is roughly constant, but misses the value predicted by the geodesic analysis by about 10\%. 
While we believe that this difference is due to the slow convergence of data toward the asymptotic OPE ansatz \eqref{eq:tLambda1Exp} and that when the number of OPE coefficients calculated is increased the numerical data will converge to the geodesic result, we currently cannot show that this is the case. 

Since we cannot conclusively say something about the most divergent term at order $x^2$ in \eqref{eq:Geodesicx2}, we cannot compare the second term (with the square bracket) which has a counterpart in \eqref{eq:SummaryProperLength}.
However, we immediately run into an interesting problem: Note that in the geodesic result, the first correction is real, while in the result from the OPE analysis, the analogous correction is multiplied by $\tau_c^2$, which is imaginary. 
Not only that, this value differs based on which of the four poles was chosen to expand the correlator around. 
We currently do not have a good understanding of this discrepancy and leave this interesting avenue for future work. 
Let us also remark that from the CFT point of view having $x\neq 0$ does not significantly alter the physics, while the behaviour of the (real) bouncing geodesic changes \cite{Fidkowski:2003nf} . It would be interesting to understand whether the difference in the prefactors is related to this phenomenon. 


\section{Black hole singularity in \texorpdfstring{$d=6$}{d=6} and \texorpdfstring{$d=8$}{d=8}}\label{a.ope6d}

In this appendix we show that the stress-tensor sector in six and eight dimensional holographic CFTs also contains singularities at the critical values predicted by bouncing geodesics. In $d+1$ spacetime dimensions the value of $\tau_c$ is given by \cite{Fidkowski:2003nf}
\begin{equation}\label{tauc}
  \tau_c^{(d+1)}  = {\beta \over 2} \pm i {\beta \over 2}  \ {\cos {\pi\over d} \over \sin {\pi\over d}} =\pm i  {\beta e^{\mp {i \pi\over d}} \over 2 \sin {\pi\over d}}\ . 
\end{equation}
Below we reproduce $\tau_c$ from the asymptotic analysis of the stress tensor OPE in $d=6$ and $d=8$.

	\subsection{Singularity in six dimensions}\label{a.6dcase}

\begin{figure}[t]
\centering
\includegraphics[width = \textwidth]{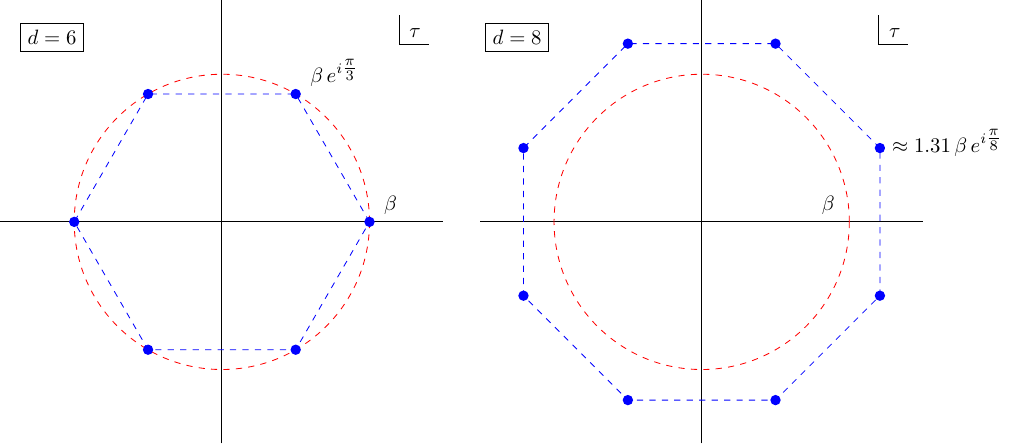}
\caption{Poles in the complex $\tau$ plane in $d=6$ and $d=8$.}
\label{fig:Polesd6d8}
\end{figure}

We start by solving the 7-dimensional scalar equations of motion in the black-hole background using the ansatz \eqref{e.theansatz}. By expanding the results and comparing them with the CPW expansion \eqref{e.resumL}, we extract the CFT data from the dual bulk theory. For example the first two OPE coefficients are
{\allowdisplaybreaks{\small{
    \begin{align}
    \Lambda_1&=\frac{8 \pi ^6 \Delta }{15309}\\
    \Lambda_2&=\frac{32 \pi ^{12} \Delta  \left(715 \Delta ^5-6930 \Delta ^4+17204 \Delta ^3-9323 \Delta ^2+26334 \Delta +9000\right)}{167571318915 (\Delta -6) (\Delta -5) (\Delta -4) (\Delta -3)}\,.
    \end{align}}}}%
Again we are interested in the large-$n$ behaviour of $\Lambda_n$.
We find that as $n$ grows, the OPE coefficients tend to
    \begin{equation}\label{e.pole6}
        \Lambda_n^a=j(\Delta)n^{2\Delta-4}\ ,
    \end{equation}
up to $1/n$ corrections which we do not here. In the above, $j(\Delta)$ is an undetermined function of the conformal dimension. An interesting difference compared to $d=4$ is the absence of an oscillating sign in $\Lambda_n^a$. Inserting these coefficients into the OPE and approximating the sum with an integral gives
    \begin{equation}\label{e.6sumint}
        G_T(\tau)=\frac{1}{\tau^{2\Delta}}\int_0^\infty\Lambda^a_n\left(\frac{\tau}{\beta}\right)^{6n}d n
        =\frac{j(\Delta)\,\Gamma(2\Delta-3)}{\tau^{2\Delta}}\left(-\log\left( \frac{\tau^6}{\beta^6}\right)\right)^{3-2\Delta}\,.
    \end{equation}
This expression has a singularity whenever the argument of the logarithm is equal to 1, which is precisely at (see left panel of Figure~\ref{fig:Polesd6d8})
	\begin{equation}\label{viva}
	\tau_c=\beta e^{i\frac{k\pi}{3}}\qq{for}k\in\mathbb{Z}\,.
	\end{equation}
Interestingly, we note that all singularities are located on a circle of radius $\beta$ and that two of these critical points coincide with the  positions where one expects KMS poles.
When comparing these critical values to the geodesic result \eqref{tauc}
    \begin{equation}
      \tau_c^{(7)}=\frac{\beta}{2}\pm i\frac{\beta}{2}\sqrt{3}=\beta e^{\pm i\frac{\pi}{3}}\ ,
    \end{equation}
we find a precise match.

	\subsection{Singularity in eight dimensions}\label{a.8dcase}

We follow the same approach in $d=8$. The first two coefficients are
{\small{
    \begin{align}
    \Lambda_1&=\frac{\pi ^8 \Delta }{184320}\\
    \Lambda_2&=\frac{\pi ^{16} \Delta}{1156266432921600 (\Delta -8) (\Delta
   -7) (\Delta -6) (\Delta -5) (\Delta -4)}\times \big(17017 \Delta ^6\nonumber\\
   &\phantom{=}-211666 \Delta ^5+681619 \Delta ^4-881554 \Delta ^3+3831472 \Delta ^2+2284352 \Delta +1317120\big)\,.
    \end{align}}}Overall
we find that the leading  asymptotic behaviour of the OPE coefficients is
	\begin{equation}\label{e.osemrozmerovLb}
	\Lambda^a_n=Q(\Delta)\frac{n^{2\Delta-5}}{W^{8n}\,e^{i\pi\,n}}\ ,
	\end{equation}
where $Q(\Delta)$ is a function of the conformal dimension and $W$ is a constant 
	\begin{equation}\label{e.Wval8d}
	W\approx1.3066\ .
	\end{equation}
We note that the $\Lambda_n^a$ have an oscillating sign,  similar to the four-dimensional case. Inserting them in the OPE and performing the integral gives
    \begin{equation}\label{e.8sumint}
        G_T(\tau)=\frac{1}{\tau^{2\Delta}}\int_0^\infty\Lambda^a_n\left(\frac{\tau}{\beta}\right)^{8\,n}d n
        =\frac{Q(\Delta)\,\Gamma(2\Delta-5)}{\tau^{2\Delta}}\left(-\log\left( \frac{\tau^8}{W\,\beta^8\,e^{i\pi}}\right)\right)^{3-2\Delta}\,.
    \end{equation}
$I.e.$ the stress-tensor sector has a singularity at 
	\begin{equation}
	\tau_c=\beta\, W\, e^{\frac{i\pi}{8}+k\frac{i\pi}{4}}\qq{for}k\in\mathbb{Z}\ ,
	\end{equation}
which are pictured on the right in Figure~\ref{fig:Polesd6d8}. This agrees with the geodesic analysis which predicts the singularities at
    \begin{equation}
       \tau_c^{(9)}=\frac{\beta}{2}\pm i\frac{\beta}{2}\cot{\frac{\pi}{8}}=\beta \frac{e^{\pm i\frac{3\pi}{8}}}{2\sin\frac{\pi}{8}}\,,
    \end{equation}
since
	\begin{equation}
	\frac{1}{2\sin\frac{\pi}{8}}\approx1.3066\approx W\ .
	\end{equation}
It is interesting to note that in $d=8$ these singularities are all located further away than where we would expect to find the KMS poles. This further strengthens the argument that the stress-tensor sector does not contain the information about the KMS poles and the double-traces are needed to recover their location.


   \section{Lowest-twist analysis}\label{app:LTanal}
   
In the main body of the paper we have examined the \textit{summed coefficients} $\Lambda_n$ defined by the expansion \eqref{e.topeL}. It is also interesting to study the individual coefficients $\lambda_{n,J'}$ corresponding to the contributions from multi-stress tensors with different spin $J'$ (respectively twist $\Delta'-J'$). We are mainly interested in the behaviour of the coefficients $\lambda_{n,2n}$ that correspond to the multi-stress tensors with the lowest twist. These coefficients are universal in holographic theories \cite{Fitzpatrick:2019zqz} and one can calculate them using bootstrap techniques (see $e.g.$ \cite{Karlsson:2019dbd}).

Alternatively, one can develop an effective method\footnote{This way one get access to the leading-twist OPE coefficients $\lambda_{n,2n}$ up to order $n\sim10000$.} to calculate $\lambda_{n,2n}$ by combining the bulk recursion relation found in \cite{Fitzpatrick:2019zqz}, with the fact that only one coefficient of the bulk ansatz contributes to the near-lightcone correlator at each order in $1/r$ expansion.\footnote{This is an analogous approach as the one used in the Appendix A of \cite{Esper:2023jeq}.} Such coefficient can be then mapped to the lowest-twist coefficient $\lambda_{n,2n}$. In practice we get this coefficient as
    \begin{equation}\label{e.baf}
        \lambda_{n,2n}=\left(-\frac{\pi^4}{4}\right)^{n}b_{n,n}\ .
    \end{equation}
where $b_{n,n}$ can be systematically calculated by solving the recursion relation \cite{Fitzpatrick:2019zqz}:
    \begin{equation}
        4(j-4i)b_{i,j}=\frac{4(1-j+i)(i-j-1+\Delta)}{j-\Delta}b_{i,j-1}-4(1+j-\Delta) b_{i-1,j+1}
    \end{equation}
where $i,j\in\mathbb{Z}$, $b_{i,j}=0$ for $j\notin[-i,i]$ and for all $i<0$, and $b_{0,0}=1$.

Analysing the coefficients $\lambda_{n,2n}$ we find their leading large-$n$ behaviour to be
    \begin{equation}\label{e.ltfo}
        \lambda_{n,2n}=A(\Delta)\frac{e^{i\pi n}n^{2\Delta-5/2}}{B^{4n}}\ ,
    \end{equation}
where $A(\Delta)$ is a function of the scaling dimension and value of the constant $B\approx0.8968...$.

To compare this with the asymptotic form of $\Lambda_n\sim n^{2\Delta-3}$, we consider the coefficient $\lambda_{n,2n}$ multiplied by $(1+J)$, yielding the behaviour $\sim n^{2\Delta-3/2}$ which grows faster than the overall summed coefficient $\Lambda_n$. Thus, terms with the subleading twists are crucial to ensure that $\Lambda_n$ has the correct overall scaling as
observed in the main part of the paper. It would be interesting to study this in more detail.


\bibliographystyle{JHEP}
\bibliography{bhope} 

\end{document}